\documentclass[reprint,superscriptaddress,pra,nofootinbib,preprintnumbers]{revtex4-2}

\usepackage{graphicx} 
\usepackage{amsfonts,amsmath,amssymb,amsthm,dsfont}

\usepackage{hyperref}
\hypersetup{colorlinks, linkcolor = [rgb]{0,0.0,0.75}, citecolor = [rgb]{0,0.0,0.75}, urlcolor = [rgb]{0,0.0,0.75}}

\usepackage[capitalise]{cleveref}
\crefname{section}{Sec.}{Secs.}
\crefname{appendix}{App.}{Apps.}
\usepackage{color}
\usepackage{placeins}
\usepackage{physics}
\usepackage{tikz}
\usepackage{makecell}
\usetikzlibrary{quantikz}
\usepackage{ulem}
\usepackage{array}
\newcolumntype{L}[1]{>{\raggedright\let\newline\\\arraybackslash\hspace{0pt}}m{#1}}
\newcolumntype{C}[1]{>{\centering\let\newline\\\arraybackslash\hspace{0pt}}m{#1}}
\newcolumntype{R}[1]{>{\raggedleft\let\newline\\\arraybackslash\hspace{0pt}}m{#1}}

\makeatletter 
    
\renewcommand\onecolumngrid{
\do@columngrid{one}{\@ne}%
\def\set@footnotewidth{\onecolumngrid}
\def\footnoterule{\kern-6pt\hrule width 1.5in\kern6pt}%
}

\renewcommand\twocolumngrid{
        \def\footnoterule{
        \dimen@\skip\footins\divide\dimen@\thr@@
        \kern-\dimen@\hrule width.5in\kern\dimen@}
        \do@columngrid{mlt}{\tw@}
}%

\makeatother    
\newcommand{\iu}{{i\mkern1mu}}
\newcommand{\me}{\mathrm{e}}
\newcommand {\unit} {\mathds{1}}

\newcommand {\sket} [1] {| #1 \rangle}
\newcommand {\bket} [1] {\bigl| #1 \bigr\rangle}
\newcommand {\sbraket} [2] {\langle #1 | #2 \rangle}
\newcommand {\saxe} [2] {| #1 \rangle\langle #2 |}

\newcommand{\fex}{\mathsf{f}}
\newcommand{\Fex}{\mathsf{F}}
\newcommand{\Hphys}{H^\text{phys}}
\newcommand{\Deltaphys}{\Delta^\text{phys}}

\begin{document}

\title{Nearly-optimal state preparation for quantum simulations of lattice gauge theories}

\author{Christopher F. Kane}
\affiliation{Department of Physics, University of Arizona, Tucson, AZ 85719, USA}

\author{Niladri Gomes}
\affiliation{Applied Mathematics and Computational Research Division, Lawrence Berkeley National Laboratory, Berkeley, California 94720, USA}

\author{Michael Kreshchuk}
\affiliation{Physics Division, Lawrence Berkeley National Laboratory, Berkeley, California 94720, USA}

\begin{abstract}
    We present several improvements to the recently developed ground state preparation algorithm based on the Quantum Eigenvalue Transformation for Unitary Matrices (QETU), apply this algorithm to a lattice formulation of U(1) gauge theory in 2+1D, as well as propose a novel application of QETU, a highly efficient preparation of Gaussian distributions.

    The QETU technique has been originally proposed as an algorithm for nearly-optimal ground state preparation and ground state energy estimation on early fault-tolerant devices.
    It uses the time-evolution input model, which can potentially overcome the large overall prefactor in the asymptotic gate cost arising in similar algorithms based on the Hamiltonian input model.
    We present modifications to the original QETU algorithm that significantly reduce the cost for the cases of both exact and Trotterized implementation of the time evolution circuit. 
    We use QETU to prepare the ground state of a U(1) lattice gauge theory in 2 spatial dimensions, explore the dependence of computational resources on the desired precision and system parameters, and discuss the applicability of our results to general lattice gauge theories.
    We also demonstrate how the QETU technique can be utilized for preparing Gaussian distributions and wave packets in a way which outperforms existing algorithms for as little as $n_q \gtrsim 2-5$ qubits.
\end{abstract}

\maketitle

\tableofcontents

\section{Introduction\label{sec:intro}}
Simulating many-particle quantum systems has always been seen as one of the most exciting potential applications of quantum computers~\cite{feynman1982simulating}.
While simulation of non-relativistic many-body physics is already by itself complex enough to be considered as a candidate for quantum advantage~\cite{lloyd1996universal,wiesner1996simulations,abrams1997simulation,zalka1998simulating,byrnes2006simulating}, quantum field theory (QFT) adds to the picture such ingredients as particle number non-conservation, a multitude of particle types and their interactions, gauge symmetry, etc.
Quantum simulation of realistic QFTs such as Quantum Chromodynamics (QCD), will likely require the access to fault-tolerant quantum computers.
In order to extract as much physics as possible from such devices, it is imperative to investigate all theoretical aspects of the quantum simulation.
The preliminary steps include constructing a model suitable for quantum computation by discretizing the original continuous theory~\cite{kogut1975hamiltonian, carena2022improved}, mapping the degrees of freedom in the discretized model onto logical qubits~\cite{Halimeh_2020, Halimeh:2020ecg, Halimeh:2021lnv, Halimeh:2021vzf, Lamm_Scott_Yukari_2020, Tran_2021, PhysRevLett.109.125302, Banerjee_2012, https://doi.org/10.48550/arxiv.2012.08620, PhysRevX.3.041018, PhysRevA.90.042305, Stannigel_2014, Stryker_2019, Anishetty_Mathur_Raychowdhury_2009, Raychowdhury_Stryker_2020, Raychowdhury_Stryker_SU2_2020, Anishetty_Mathur_Raychowdhury_SU3_2009, doi:10.1063/1.3464267, doi:10.1063/1.3660195, Raychowdhury_thesis, Alexandru_2019, PhysRevD.91.054506, PhysRevA.95.023604, PhysRevD.102.114513, Alexandru:2021jpm, Gustafson:2022xdt, PhysRevA.99.062341, Kreshchuk_2022, e23050597, Kaplan_2003, Buser_2021, Bender_2020, Unmuth-Yockey:2018xak, bauer2021efficient, Kaplan:2018vnj, Haase:2020kaj, bauer2023new, Klco:2019evd}
as well as mapping the logical qubits onto the physical ones within an error correcting procedure~\cite{shor1995scheme,kitaev1997quantum,kitaev2003fault,steane1996multiple,lidar2013quantum,litinski2019game,brown2017poking,bombin2023logical,rajput2023quantum}.\footnote{Further optimization is likely to be achieved via cross-layer design, e.g., mapping physical degrees of freedom directly onto physical qubits~\cite{landahl2021logical}.}
Those are followed by the quantum simulation itself, which typically involves preparing on a quantum computer states of specific form \cite{Jordan:2011ci, Garcia-Alvarez:2014uda, Jordan:2014tma, Jordan:2017lea, HamedMoosavian:2017koz, Moosavian:2019rxg, Klco:2019xro, Klco:2019yrb, Harmalkar:2020mpd, Ciavarella:2022qdx,Kokail:2018eiw, Roggero:2019myu,Kaplan:2017ccd,Davoudi:2022uzo} (eigenstates, wave packets, thermal states), and evolving those in time.
Lastly, once the desired final state is prepared, it is used for measuring the observables of interest.

Most early approaches to simulating time evolution and ground state preparation were based on product formul\ae~\cite{lloyd1996universal,abrams1997simulation,wiebe2010higher}.
These methods do not require ancillary qubits and, due to the overall simplicity of quantum circuit design, could be readily applied to simple models.
However, they are likely to not take advantage of neither noisy near-term devices nor fault-tolerant ones.
The former direction has been revolutionized by the invention of variational~\cite{peruzzo2014variational,mcclean2016theory,grimsley2019adaptive,shlosberg2023adaptive,tilly2022variational,fedorov2022vqe,cerezo2021variational} and subspace~\cite{mcclean2017hybrid,mcclean2020decoding,takeshita2020increasing,yoshioka2022variational,yoshioka2022generalized} methods, while the exploration of the latter one has lead to the development of quantum simulation algorithms optimal and nearly-optimal in problem parameters~\cite{berry2015hamiltonian,low2017optimal,low2019hamiltonian,HHKL18,low2018hamiltonian,kieferova2019simulating,low2019hamiltonian,berry2020time,Motlagh:2023oqc}.
The asymptotic cost of these nearly-optimal algorithms is typically given in terms of the number of calls to subroutines that provide information about the physical Hamiltonian.
Most of such algorithms rely on the usage of \emph{block encoding} subroutines whose construction comes at a great expense and leads to large prefactors in the final gate cost.
A number of algorithms, however, use instead the Hamiltonian time evolution input model~\cite{somma2019quantum,choi2021rodeo,qian2021demonstration,lin2022heisenberg}, in which case it is an approximate time evolution circuit that serves as an elementary block for the circuit construction. 
The development of this idea has ultimately lead to the Quantum Eigenvalue Transformation for Unitary matrices (QETU) algorithm~\cite{dong2022ground}, which enables one to assemble circuits applying a large class of polynomial transformations of a unitary operator to a given state, including those that implement non-unitary dynamics~\cite{chan2023simulating}.

In this work, we present the first study using QETU for preparing the ground state of a lattice gauge theory. 
The test theory we consider is a particular formulation of a U(1) lattice gauge theory in two spatial dimensions~\cite{bauer2021efficient,Grabowska:2022uos,Kane:2022ejm}.
Our study includes an analysis of how the parameters of the QETU algorithm, as well as the Trotter error from approximating the time evolution circuit, scale with the system size, number of qubits per site, and gauge coupling. 
Armed with the intuition from this numerical study, we provide a general discussion about how the cost of QETU for ground state preparation of a QCD-like lattice gauge theory will scale with the system parameters.

We also present a novel application of QETU for preparing Gaussian states in quantum mechanical systems. 
We show that, while a na\"ive application of QETU to this problem results in the error of the approximation decreasing polynomially with the number of calls to the time-evolution circuit, this scaling can be made exponential with simple modifications to the QETU procedure.
Using our improved methods, we show that the cost of preparing Gaussian states with QETU outperforms existing state preparation methods for states represented using $n_q > 2-5$ qubits.

In addition to these more specialized studies, we developed modifications to the original QETU algorithm that can significantly reduce the cost for arbitrary Hamiltonians. These modifications can be applied to both the scenario when the time evolution circuit is prepared exactly, and when prepared approximately using Trotter methods. 

The rest of this work is organized as follows. In Sec.~\ref{ssec:qetu_review} we review the general QETU algorithm. 
From there, in Sec.~\ref{ssec:qetu_gs_prep} we review how QETU can be used for ground state preparation.
Next, in Sec.~\ref{sssec:tau_neq_1} we discuss our modifications to the original QETU algorithm, and provide numerical demonstrations of the achievable cost reduction.
In Sec.~\ref{ssec:wp_intro} we discuss how to use QETU to prepare Gaussian states. 
Section~\ref{sec:u1} reviews the details of the test theory we consider, which is a particular formulation of a U(1) lattice gauge theory in two spatial dimensions. 
Numerical results for ground state preparation of this U(1) lattice gauge theory using QETU are presented in Sec.~\ref{ssec:u1_numerical}.
From there, we present the results of our study using QETU to prepare Gaussian states in Sec.~\ref{ssec:numerics_wp}.
In Sec.~\ref{sec:gsprep}, we provide a general discussion of the scaling expected when extending QETU for ground state preparation of a general lattice gauge theory with the same qualititive properties as QCD.
Our conclusions, as well as a discussion of future applications, are presented in Sec.~\ref{sec:conclusion}.
The main notations used throughout the paper are listed in~Tab.~\ref{tab:notations}.

\section{QETU}
\renewcommand{\arraystretch}{1.3}
\begin{table*}[p]
    \centering
    \begin{ruledtabular}
    \begin{tabular}{L{0.35\textwidth }R{0.635\textwidth}}
       \multicolumn{2}{c}{QETU algorithm}
       \\ \hline \hline 
       $\fex(x)$  & Function to be approximated by QETU
       \\ \hline 
       $\Fex(x)$  & $\fex(2\arccos(x))$
       \\ \hline 
       $F(x)$ & Polynomial approximation to $\Fex(x)$
       \\ \hline  
       $f(x)$ & $F(\cos(x/2))$
       \\ \hline  
       $U$ & QETU input operator
       \\ \hline
       $\Hphys = \sum_n E^\text{phys}_n \saxe{\psi_n}{\psi_n}$ & Physical Hamiltonian
       \\ \hline  
       $H = \sum_n E_n \saxe{\psi_n}{\psi_n}$ & Rescaled Hamiltonian
       \\ \hline 
       $c_{1,2}$ & Constants in operator rescaling,~\cref{eq:c1c2,eq:c1c2X} 
       \\ \hline  
       $P_{<\mu} = \saxe{\psi_0}{\psi_0}$ &  Projector onto ground state of $H$
       \\ \hline  
       $\sket{\psi_{\rm init}}$ & Initial state
       \\ \hline  
       $\sigma_\pm$, $\sigma_{\rm min}$, $\sigma_{\rm max}$ & Parameters of shifted sign function,~\cref{eq:sigmas}
       \\ \hline  
       $\Deltaphys$ & Initial (approximate) knowledge of $(E_1^\text{phys}-E_0^\text{phys})/2$
       \\ \hline  
       $\Delta$ & Initial (approximate) knowledge of $(E_1-E_0)/2$
       \\ \hline  
       $\mu$ & Initial (approximate) knowledge of $(E_1+E_0)/2$
       \\ \hline  
       $\tau$ & Evolution parameter of 
       \\ \hline
       $N_\text{steps}$ & Number of steps in Trotter implementation of $U$
       \\ \hline  
       $\delta\tau$ & $\tau / N_\text{steps}$, Trotter step size
       \\ \hline  
       $\tau_{\rm max}$ & Largest value of $\tau$ guaranteeing isolation of the ground state
       \\ \hline  
       $\eta$ & Parameter defining $H$ spectrum bounds,~\cref{eq:c1_c2_def}
       \\ \hline  
       $\eta_{P_{< \mu}}$ & Parameter of the shifted sign function,~\cref{eq:eta1,eq:eta2}
       \\ \hline  
       $\gamma={\abs{\sbraket{\psi_\text{init}}{\psi_0}}\geq0}$ & Overlap between initial guess and true ground state 
       \\ \hline  
       $d$ & \makecell[r]{Degree of (even) $F(x)$ polynomial\\Total number of calls to both $U$ and $U^\dagger$}
       \\ \hline  
       $n_\text{Ch}$ & Number of Chebyshev polynomials used to represent (even) $F(x)$
       \\ \hline  
       $N_\text{steps}$ & Number of Trotter steps in single call to $U$
       \\ \hline  
       $N_\text{tot}$ & $d\times N_\text{steps}$, total number of Trotter steps in QETU circuit
       \\ \hline  
       $\text{FT}$ & Discrete Fourier transformation matrix
       \\ \hline \hline 
       \multicolumn{2}{c}{U(1) gauge theory}
       \\ \hline \hline  
       $g$ & Coupling constant
       \\ \hline  
       $a$ & Lattice spacing 
       \\ \hline  
       $V$ & Physical volume
       \\ \hline
       $N_s$ & Number of sites in each dimension of a lattice
       \\ \hline  
       $N_p$ & Number of independent plaquettes
       \\ \hline  
       $b_\text{max,$p$}$, $\delta b_p$ & \makecell[r]{Maximum value of magnetic field operator and\\discretization step for $p^\text{th}$ plaquette,~\cref{eq:Bdisc}}
       \\ \hline  
       $r_{\text{max}, p}$, $\delta r_p$ & \makecell[r]{Maximum value of rotor operator and\\discretization step for $p^\text{th}$ plaquette,~\cref{eq:Rdisc}}
       \\ \hline  
       $n_q$ & Number of qubits per lattice site
       \\ \hline \hline 
       \multicolumn{2}{c}{Wavepacket preparation}
       \\ \hline \hline 
       $x_0$, $p_0$ & Expectation values of position and momentum operators
       \\ \hline  
       $\sigma_x$ & Wavepacket width
       \\ \hline
       $\hat{x}_\text{sh}$ & Shifted position operator
       \\ \hline
       $x_0^\text{QETU}$ & $c_1 x_0 + c_2$
       \\ \hline
       $\sigma_\text{QETU}$ & $c_1 \sigma_x$
       \\ \hline
       $\Fex_+(x), \Fex_-(x)$ & Even and odd parts of $\Fex(x)$
    \end{tabular}
    \end{ruledtabular}
    \caption{Main notations used in the paper.}
    \label{tab:notations}
\end{table*}
\renewcommand{\arraystretch}{1}

We begin this section by reviewing the QETU algorithm in its most general form, as well as its application to ground state preparation.
Next, we present modifications to the original algorithm that reduce the cost of ground state preparation.
Lastly, we consider a novel application of QETU to the preparation of Gaussian distributions and wave packets.

\subsection{Algorithm review \label{ssec:qetu_review}}

Similarly to the Quantum Eigenvalue Transformation~\cite{gilyen2019quantum,lin2022lecture} algorithm, the QETU circuit realizes a polynomial transformation of $\me^{-i\tau H}$, which, in turn, can be used for implementing a wide class of functions of $H$.
While for a given Hermitian operator $H$ constructing the exact circuit for $U=\me^{-\iu \tau H }$ is, generally, a non-trivial problem~--- even for short times $\tau$~--- QETU can potentially render useful results for approximate implementations of $U$.
The impact of such approximations on the method's performance will be explored in future sections. 

Preparing the ground state of a Hamiltonian with the aid of the QETU algorithm is based on the concept of \emph{filtering}~\cite{poulin2009preparing,ge2019faster,lin2020,lin2022heisenberg,dong2022ground,cohen2023optimizing}, which implies constructing a circuit approximately implementing the action of a projector $P_{<\mu} = \saxe{\psi_0}{\psi_0}$ onto the ground state of $H = \sum_n E_n \saxe{\psi_n}{\psi_n}$ and applying this circuit to a state $\sket{\psi_\text{init}}$ having significant overlap ${\abs{\sbraket{\psi_0}{\psi_\text{init}}}=\gamma\geq0}$ with the ground state:
\begin{equation}
    \label{eq:proj}
    P_{<\mu} \sket{\psi_\text{init}} \propto \sket{\psi_0}\,.
\end{equation}

In Ref.~\cite{ge2019faster} it was observed that $\cos^M H$ for large values of $M$ is approximately proportional to the projector onto the ground state of $H$ for a properly normalized $H$.
The circuit for $\cos^M H$ in this approach is expressed as ${\cos^M \! H= (\me^{\iu H}+\me^{-\iu H})^M/2^M}$, where the exponents are implemented using any available time evolution algorithm, while the linear combination of terms is obtained via the Linear Combination of Unitaries (LCU) algorithm~\cite{berry2015simulating,babbush2015exponentially,babbush2017exponentially,babbush2018encoding,HHKL18,LC19,lin2022lecture,kirby2022exact}.
While Ref.~\cite{ge2019faster} combined the ideas of the Hamiltonian time evolution input model~\cite{somma2019quantum,choi2021rodeo,qian2021demonstration,lin2022heisenberg} (i.e., utilizing $\me^{-\iu H}$ as a building block for the algorithm), this approach did not have optimal performance.
In Ref.~\cite{lin2020} algorithms based on the filtering concept have been developed based on the Hamiltonian oracle access model and QET technique~\cite{low2017optimal}.
Despite the asymptotically nearly-optimal performance of algorithms in Ref.~\cite{lin2020}, they suffer from a high cost of the oracle subroutines.

\begin{figure*}[t]
    \large
    \centering
    \begin{quantikz}[row sep={1cm,between origins}, column sep=0.25cm]
        \lstick{$\ket{0}$} & \qw & \gate{\me^{i \varphi_{0} X}} & \ctrl{1} & \gate{\me^{i \varphi_{1} X}} & \ctrl{1} & \qw & \ldots\ & \qw &  \ctrl{1} & \gate{\me^{i \varphi_{1} X}} & \ctrl{1} & \gate{\me^{i \varphi_{0} X}} &\qw & \meter{} & = 0
        \\
        \lstick{$\ket{\psi}$} & \qw & \qw & \gate{U} & \qw & \gate{U^\dagger} & \qw & \ldots\ & \qw & \gate{U} & \qw & \gate{U^\dagger} & \qw & \qw & \qw & \frac{F(\cos(H/2)) \ket{\psi}}{||F(\cos(H/2)) \ket{\psi}||}
    \end{quantikz}
    \caption{QETU circuit diagram. The top qubit is the control qubit, and the bottom register is the state that the matrix function is applied to.
    Here $U = \me^{-i H}$ is the time evolution circuit.
    Upon measuring the ancillary qubit to be in the zero state, one prepares the normalized quantum state $F(\cos(H/2))\ket{\psi}/||F(\cos(H/2)) \ket{\psi}||$ for some polynomial $F(x)$.
    For symmetric phase factors $(\varphi_0, \varphi_1, \dots, \varphi_1, \varphi_0) \in \mathds{R}^{d+1}$, then $F(\cos(H/2))$ is a real even polynomial of degree $d$. 
    The probability of measuring the control qubit in the zero state is $p = ||F(\cos(H/2)) \ket{\psi}||^2$.}
    \label{fig:qetu_circ}
\end{figure*}
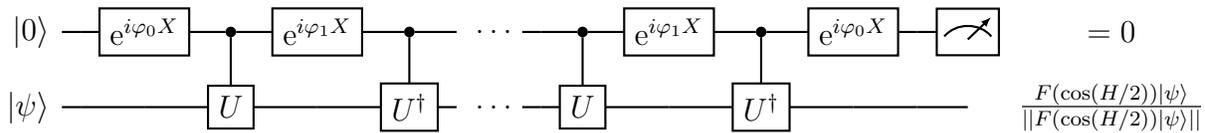

Finally, nearly-optimal algorithms for ground state energy estimation~\cite{lin2022heisenberg} and ground state preparation~\cite{dong2022ground} based on the Hamiltonian time evolution input model have been proposed.
While efficient ground state energy estimation (without ground state preparation) could be achieved with as little as a single implementation of $\me^{-\iu\tau H}$ for several values of $\tau$~\cite{lin2022heisenberg}, preparing the ground state in QETU amounts to using a circuit similar to the circuit used in Quantum Eigenvalue Transformation (QET)~\cite{gilyen2019quantum,lin2022lecture,dong2022ground}, see Fig~\ref{fig:qetu_circ}.

The QETU Theorem~\cite{dong2022ground} assumes the access to a circuit implementing $U = \me^{-i H}$ for an $n$-qubit Hermitian operator $H$. 
It states that for any even real polynomial $F(x)$ of degree $d$ satisfying $|F(x)| \leq 1, \, \forall x \in [-1,1]$, one can find a sequence of symmetric phase factors $(\varphi_0, \varphi_1, \dots, \varphi_1, \varphi_0) \in \mathds{R}^{d+1}$, such that the circuit in Fig.~\ref{fig:qetu_circ} denoted by $\mathcal{U}$ satisfies $\left( \bra{0} \otimes \mathds{1}_n \right) \mathcal{U} \left( \ket{0} \otimes \mathds{1}_n \right) = F(\cos(H/2))$ \cite{dong2022ground}.
In practice, the QETU circuit is used for approximately implementing $\fex(H)$ by realizing a transformation $F(\cos (H/2))$, where $F(x)$ is a polynomial approximation to $\Fex(x)\equiv \fex(2\arccos (x))$.

Note that there also exists a control-free version of the QETU circuit that avoids having to implement controlled calls to the $\me^{-i H}$ circuit~\cite{dong2022ground}.
The procedure for using the control-free implementation involves grouping the terms of $H$ into groups $H=\sum_j H^{(j)}$ such that each term in $H^{(j)}$ anticommutes with the Pauli operator $K_j$. 
It can be shown that, once this grouping is found, only the $K_j$ operators must be controlled.
If one finds $l$ such groups, instead of having to control each term in $\me^{-i H}$, the control-free version of QETU requires only an additional $l$ controlled operations.
Using this control-free circuit implements instead the mapping $F(\cos(H))$. 
More details about the control-free implementation, including the quantum circuit, can be found in App.~\ref{app:ctrl_free}.

The original algorithm in Ref.~\cite{dong2022ground} proposed the use of $\me^{-i H}$ in the QETU circuit. It is possible, however, to use $\me^{-i \tau H}$ instead. Doing so leads to a modification of the Lemma above, where the mapping is now $F(\cos(\tau H/2))$. Because $\cos(\tau x/2)$ is periodic, one must be careful when choosing $\tau$ to ensure construction of the desired function $\fex(x)$. In the following section, we discuss how using $\tau \neq 1$ can reduce the cost of state preparation. In Sec.~\ref{ssec:numerics_wp} we discuss how $\tau \neq 1$ can reduce the cost of using QETU for constructing Gaussian states.

\subsection{Ground state preparation via QETU \label{ssec:qetu_gs_prep}}

In this section, we review the algorithm proposed in Ref.~\cite{dong2022ground} for ground state preparation using QETU.
We first state the necessary assumptions and scaling of the algorithm.
After reviewing the original algorithm in more detail, we discuss how using instead $\me^{-i \tau H}$ as a building block in the QETU circuit can lead to significant cost reductions, both when implementing $\me^{-i \tau H}$ exactly as well as using product formulas.
We conclude by discussing how QETU can be used to prepare $n_q$-qubit Gaussian states with a cost that is linear in $n_q$.

\subsubsection{Original algorithm}
\begin{figure}[h]
    \centering
    \includegraphics[width=0.48\textwidth]{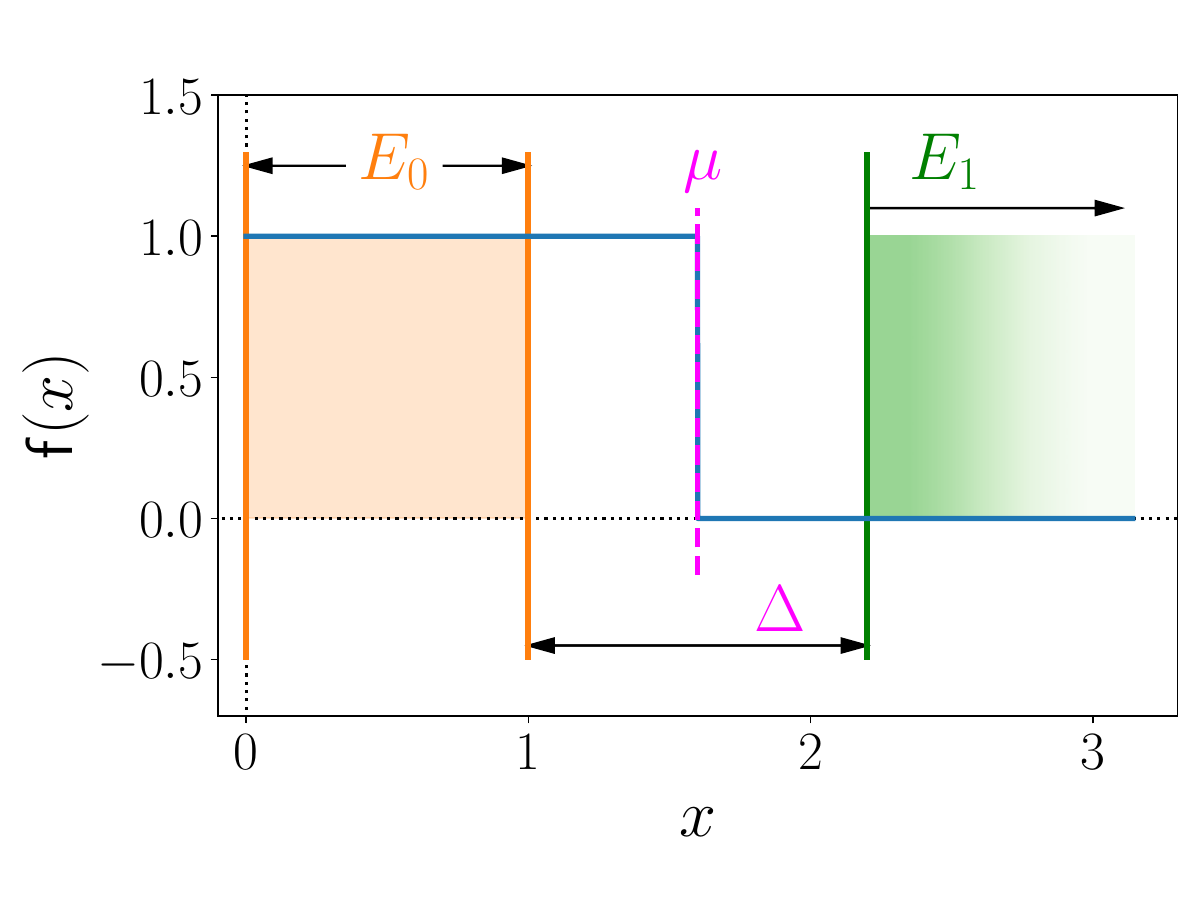}
    \caption{
    Plot of the step function $\fex(x)=1-\theta(x-\mu)$ encoding partial information known about $E_0$ and $E_1$.
    The the x-axis variable $x$ corresponds to the energy $E$.
    The ground state energy $E_0$ is known to be in the shaded orange region between the orange vertical lines. 
    The first excited state $E_1$ is known to be larger than the value indicated by the green vertical line. 
    The gap $\Delta$ used is the difference between the lower bound of $E_1$ and the upper bound of $E_0$. 
    The central value $\mu$ is chosen to be halfway between the upper bound of $E_0$ and the lower bound of $E_1$.
    Isolating the ground state is possible without exact knowledge of $E_0$ and $E_1$.
    } 
    \label{fig:step_fun}
\end{figure}

Suppose one has access to a Hamiltonian $H$ via $U = \me^{-i H}$, and that the spectrum of $H$ is in the range $[\eta, \pi-\eta]$ for some $\eta > 0$.
Furthermore, assume one has knowledge of parameters $\mu$ and $\Delta$ such that
\begin{equation}
    E_{0} \leq \mu - \Delta/2 \leq \mu + \Delta/2 \leq E_{1}\,,
\end{equation}
where $E_{i}$ is the $i^\text{th}$ excited state of $H$. 
Here $\mu$ represents the knowledge of the precise values of the energies, and $\Delta$ is a lower bound of the excited state energy gap $E_{1} - E_{0}$.
Figure~\ref{fig:step_fun} shows an example of the exact projector onto the ground state, given by the step function $1-\theta(x-\mu)$, that isolates the ground state even with only partial knowledge of $E_0$ and $E_1$.
Lastly, assume one has an initial guess $\ket{\psi_\text{init}}$ for the ground state with overlap satisfying $|\bra{\psi_0} \ket{\psi_\text{init}}| \geq \gamma$. 
Under these assumptions, one can prepare a state $|\widetilde{\psi}_0\rangle$ such that $|\langle \psi_0| \widetilde{\psi}_0\rangle| \geq 1-\epsilon$ using $\widetilde{\mathcal{O}}(\gamma^{-2} \Delta^{-1} \log(1/\epsilon))$ controlled calls to the time evolution circuit for $U$.
We will spend the rest of this section describing the details of the algorithm, including how the scaling with $\gamma, \Delta$ and $\epsilon$ arise. 

\begin{figure}[h]
    \centering
    \includegraphics[width=0.48\textwidth]{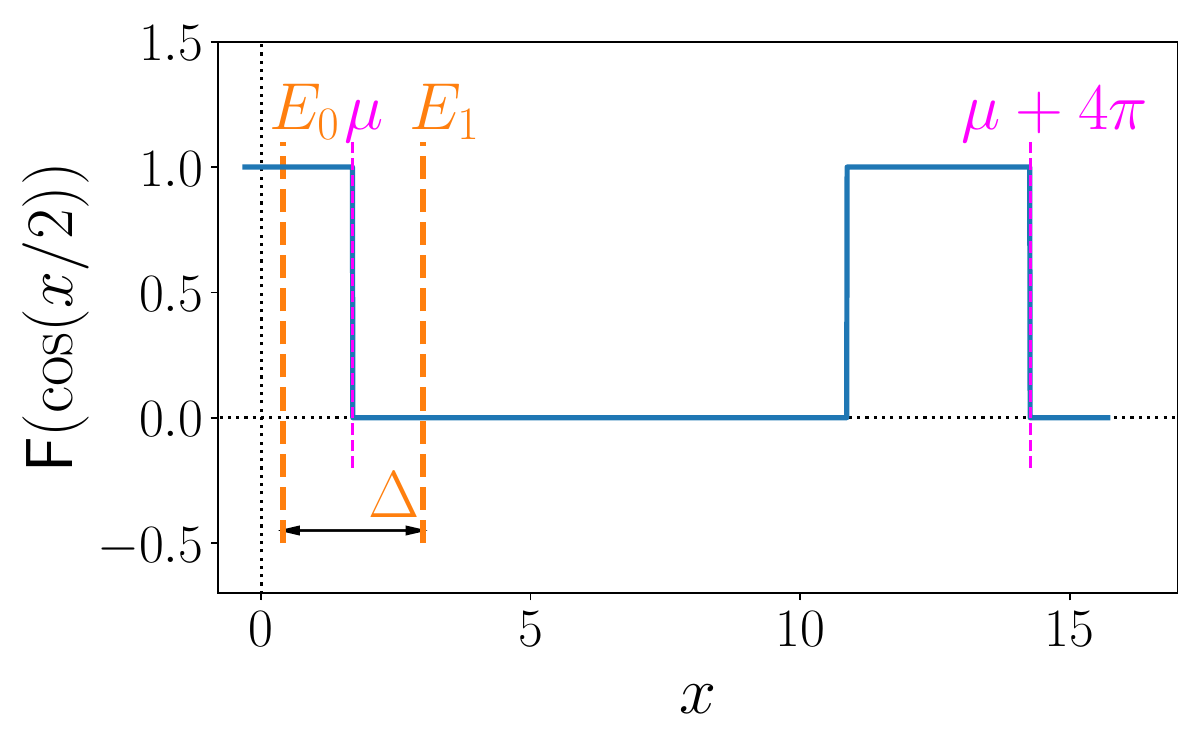}
    \caption{
    In order to construct the approximate projector $P_{<\mu}$ onto the ground state, QETU implements an approximation of the function $\Fex(\cos(H/2))$, where $\Fex(x) = \fex(2\arccos(x))$ and $\fex(x)$ is a unit step function $\fex(x) = 1-\theta(x-\mu)$.
    The the x-axis variable $x$ corresponds to the energy $E$.
    Because $\cos(x/2)$ is periodic, the step function repeats itself with period $4\pi$. 
    Unless the spectrum of the Hamiltonian is in a limited range, one cannot guarantee that all the excited states will be filtered out.}
    \label{fig:periodic_step_fun}
\end{figure}

We begin by addressing the fact that the spectrum of our target Hamiltonian must be in the range $[\eta, \pi-\eta]$ for some $\eta>0$. 
Because QETU returns a function with a periodic argument, i.e., $F(\cos(H/2))$, any projector constructed with QETU will repeat itself for large enough energies, as illustrated in Fig.~\ref{fig:periodic_step_fun}. 
From this we see that unless the spectrum of $H$ is in a limited range, we cannot guarantee that higher excited states will be filtered out.
To avoid this problem, in Ref.~\cite{dong2022ground} the Hamiltonian was first scaled so that its spectrum was in the range $[\eta, \pi-\eta]$ for some $\eta>0$.
We find that this constraint can be somewhat relaxed, which is described in detail in Sec.~\ref{sssec:tau_neq_1}.
If the physical, unshifted Hamiltonian is given by $\Hphys$ with energies $E_{i}^{\text{phys}}$, the Hamiltonian with the shifted spectrum is given by
\begin{equation}
    \label{eq:c1c2}
    H = c_1 \Hphys + c_2 \unit \,,
\end{equation}
where
\begin{align}
    c_1 &= \frac{\pi - 2\eta}{E_{\text{max}}^{\text{phys}} - E_{0}^{\text{phys}}} \,,
    \quad c_2 = \eta - c_1 E_{0}^{\text{phys}}.
    \label{eq:c1_c2_def}
\end{align}
Alternatively, one could replace $E^\text{max}_\text{phys}$ with an upper bound on the maximum eigenvalue. 
One important consequence of shifting the spectrum is that the energy gap of the shifted Hamiltonian shrinks as well.
If $\Deltaphys$ is the energy gap of the physical Hamiltonian $\Hphys$, the shifted gap is given by $\Delta = c_1 \Deltaphys$. This tells us that $\Delta \approx \Deltaphys / E_\text{max}$. 
Because the maximum eigenvalue of a Hamiltonian generally grows with the number of terms, the gap $\Delta$ used in the QETU algorithm will generally shrink with the number of lattice sites.
Therefore, one generally expects $\Delta \sim 1/N_\text{sites}$, where $N_\text{sites}$ denotes the number of lattice sites use in a simulation of some lattice gauge theory.
We will discuss this scaling in more detail in Secs.~\ref{ssec:u1_numerical}~and~\ref{sec:gsprep}.

Given some Hamiltonian $H$ and the associated parameters $\mu$ and $\Delta$, one can construct an approximate projector onto the ground state by constructing an approximation $f(x)$ to $\fex(x)=1-\theta(x-\mu)$ satisfying
\begin{equation}
\begin{split}
    |f(x) - c| &\leq \epsilon\,, \quad \forall x \in [\eta, \mu-\Delta/2]\,,
    \\
    |f(x)| &\leq \epsilon\,, \quad \forall x \in [\mu+\Delta/2, \pi-\eta]\,.
\end{split}
\end{equation}
Furthermore, because a unitary matrix must have entries with magnitude less than or equal than one, we require $|f(x)| \leq 1, \forall x$. 
The parameter $c$ is chosen to be slightly smaller than 1 to avoid numerical overshooting when finding a Chebyshev polynomial satisfying the above constraints, and is discussed in more detail later in the section.

In order to approximately implement the action of $\fex(H)$ with QETU, we need the polynomial $F(x)$ to approximate the function $\fex(2\arccos(x))$.
The presence of the $\arccos(x)$ implies that $F(x)$ can only be defined for $x \in [-1,1]$.  
Taking the cosine transform into account, we wish to construct $F(x)$ such that
\begin{equation}
\begin{split}
    |F(x) - c| &\leq \epsilon\,, \quad \forall x \in [\sigma_+, \sigma_\text{max}]\,,
    \\
    |F(x)| &\leq \epsilon\,, \quad \forall x \in [\sigma_\text{min}, \sigma_-]\,,
    \label{eq:Fx_stepfun}
    \\
    |F(x)| & \leq c\,, \quad \forall x \in [-1,1]\,.
\end{split}
\end{equation}
where
\begin{equation}
\label{eq:sigmas}
\begin{split}
    \sigma_\pm &= \cos\frac{\mu \mp \Delta/2}{2}\,,
    \\
    \sigma_\text{min} &= \cos\frac{\pi-\eta}{2}\,, 
    \\
    \sigma_\text{max} &= \cos\frac{\eta}{2}\,.
\end{split}
\end{equation}
For a non-zero gap $\Delta$ and a non-zero error $\epsilon$, one candidate function for $F(x)$ is the shifted error function, which has the important property that Chebyshev approximations of it converge exponentially with the degree of the polynomial \cite{dong2022ground}. 
In principle, one could solve for $F(x)$ by choosing a specific error function such that the Chebyshev approximation to it has a constant error for all $x$. 
While it was shown in Ref.~\cite{low2017hamiltonian} that this procedure prepares the ground state of a system with a gap $\Delta$ to a precision $\epsilon$ using a polynomial with degree scaling as $\mathcal{O}(\Delta^{-1} \log \epsilon^{-1})$, this method can lead to numerical instabilities if not performed carefully \cite{dong2022ground}.
Instead, we follow a different procedure, also described in Ref.~\cite{dong2022ground}, that avoids the need to first choose a shifted error function while still producing a near-optimal approximation.  
This procedure is discussed in detail later in this section.

\begin{figure}[h]
    \centering
    \includegraphics[width=0.48\textwidth]{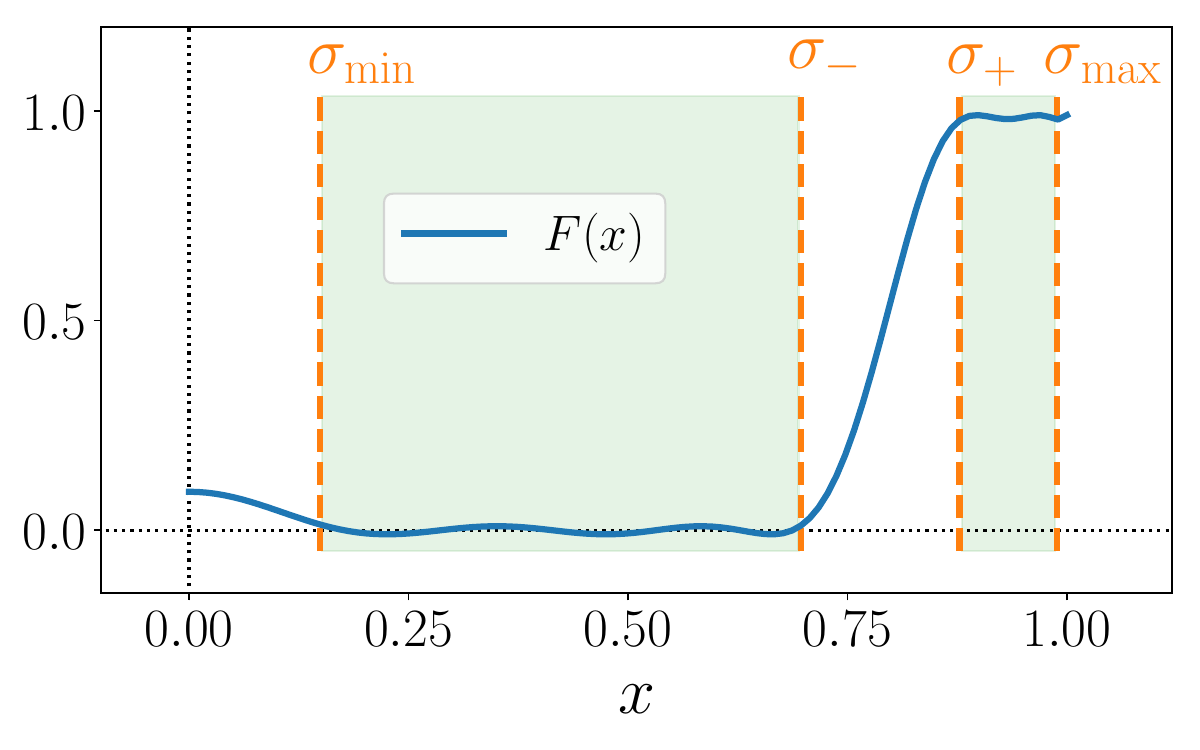}
    \caption{The blue curve shows $F(x)$, where $F(x)$ is a Chebyshev approximation of the shifted error function. The orange vertical dashed lines indicate the values $\sigma_\text{min}, \sigma_\text{max}, \sigma_\pm$. The green shaded regions indicate $x$ values that are included when solving for the Chebyshev expansion of the cosine transform shifted error function. For values of $x$ outside the shaded green regions, the function $F(x)$ can take on any value as long as $|F(x)| \leq 1$.}
    \label{fig:cos_transformed_step_fun}
\end{figure}

For the sake of argument, if one did choose $\Fex(x)$ to be a shifted error function, would exist a Chebyshev polynomial approximation to it with fidelity $1-\epsilon$ where the degree of the polynomial is $\mathcal{O}(\Delta^{-1} \log(\epsilon^{-1}))$ \cite{dong2022ground}. 
While we do not in practice choose $\Fex(x)$ to be a shifted error function manually, the convex-optimization method we use to determine the Chebyshev approximation of $\Fex(x)$ has this same scaling.
The scaling with $\Delta$ can be understood by first recognizing that for a smaller $\Delta$, the shifted error function rises more quickly. 
A steeper rising error function requires a higher degree polynomial to approximate the function to the same precision than a more slowly rising one.
By inverting $\mathcal{O}(\Delta^{-1} \log(\epsilon^{-1}))$, the error $\epsilon$ scales as $\epsilon = \mathcal{O}(\me^{-b \Delta d})$, where $b$ is a constant. 
Figure~\ref{fig:cos_transformed_step_fun} shows an example $F(x)$ for parameter values $\eta=0.3$, $c=0.999$, $E_0=1$, $E_1=1.6$, $\mu = (E_0+E_1)/2$, $\Delta=(E_1-E_0)$, corresponding to $\sigma_\text{min} = 0.15$, $\sigma_- = 0.70$, $\sigma_+=0.88$, and $\sigma_\text{max}=0.99$. 
The function $F(x)$ is represented using a $d=22$ degree polynomial, resulting in error $\epsilon \approx 0.0096$.

The final piece of the scaling has to do with the overlap of the initial guess for the ground state.
Because the probability to measure zero in the ancillary register of the QETU circuit is given by $||F(\cos(H/2)) \ket{\psi_\text{init}}|| = \gamma^2$, the number of times the circuit must be prepared in order to measure zero increases for a poor initial guess.
It can therefore be beneficial to dedicate substantial resources to prepare a high quality initial guess $\ket{\psi_\text{init}}$.

We now review the procedure in Ref.~\cite{dong2022ground} for finding a degree $d$ even Chebyshev polynomial for $F(x)$.
The goal is to solve for $n_\text{Ch}=d/2+1$ Chebyshev coefficients $c_k$ such that 
\begin{equation}
    F(x) = \sum_{k=0}^{n_\text{Ch}-1} c_{2k} T_{2k}(x)\,.
\end{equation}
To do so, one first samples $F(x)$ at a set of $M$ discrete points. It is known that polynomial interpolation sampling at equidistant points is exponentially ill-conditioned \cite{trefethen_apprx_theory}. To avoid this problem, $F(x)$ is instead sampled using the roots of Chebyshev polynomials $x_j = -\cos(\frac{j \pi}{M-1})$ for $j=0, \dots, M-1$ for some large value of $M$. 
Because the region where $F(x) = c$ shrinks with $\Delta$, one has to be careful to choose a value of $M$ large enough to resolve this region. 
Once we choose the values of $x_j$ to sample $F(x)$, we define a coefficient matrix $A_{jk} = T_{2k}(x_j)$ such that $F(x_j) = \sum_{k} A_{jk} c_{2k}$. 
The coefficients are then determined by solving the following optimization problem:
\begin{equation}
    \label{eq:minmax}
    \min_{\{c_k\}}\, \max \left\{\max_{x_j \in [\sigma_\text{+}, \sigma_\text{max}]} |F(x_j)-c|, \max_{x_j \in [\sigma_\text{min}, \sigma_-]} |F(x_j)|\right\}
\end{equation} 
This is a convex optimization problem, and can be solved using, e.g., Matlab code CVX~\cite{cvx} or Python code CVXPY~\cite{diamond2016cvxpy}. 
One chooses $c$ smaller than one to allow the Chebyshev approximation to overshoot the step function, which, by the equioscillation theorem, is necessary to achieve an optimal Chebyshev approximations \cite{trefethen_apprx_theory}.
In practice, one can choose $c$ to be sufficiently close to one such that the effect is negligible.
Our numerical investigations showed that the cost of solving for the coefficients generally scales linearly or better in the degree of the polynomial.

Once the Chebyshev expansion for $F(x)$ is known, the final step is to calculate the phases $\{\varphi_j\}$ used in the QETU circuit.
Note that multiple conventions for defining the phases $\{\varphi_j\}$ exist, and throughout this work we use the so-called \emph{W-convention}.
Here we mostly follow the pedagogical discussion given in Ref.~\cite{lin2022lecture}.
To calculate the phases $\{\varphi_j\}$, one needs to combine a number of techniques, including quantum signal processing (QSP)~\cite{low2017optimal}, qubitization~\cite{LC19}, and QETU~\cite{dong2022ground}.
QSP is the theory of the unitary representation of scalar polynomials $P(x)$. 
Put more concretely, QSP tells us that, for some set of symmetric phase factors $(\phi_0, \phi_1, \dots, \phi_1, \phi_0) \in \mathbb{R}^{d+1}$, one can construct an even Chebyshev polynomial $g(x, \{\phi_j\})$ of degree $d$ in the following way,
\begin{equation}
\begin{split}
    g(x, \{\phi_j\}) = \text{Re} [\bra{0} &\me^{i \phi_0 Z} \me^{i \arccos(x) X} \me^{i \phi_1 Z} \me^{i \arccos(x) X} \
    \\
    & \dots \me^{i \phi_1 Z} \me^{i \arccos(x) X} \me^{i \phi_0 Z} \ket{0} ],
\end{split}
\end{equation}
where $X$ and $Z$ are the usual Pauli matrices. Qubitization can then be used to lift the above SU(2) representation to matrices of arbitrary dimensions. 
Finally, the quantum eigenvalue transformation for unitary matrices (QETU) tells us that, if we choose $\{\phi_j\}$ such that $g(x, \{\phi_j\}) = F(x)$, the circuit in Fig.~\ref{fig:qetu_circ} implements a block encoding of $F(\cos(H/2))$, where $\varphi_j = \phi_j + (2-\delta_{j\, 0}) \pi/4$ (see App. B in Ref.~\cite{dong2022ground}).
Combining these techniques leads to a procedure for calculating the phases $\{\phi_j\}$ completely in terms of SU(2) matrices, which we describe now.

Because we have $n_\text{Ch}$ independent phases, to produce exactly the Chebyshev polynomial $F(x)$ we must sample at $n_\text{Ch}$ values, which are taken to be the positive roots of the Chebyshev polynomial $T_{2 n_\text{Ch}}(x)$ given by $x_j = \cos(\pi \frac{2k-1}{4 n_\text{Ch}})$.
If we define the functional $\mathcal{F}(\{\phi_j\})$ as 
\begin{equation}
    \mathcal{F}(\{\phi_j\}) = \frac{1}{n_\text{Ch}} \sum_{j=1}^{n_\text{Ch}} |g(x_j, \{\phi_j\}) - F(x_j)|^2\,,
\end{equation}
then the phases $\{\phi_j\}$ that produce $F(x)$ can be found by solving $\mathcal{F}(\{\phi_j\}) = 0$. 
It has been found in Ref.~\cite{PhysRevA.103.042419} that using a quasi-Newton method with a particular initial guess of
\begin{equation}
    \{\phi^{(0)}_j\} = \left(\frac{\pi}{4}, 0, 0, \dots, 0, 0, \frac{\pi}{4}\right)
\end{equation}
one can robustly find the symmetric phase factors for values of $d \sim 10000$.
An example code to solve for the Chebyshev coefficients $c_k$ and the associated phase factors has been implemented in QSPPACK~\cite{qsppack}. Through numerical studies, we find that the cost of finding the phase factors scales roughly quadratically with the number of phases.

To summarize, using QETU to prepare the ground state of a target Hamiltonian $\Hphys$ can be done according to the following steps:
\begin{enumerate}
    \item Construct $H = c_1 \Hphys + c_2$ such that the spectrum of $H$ is in $[\eta, \pi-\eta]$ for some $\eta>0$,
    \item Determine $\mu$ and $\Delta$,
    \item Solve for the Chebyshev approximation $F(x)$,
    \item Solve for the phase factors $\{\phi_j\}$,
    \item Implement the circuit in Fig.~\ref{fig:qetu_circ}.
\end{enumerate}

In the next section, we describe how the constraint that the spectrum of $H$ should be in the range $[\eta, \pi-\eta]$ can be relaxed.

\subsubsection{Modified ground state preparation}
\label{sssec:tau_neq_1}

In the previous section, we reviewed the original QETU algorithm for ground state preparation in which the Hamiltonian spectrum was assumed to be in the range $[\eta, \pi-\eta]$. 
This assumption was necessary because the function $\cos(x)$ is monotonic in the range $x \in [0, \pi]$. 
While this is true, QETU actually returns $F(\cos(x/2))$\footnote{The control-free version of QETU, considered below in Sec.~\ref{ssec:u1_numerical} and in App.~\ref{app:ctrl_free}, implements instead $F(\cos(x))$. The analysis of this section can be adapted to that case upon replacing $F(\cos(x/2)) \to F(\cos(x))$. }, with $\cos(x/2)$ being monotonic in the range $x \in [0, 2\pi]$.
This observation can be leveraged to increase the allowed range of the spectrum of $H$. 
This is useful because a larger range leads to a larger energy gap used in the QETU algorithm, which reduces the overall cost of the simulation. 
For the modified algorithm, we introduce the variable parameter $\tau$ to characterize the increased spectrum range. 
This adjustment to the original QETU algorithm can be viewed from two perspectives: one can either continue to use $\me^{-i H}$ as a building block and change the spectrum of $H$ to be in the range $[\eta, \tau(\pi-\eta)]$, or, equivalently, consider $\me^{-i \tau H}$ as a building block with the spectrum of $H$ in the original range $[\eta/\tau, \pi-\eta]$.
We choose the latter perspective, and in what follows derive the largest value of $\tau$ one can use while still guaranteeing isolation of the ground state.
We conclude with a discussion of how using $\tau \neq 1$ can reduce the cost in the scenario when one implements $\me^{-i \tau H}$ using product formulas.

We now discuss the modifications required when using $\tau \neq 1$.
First, while the initial algorithm in Ref.~\cite{dong2022ground} proposed using the same value of $\eta$ when constructing the shifted sign function and when shifting the spectrum of the Hamiltonian, it is in principle possible to use different values.
We continue to use $\eta$ to denote the value used for shifting the spectrum of $H$ to be in the range $[\eta, \pi-\eta]$, and introduce $\eta_{P_{< \mu}}$ to denote the value used when constructing the shifted error function.
Note that, in order to avoid the scenario where the ground state is filtered out and excited states are not, one must ensure $\eta_{P_{< \mu}} \leq \eta$. 
The parameters of the shifted sign function for general $\tau$ are then given by
\begin{align}
    \sigma_{\pm}(\tau) &= \cos \left(\tau \frac{\mu \mp \Delta/2}{2} \right),
    \\
    \label{eq:eta1}
    \sigma_\text{min}(\tau) &= \cos \left(\tau \frac{\pi-\eta_{P_{< \mu}}}{2} \right),
    \\
    \label{eq:eta2}
    \sigma_\text{max}(\tau) &= \cos \left(\tau \frac{\eta_{P_{< \mu}}}{2}\right).
\end{align}
If we implement $F(x)$ with error $\epsilon$, on the range $-1 \leq x \leq 1$ we would have
\begin{equation}
    F(x) = \begin{cases} 
      \epsilon & \hspace{0.13in} \sigma_\text{min} \leq x \leq \sigma_- 
      \\
      \epsilon & \hspace{0.12in} -\sigma_- \leq x \leq -\sigma_\text{min} 
      \\
      c-\epsilon & \hspace{0.23in} \sigma_+ \leq x \leq \sigma_\text{max}
      \\
      c-\epsilon & -\sigma_\text{max} \leq x \leq -\sigma_+\,,
   \end{cases}
\end{equation}
where we have used the fact that $F(x)$ is constructed using even polynomials. 
Note that while there are no theoretical issues with using $\tau < 1$, doing so shrinks the energy gap and is not beneficial in the context of exact implementations of $U$. 
We are now in a position to discuss the two possible pitfalls that can occur when using $\tau \neq 1$ and how to overcome them.

The first caveat is that, for $\tau > 1$, it is possible that an excited state energy falls into the region $-\sigma_\text{min} \leq x \leq \sigma_\text{min}$ where $F(x)$ can take on values larger than $\epsilon$. 
This can be avoided by implementing the step function with $\eta_{P_{< \mu}} = 0$, i.e., set $\sigma_\text{min}=0$. 
Our numerical studies indicate the quality of the approximation is largely independent of $\eta_{P_{< \mu}}$, with the choice $\eta_{P_{< \mu}} = 0$ resulting in close to the smallest error in the range $0 \leq \eta_{P_{< \mu}} \leq \eta$.

\begin{figure}[h]
    \centering
    \includegraphics[width=0.45\textwidth]{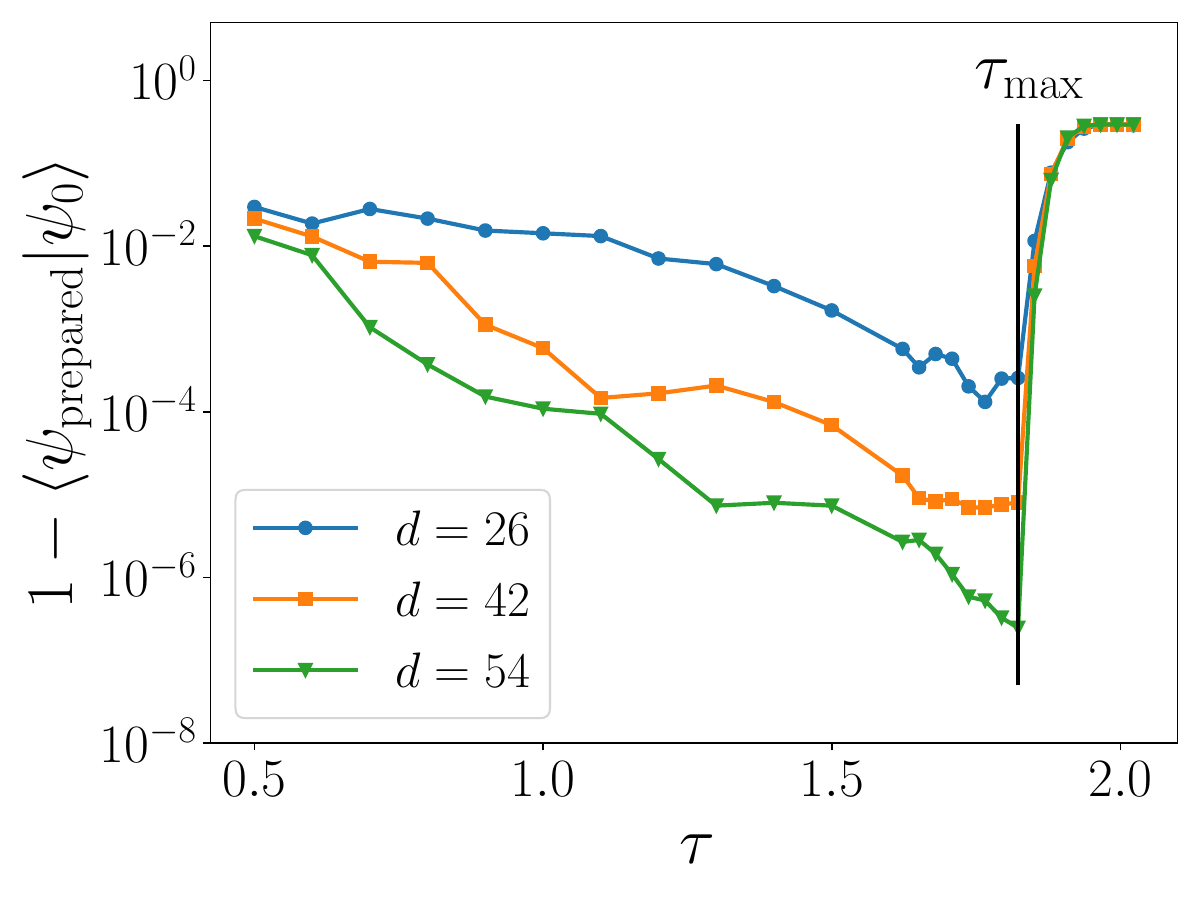}
    \caption{Error of the state prepared using QETU as a function of $\tau$ for a simple harmonic oscillator system. The parameters used are $g=1$, $\eta_{P_{< \mu}} = 0$, $\eta = 0.05$, $\mu = 0.233$, and $\Delta = 0.244$.   Different colored/shaped points indicated different degree polynomial approximations to the shifted error function. The horizontal black line indicates the maximum value of $\tau$ that still guarantees isolation of the ground state, which, for this choice of parameters, is $\tau_\text{max} = 1.823$. Using $\tau = \tau_\text{max}$ leads to a significant reduction in the error compared to using $\tau = 1$, while using values of $\tau > \tau_\text{max}$ leads to significant excited state contamination in the prepared state.
    }
    \label{fig:tmax_demo}
\end{figure}

The second caveat is that one must ensure that higher excited states fall in regions where they are filtered out by the approximate step function $F(x)$. 
Consider first using $\tau = 1$. 
Taking the cosine transform into account, the locations of the shifted energies $E_i$ are $\cos(E_i/2)$, where $E_i \in [\eta, \pi-\eta]$. 
Because $\cos(x)$ is monotonically decreasing on $x \in [0, \pi]$, the cosine transformed energies get successively closer to zero for larger energies, with $\cos(E_\text{max}/2) = \cos((\pi-\eta)/2)$ which is close to zero for small $\eta$.
If one instead uses $\tau > 1$, the transformed energies are $\cos(\tau E_i/2)$.
Taking advantage of the fact that the step function is even, we see that as long as $\cos(\tau E_i / 2) > -\sigma_-(\tau)$, one still guarantees isolation of the ground state. 
This leads to a maximum value
\begin{equation}
    \tau_\text{max} = \frac{2\pi}{\pi - \eta + (\mu+\Delta/2)}\,.
\end{equation}
Figure~\ref{fig:tmax_demo} shows the fidelity of the prepared ground state of a simple harmonic oscillator as a function of $\tau$. 
The Hamiltonian is given by
\begin{equation}
    \widehat{H} = \frac{g^2}{2} \hat{p}^2 + \frac{1}{2g^2} \hat{x}^2\,,
\end{equation}
and is represented using $n_q$ qubits.
Working in the digitized eigenbasis of the position operator, we choose to sample its eigenvalues as
\begin{equation}
	x_j = -x_\text{max} + (\delta x) j, \quad j = 0,1,\dots, 2^{n_q}-1.
\end{equation}
where $\delta x = 2 x_\text{max}/(2^{n_q}-1)$. Using fact that $[\hat{x}, \hat{p}]=i$, the momentum operator is implemented as
\begin{align}
	p^{(\hat{p})}_j &= -p_\text{max} + (\delta p) j, \quad j = 0,1,\dots, 2^{n_q}-1,
	\\
	\hat{p}^{(\hat{x})} &= \text{FT}^\dagger \hat{p}^{(\hat{p})} \text{FT},
\end{align}
where $p_\text{max} = \pi/(\delta x)$, $\delta p = 2\pi / (2^{n_q} \delta x)$, and $\text{FT}$ is the discrete Fourier transformation matrix. The superscripts $(\hat{p})$ and $(\hat{x})$ indicate that the momentum operator is written in the momentum and position basis, respectively. It will be useful for the circuit construction discussion to review the cost of exponentiation of $\hat{x}$ and $\hat{p}$. Because $\hat{x}$ and $\hat{p}^{(\hat{p})}$ are diagonal matrices with evenly spaced eigenvalues, $\me^{i p_0 \hat{x}}$ and $\me^{i x_0 \hat{p}^{(\hat{p})}}$ can be implemented using zero CNOT gates and $n_q$ rotation gates~\cite{klco2019digitization}. One can then use the efficient quantum Fourier transform circuit to construct $\me^{i x_0 \hat{p}^{(\hat{x})}} = \text{FT}^\dagger \me^{i x_0 \hat{p}^{(\hat{p})}} \text{FT}$, with each Fourier transform requiring $\mathcal{O}(n_q^2)$ gates~\cite{nielsen_chuang_2010}.

The results in Fig.~\ref{fig:tmax_demo} are for a three qubit system with $g=1$, $\eta_{P_{< \mu}} = 0$, $\eta = 0.05$, $\mu = 0.233$, and $\Delta = 0.244$. 
The maximum value of $\tau$ for these parameters is $\tau_\text{max} = 1.823$.
For various degree polynomials, we see that the fidelity in general improves for increasing $\tau$ up to $\tau_\text{max}$, where increasing further leads to an increase in error. 
Relative to using $\tau=1$, using $\tau=\tau_\text{max}$ leads to a general improvement in precision by a factor of $\mathcal{O}(\exp(d \Delta(\tau_\text{max}-1)))$.
We now discuss how using $\tau \neq 1$ can improve simulations when the approximating $\me^{-i \tau H}$ using product formulas. 
We denote by $N_\text{steps}$ the number of Trotter steps per time evolution circuit, so that the building block of the QETU circuit is $\bigl(\me^{-\iu \tau H / N_{\text{steps}}}\bigr)^{N_{\text{steps}}}$.
Each step is approximated using a first-order Trotter formula. If $H_x = \hat{x}^2/(2 g^2)$ and $H_p = g^2 \text{FT}^\dagger (\hat{p}^{(\hat{p})})^2 \text{FT}/2$, then
\begin{equation}
\begin{split}
    \me^{-\iu \delta \tau H} &\approx \me^{-\iu \delta \tau H_x} \me^{-\iu \delta \tau H_p} 
    \\
    &= \me^{-\iu \delta \tau H_x} \text{FT}^\dagger \me^{-\iu \delta \tau H_p^{(\hat{p})}} \text{FT},
    \label{eq:sho_trotter}
\end{split}
\end{equation}
where $\delta \tau = \tau / N_\text{steps}$. 
Recall that $d$ denotes the total number of calls to the time evolution circuit 
(see~Fig.\ref{fig:qetu_circ}), which corresponds to the polynomial of degree $d$.
The total number of calls to the elementary time evolution circuit is then $N_\text{tot}=N_\text{steps} \times d$.

The error from approximating the step function and from a finite Trotter step size are denoted as $\epsilon_\text{QETU}$ and $\epsilon_\text{Trotter}$, respectively; the parameters are defined as before, such that the prepared ground state $|\widetilde{\psi}_0\rangle$ has overlap with the exact ground state given by $|\langle \widetilde{\psi}_0 | \psi_0\rangle|= 1 - \epsilon_\text{QETU} - \epsilon_\text{Trotter}$.
For concreteness, we assume the errors take the forms
\begin{equation}
\begin{split}
    \epsilon_\text{QETU} &= a \me^{-b (\tau \Delta) d} 
    \\
    \epsilon_\text{Trotter} &= c (\tau / N_\text{steps})^p,
\end{split}
\end{equation}
where the first line is again obtained by inverting $d = \widetilde{\mathcal{O}}(\Delta^{-1} \log(1/\epsilon_\text{QETU}))$, and the second line is the standard form for the error when using a $p^\text{th}$ order Trotter formula. 
The parameters $a, b$ and $c$ are constants. 
These expressions are expected to be correct to leading order. 
Replacing $d = N_\text{tot}/N_\text{steps}$, our total error is
\begin{equation}
\begin{split}
    \epsilon &= a \me^{-b(\tau \Delta) N_\text{tot} / N_\text{steps}} + c (\tau / N_\text{steps})^p,
    \\
    &= a \me^{- b \Delta N_\text{tot} \delta \tau} + c (\delta \tau)^p.
\end{split}
\label{eq:error_tot}
\end{equation}
Notice that the total error $\epsilon$ only depends on $N_\text{tot}$ and $\delta \tau$.
Solving for $N_\text{tot}$ gives
\begin{equation}
\begin{split}
    N_\text{tot} = \frac{1}{b \Delta \delta \tau}\log \left( \frac{a}{\epsilon-c (\delta \tau)^p}\right).
    \label{eq:Ntot}
\end{split}
\end{equation}
With this, we can now ask the question of what value of $\delta \tau$ minimizes the cost $N_\text{tot}$ for some fixed error threshold $\epsilon$. 
Before doing so, we emphasize the fact that $N_\text{tot}$ depends only on $\delta \tau$, and not the individual values for $N_\text{steps}$ and $\tau$. 
One might expect that reducing $\tau$ would decrease the overall cost by reducing the number of Trotter steps per time evolution circuit $N_\text{steps}$ for the same $\delta \tau$.
However, reducing $\tau$ also shrinks the gap, and this increase in cost exactly cancels out the savings from decreasing $N_\text{steps}$. 

To solve for $\delta \tau^*$ that minimizes $N_\text{tot}$, we first study the constraints on our parameters from the form of Eq.~\eqref{eq:Ntot}. 
The parameter $N_\text{tot}$ is a positive real integer. 
Because the argument of the logarithm must be positive, we find $\delta \tau < (\epsilon/c)^{(1/p)}$. In words, one must choose a value of $\delta \tau$ such that the Trotter error is below the total target error $\epsilon$. 
Furthermore, notice that the logarithm can return a negative result. 
A negative $N_\text{tot}$ implies that one can achieve a precision of $\epsilon$ by setting $N_\text{tot}=1$ and decreasing $\delta \tau$ until the target precision is achieved.

In App.~\ref{app:optimal_dtau}, a perturbative solution for $\delta \tau^*$ is presented. 
Setting $\dv{N_\text{tot}}{\tau}$ to zero and expanding the log to lowest order, we find $(\delta \tau^*)^p \approx \frac{\epsilon}{c}(1-p/(\log(\frac{a}{\epsilon})+p))$.
Because $a > \epsilon$ in general, the lowest order result for $\delta \tau^*$ is slightly below the maximum value of $(\epsilon/c)^{1/p}$.
From this we learn that, for a given choice of $N_\text{tot}$, one should choose a time-step $\delta \tau$ such that most of the error comes from the Trotter error. 
This choice can be understood intuitively by comparing the rate of convergence of the two sources of error. 
Because the Trotter error converges as some power of $\delta \tau$, while the QETU error converges exponentially in $N_\text{tot}$, it is generally more cost effective for QETU to produce a smaller error than the Trotter error.
Therefore, having the Trotter error be the majority of the error results in the smallest value for $N_\text{tot}$. 

Once the value of $\delta \tau^*$ has been chosen, one must choose values for $N_\text{steps}$ and $\tau$.
One option would be to always set $N_\text{steps}=1$ and $\tau = \delta \tau^*$. 
Another option would be to choose $\tau$ minimizing the degree of the Chebyshev polynomial achieving the target error $\epsilon$, therefore reducing the classical cost of determining the angles $\{\varphi_j\}$. 
Since $N_\text{tot}=d\times N_\text{steps}$ is fixed, in order to decrease $d$, one needs to increase $N_\text{steps}$ while ensuring that $\tau=N_\text{steps}\times\delta\tau^*\leq\tau_\text{max}$.
This leads to the choice of $\tau$ being the largest natural number multiple of $\delta \tau^*$ that is less than $\tau_\text{max}$. 
One possible caveat to keep in mind is that, because approximate time evolution using product formulas can be viewed as exact time evolution according to an effective Hamiltonian $H_\text{eff}$~\cite{Carena:2021ltu}, one actually implements $F(\cos(\tau H_\text{eff}/2))$. Because the spectrum of the effective Hamiltonian will in general be different from the exact Hamiltonian by $\mathcal{O}(\delta \tau^p)$, if one has shifted the spectrum of $H$ to be in the range $[\eta/\tau, \pi-\eta]$, the spectrum of $H_\text{eff}$ will be in the range $[\eta/\tau \pm \mathcal{O}(\delta \tau^p), \pi-\eta\pm \mathcal{O}(\delta \tau^p)]$.
To avoid systematic errors from the possibility that the maximum energy of $H_\text{eff} > \pi-\eta$, one should decrease $\tau_\text{max}$ by $\mathcal{O}(\delta \tau^p)$.

\begin{figure}[h]
    \centering
    \includegraphics[width=0.48\textwidth]{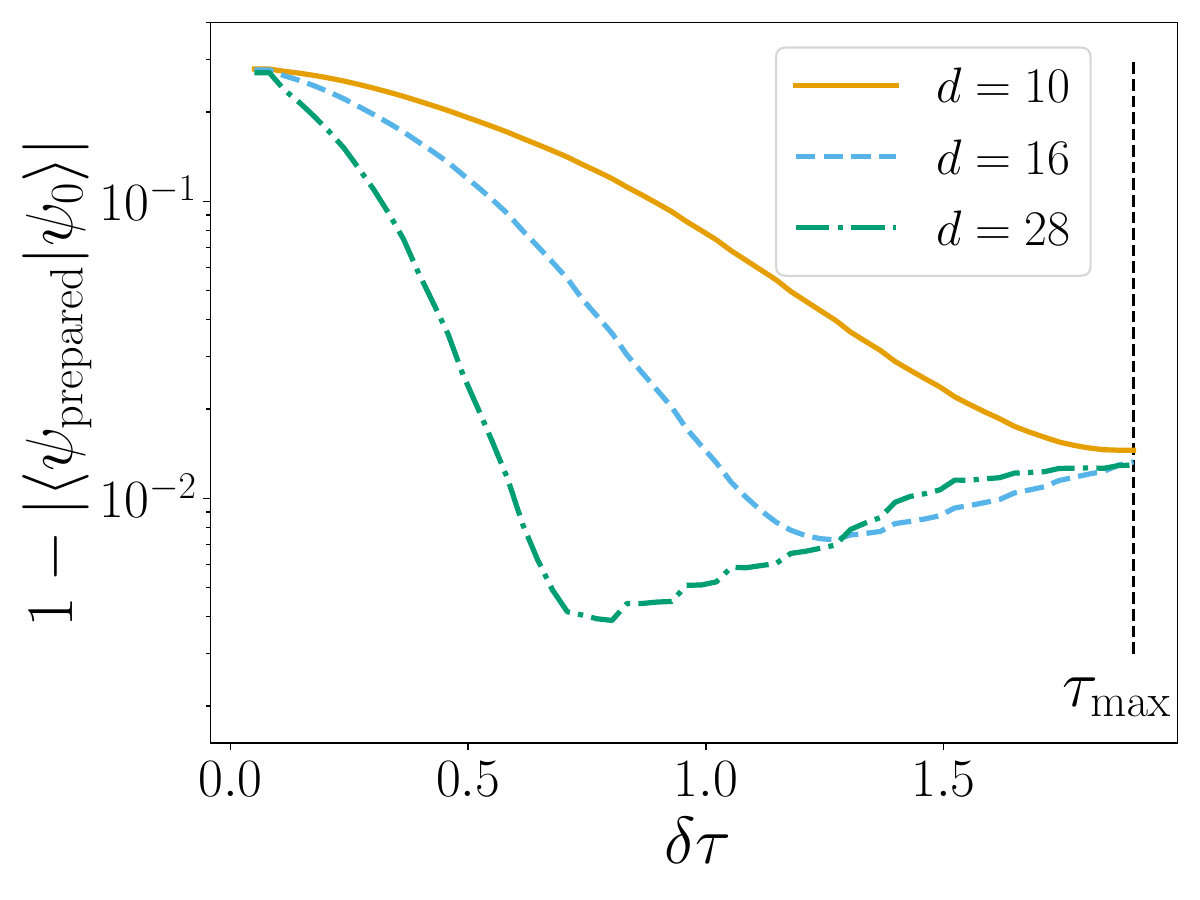}
    \includegraphics[width=0.48\textwidth]{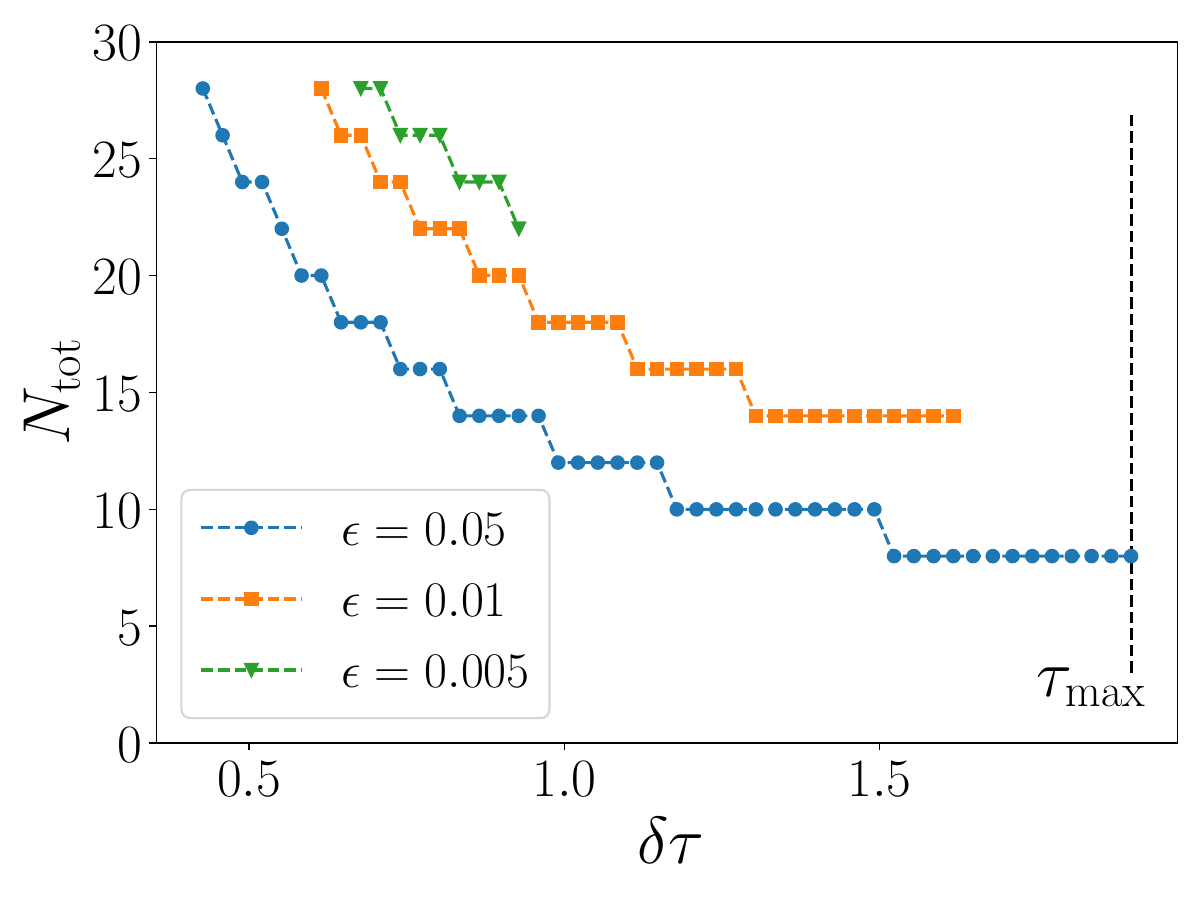}
    \caption{Both plots show results for a compact U(1) lattice gauge theory in the waved basis for $N_p=3$ and $n_q=2$ with $g=1$. The time evolution circuit was implemented using a single Trotter step with $\delta \tau = \tau$. The vertical black line indicates the value of $\tau_\text{max}$ for this system, with a value $\tau_\text{max} = 1.90$. Top: Fidelity of prepared ground state as a function of $\delta \tau$. Different colored lines indicate different degree polynomial approximations to the shifted error function. This figure demonstrates that choosing an optimal value of $\delta \tau$ results in significant precision improvements for a given computational cost. Bottom: Number of total calls to the circuit for a single first order Trotter step as a function of $\delta \tau$. Each point is the smallest value of $N_\text{tot}$ that achieves a given error $\epsilon$, with different $\epsilon$ indicated by different colors/markers. The figure demonstrates that choosing an optimal value of $\delta \tau$ results in significant cost reductions for a fixed target precision. 
    }
    \label{fig:error_and_Ntot_vs_dtau}
\end{figure}

Figure~\ref{fig:error_and_Ntot_vs_dtau} shows plots demonstrating the improvements one can achieve by choosing an optimal value of $\delta \tau$ from two perspectives, namely fixed computational cost, and fixed target precision.
The system studied is a compact U(1) lattice gauge theory in the weaved basis for a $2\times 2$ lattice with $n_q=2$ qubits per site using gauge coupling $g=1$. 
The digitization scheme used is reviewed in Sec.~\ref{sec:u1}. 
The relevant parameters for state preparation are $\eta=0.05, \eta_{P_{< \mu}}=0, \mu=0.1498$ and $\Delta = 0.1330$. 
The time evolution operator was implemented using a first order Trotter method with $N_\text{steps}=1$ Trotter steps, and therefore $\tau = \delta \tau$. 
The top plot in Fig.~\ref{fig:error_and_Ntot_vs_dtau} shows $\epsilon$ as a function of $\delta \tau$ for different degree polynomial approximations.
We see that for a fixed computational cost, significant precision improvements can be achieved by using an optimal choice for $\delta \tau$.
The bottom plot in Fig.~\ref{fig:error_and_Ntot_vs_dtau} studies the total computational cost needed to achieve a fixed target precision as a function of $\delta \tau$. 
The cost is given in terms of the total number of calls to the circuit that implements a single Trotter time-step. 
For all choices of fixed target precision, choosing an optimal value of $\delta \tau$ results in significant cost reductions.

We conclude this section with a discussion of additional improvements which can be gained with the aid of \emph{zero error extrapolation} in cases when $\me^{-i \tau H}$ is implemented approximately.
For implementations based on product formul\ae, this would require running simulations at multiple values of $\epsilon$ and $\tau$ and extrapolating to $\epsilon \to 0$ and $\delta \tau \to 0$.
Note that if the optimal value of $\delta \tau^*$ is chosen, the errors from QETU and from Trotter are comparable in magnitude. 
Therefore, extrapolating to $\epsilon \to 0$ and $\delta \tau \to 0$ would involve a fitting function of the form similar to that given in Eq.~\eqref{eq:error_tot}. 
While this is in principle possible, such a functional form is relatively complicated and undesirable.
To simplify the extrapolation, one could instead work in a regime where the QETU error is negligible compared to the Trotter error.
This would allow one to use a fitting function of simple form $c (\delta \tau)^p$.
Because of QETU's fast convergence, it is possible that such slight increase in computational cost could be worth the trade-off of better control over systematic errors.

\subsection{Wavepacket construction via QETU}
\label{ssec:wp_intro}
While the original application of the QETU algorithm in Ref.~\cite{dong2022ground} was ground state preparation, QETU is a flexible algorithm that can construct general matrix functions of any Hermitian operator. 
In this section, we provide a procedure for constructing Gaussian wavepackets in the position basis of a quantum mechanical system.
We first describe the procedure at a high level.
From there, we discuss the procedure for constructing an approximation to the Gaussian filter operator $\me^{-\hat{x}^2/(2\sigma_x^2)}$ using QETU. 
We first discuss how the na\"ive application of QETU to this problem leads to the error decreasing only polynomially with the number of Chebyshev polynomials. 
We then discuss a number of modifications to the QETU procedure that achieve an exponential scaling in the error for any desired value of the width.
Using the improved methodology, we present a method that allows one to prepare the Gaussian state to high precision while avoiding the costly implementation of LCU, and find that this method outperforms existing methods for preparing Gaussian states for values of $n_q \gtrsim 2-5$. 
Throughout this section, we will only discuss qualitative results; detailed scaling of the precision, as well as gate cost comparisons to direct state preparation, 
are presented in Sec.~\ref{ssec:numerics_wp}.
Note that the general method we employ is similar to that proposed in Ref.~\cite{Motlagh:2023oqc} for exactly constructing functions of diagonal operators.

The state we wish to prepare is given by
\begin{equation}
	\ket{\psi} = \frac{1}{\sqrt{\mathcal{N}}} \sum_{j=0}^{2^{n_q}-1} \me^{-\frac{1}{2} (\frac{x_j-x_0}{\sigma_x})^2} \me^{i p_0 x_j}  \ket{x_i},
    \label{eq:wavepacket}
\end{equation}
where $n_q$ is the number of qubits used to represent hte state, $\hat{x} \ket{x_i} = x_i \ket{x_i}$, $\mathcal{N}$ is the normalization factor, $x_0$ is the central value of the Gaussian, $p_0$ is the expectation value of the momentum, and $\sigma_x$ is the width of the wavepacket in position space. To take advantage of the relative simplicity the $\hat{x}$ and $\hat{p}^{(\hat{x})}$ operators, we break the construction of the wavepacket in Eq.~\eqref{eq:wavepacket} into three steps. The first, and the most costly step, is constructing a Gaussian with width $\sigma_x$ centered at $x=0$ using QETU. This will be done by applying an approximate implementation of the Gaussian filter operator $\me^{-\frac{1}{2} \hat{x}^2 / \sigma_x^2}$ to the state that is an equal superposition of all position eigenstates, i.e., $\ket{\psi_\text{init}} = \frac{1}{\sqrt{2^{n_q}}} \sum_{j=0}^{2^{n_q}-1} \ket{x_i}$. The wavepacket is then shifted in position space by $x_0$ and in momentum space by $p_0$ with the aid of the operators $\me^{-i x_0 \hat{p}^{(\hat{x})}}$ and $\me^{-i p_0 \hat{x}}$, respectively.

To construct the approximate Gaussian filter operator, which is a matrix function of the $\hat{x}$ operator, we will use QETU. 
The building block used in the QETU circuit is $\me^{-i \tau \hat{x}_\text{sh}}$, where $\hat{x}_\text{sh}$ is the shifted/scaled position whose spectrum is in the range $[\eta, \pi-\eta]$. 
Because $\hat{x}_\text{sh}$ is also a diagonal matrix with evenly spaced eigenvalues, as previously discussed, $\me^{-i \tau \hat{x}_\text{sh}}$ can be implemented exactly, using zero CNOT gates and $n_q$ rotation gates. 
The controlled version can therefore be implemented using $n_q$ CNOT and $n_q$ rotation gates, leading to an asymptotic scaling linear in the number of qubits used to represent the operator, albeit with a large overall prefactor.

The operator $\hat{x}_\text{sh}$ is given by
\begin{equation}
    \label{eq:c1c2X}
    \hat{x}_\text{sh} = c_1 \hat{x} + c_2,
\end{equation}
where $c_1$ and $c_2$ are the same as in Eq.~\eqref{eq:c1_c2_def}, upon replacing $E_0^\text{phys} \to \min(\hat{x})$ and $E^\text{phys}_\text{max} \to \max(\hat{x})$.
Note that because the spectrum of $\hat{x}$ is known, one does not have to use upper limits as is generally the case with state preparation where the spectrum of the Hamiltonian is not known \textit{a priori} (this fact will also allow for other improvements, and will discussed in more detail later in this section). 

Let $\fex(\hat{x}) = c\, \me^{-\frac{1}{2} \hat{x}^2 / \sigma_x^2}$ denote the exact Gaussian filter operator we wish to approximate using QETU. 
Again, the parameter $c$ is chosen to be slightly less than 1 to allow overshooting of the Chebyshev approximation.
In order to produce a Gaussian centered at $x_0=0$ with width $\sigma_x$ using $\me^{-i \tau \hat{x}_\text{sh}}$ as a building block, one must approximate the function
\begin{equation}
    \Fex(x) = c\, \exp(-\frac{1}{2\sigma_\text{QETU}^2} \left(\frac{2}{\tau} \arccos(x)-x_0^{\text{QETU}}\right)^2),
    \label{eq:Fx_wavepacket}
\end{equation}
where $x_0^{\text{QETU}} = c_1 x_0 + c_2$ and $\sigma_\text{QETU} = c_1 \sigma_x$.
Note that because the function in Eq.~\eqref{eq:Fx_wavepacket} does not have definite parity, one must use QETU to first prepare approximations to the even ($\Fex_+(x)$) and odd ($\Fex_-(x)$) parts separately, and then add them using, e.g., linear combination of unitaries (LCU)~\cite{childs2012hamiltonian}. While performing LCU does not change the overall scaling with $n_q$, it does lead to a larger constant factor in the asymptotic cost of preparing the Gaussian state.

We now describe the procedure to determine the Chebyshev approximation of $\Fex_+(x)$.
The procedure for $\Fex_-(x)$ is analogous, except that one uses odd Chebyshev polynomials. 
Let $F_+(x) = \sum_{k=0}^{d/2} c_{2k} T_{2k}(x)$ denote the Chebyshev approximation to $\Fex_+(x)$.
As before, we define the coefficient matrix $A_{jk} = T_{2k}(x_j)$, such that $F_+(x_j) = \sum_{k} c_{2k} A_{jk}$.
The coefficients $c_k$ are then determined by solving the following convex optimization problem:
\begin{equation}
    \min_{\{c_k\}}\, \max_{x_j \in [\sigma_\text{min}, \sigma_\text{max}]}  |F_+(x_j)-\Fex_+(x_j)|\,,
    \label{eq:minmax_wp_naive}
\end{equation}
where the functions are sampled using the roots of Chebyshev polynomials $x_j = -\cos \left(\frac{j \pi}{M-1}\right)$ for some large value of $M$.
The main difference between this optimization problem and the one in~\cref{eq:minmax} is that the error in~\cref{eq:minmax_wp_naive} is minimized over the entire range of $x$ values in $[\sigma_\text{min}, \sigma_\text{max}]$.

We now discuss how the error is expected to scale as the degree of the Chebyshev approximation is increased.
Unlike the case of constructing the shifted error function for ground state preparation, the functions $\Fex_\pm(x)$ are not infinitely differentiable on the interval $x \in [-1, 1]$ due to the presence of the $\arccos(x)$.  
It is well known that Chebyshev approximations of functions that are not infinitely differentiable converge polynomially to the true function with the typical rate $(1/n_\text{Ch})^m$, where $m$ is the number  of times the function is differentiable on $[-1, 1]$, and $n_\text{Ch}$ is the number of Chebyshev polynomials used in the approximation \cite{trefethen_apprx_theory}.
We therefore expect the convergence of our Chebyshev approximations to be only polynomial. 

This observation, however, leads to a natural method for avoiding the polynomial scaling.
As one increases the value of $\eta$, the interval $x_j \in [\sigma_\text{min}, \sigma_\text{max}]$ will move farther from the non-differentiable points at $x = \pm 1$, and improve the precision of approximation for a given degree polynomial.
However, as $\eta$ is increased, the functions $\Fex_\pm(x)$ become more sharply peaked, and will eventually require more Chebyshev polynomials to achieve the same precision.
We therefore expect that, for a given degree approximation, there will be a value of $\eta$ that results in the smallest error.
We find that this procedure of varying $\eta$ results in the error decreasing more favorably with the degree of the polynomial.
This scaling can made exponential, however, by also varying the $\tau$ parameter.
By varying $\tau,$ the shape of the exact functions $\Fex_\pm$ change, and can improve the quality of the approximation. 
Note that, unlike the case of constructing the filter operator, there are no theoretical issues with using a value of $\tau > \tau_\text{max}$. 
This is due to the fact that we know exactly the values of the operator we sample. 
The optimization problem that determines the Chebyshev coefficients using this modification is
\begin{equation}
    \min_{\{c_k, \eta, \tau \}}\, \max_{x_j \in [\sigma_\text{min}, \sigma_\text{max}]}  |F_+(x_j)-\Fex_+(x_j)|\,,
    \label{eq:minmax_allX_vary_eta_tau}
\end{equation}
where the parameters $\sigma_\text{min}$ and $\sigma_\text{max}$ are functions of $\eta$ and $\tau$.
Note that the optimization problem in this form is no longer a convex optimization problem.
The parameters $\eta$ and $\tau$ are found by passing the convex optimization problem in Eq.~\eqref{eq:minmax_wp_naive} to a numerical minimization procedure that solves for $\eta$ and $\tau$.

While varying $\eta$ and $\tau$ already results in the error decreasing exponentially with the degree of the polynomial, we can further improve the rate of convergence, in particular for small values of $n_q$.
Because we know the spectrum of the operator $\hat{x}_\text{sh}$ exactly, we only need the Chebyshev expansion to approximate $\Fex(x)$ at those points, not the entire range $x \in [\sigma_\text{min}, \sigma_\text{max}]$.
Taking into account the cosine transformation, the values of $x$ we need to faithfully approximate $\Fex(x)$ are $\tilde{x}_j = \cos(\tau x_{\text{sh}, j}/2)$, where $x_{\text{sh}, j}$ is the $j^\text{th}$ eigenvalue of the $\hat{x}_\text{sh}$ operator. 
If we denote the set of all $\tilde{x}_j$ values as $\tilde{\chi}$, then the Chebyshev coefficients are found by solving the modified optimization problem 
\begin{equation}
    \min_{\{c_k, \eta, \tau \}}\, \max_{\tilde{x}_j \in \tilde{\chi}}  |F_+(x_j)-\Fex_+(x_j)|\,.
    \label{eq:minmax_only_sampled}
\end{equation}
We find that, in general, this method has slightly better performance compared to sampling all $x$ values, but for certain values of $\sigma_x/x_\text{max}$ it can improve the rate of convergence significantly. 
Additionally, because we only include $2^{n_q}$ of $\tilde{x}_j$ values in the optimization, the error will be zero when the degree of the polynomial is equal to the number of points.
Furthermore, by increasing $\tau$, one can sample $\tilde{x}_j$ at negative values.
This can be used to exploit the parity of the Chebyshev expansions, reducing the effective number of points we need to approximate $\Fex(\tilde{x}_j)$.
One important consideration when using this method is that, as one varies $\eta$ and $\tau$, care must be taken to ensure that $|F_\pm(x)| \leq 1$ for all $x \in [-1,1]$, and not just for $\tilde{x}_i$.
If this condition is not satisfied, there are no possible values of phases $\{\varphi_j\}$ that implement the desired function $F(x)$.

Although we have developed a procedure using QETU that can be used to implement Gaussian states with a cost $\mathcal{O}(n_q \log(1/\epsilon))$, the cost of performing LCU to add the even and odd pieces introduces a large overall constant factor in the asymptotic cost.
If one could modify the procedure to avoid using LCU, the gate count reduction would be reduced by a factor of 10 or more. 
We now discuss how to prepare Gaussian states using only the even component, and eliminate the need for LCU altogether. The main idea is to choose $\tau$ such that the function $\Fex(x)$ becomes a purely even function. To start, note that our choice of digitizing the $\hat{x}$ operator results in the parameter $c_2 = \pi/2$. This, combined with the fact that $x_0 = 0$, leads to $x_0^\text{QETU} = \pi/2$. With this, we see that setting $\tau = 2$ leads to $\Fex(x)$ becoming
\begin{equation}
\begin{split}
    \Fex(x) &= c\, \me^{-\left(\arccos(x)-\pi/2\right)^2/(2 \sigma_\text{QETU}^2)}
    \\
    &= c\, \me^{-\left(\arcsin(x)\right)^2/(2 \sigma_\text{QETU}^2)}\,,
\end{split}
\end{equation}
where we have used the relation $\arccos(x)-\pi/2 = -\arcsin(x)$.
Because $\arcsin(x)$ has definite parity, $\Fex(x)$ is an even function for $\tau = 2$. 
After setting $\tau = 2$, the parameter $\eta$ and the Chebyshev coefficients are found by solving an optimization problem similar to that in Eq.~\eqref{eq:minmax_only_sampled}, except that $\tau$ is fixed to 2.
We find that this method offers the best precision for a given fixed gate count cost, and outperforms existing methods for Gaussian state preparation for values of $n_q > 2-3$.
More details of this comparison are given in Sec.~\ref{ssec:numerics_wp}.

We conclude by noting that, due to the simplicity of the building block $\me^{-i \tau \hat{x}_\text{sh}}$, the number of controlled gates in control-free version of QETU is the same as in the original version.
We will therefore use the original version when showing numerical results in~\cref{ssec:numerics_wp}.

\section{Lattice formulation of U(1) gauge theory\label{sec:u1}}

In this section, we review a formulation of a compact U(1) gauge theory in 2 spatial dimensions with a Hilbert space that has been constrained to satisfy Gauss' law in the charge density $\rho(x)=0$ sector. 
We also review the representation of the magnetic and electric operators introduced in Ref.~\cite{bauer2021efficient}, which can be used at all values of the gauge coupling. 
From there, we discuss the basis change introduced in Ref.~\cite{Grabowska:2022uos}, which is necessary to break the exponential volume scaling in the gate cost when performing time-evolution using Trotter methods. 
We conclude with a discussion of how the digitization of the theory in the weaved basis must be modified in order to maintain the efficiency of the representation \cite{Kane:2022ejm}.
Note that we provide only the details necessary to understand the numerical results presented in Sec.~\ref{ssec:u1_numerical}.
Further details can be found in the original references.

We chose the apply the QETU algorithm to the compact U(1) theory because it shares a number of important features with QCD, including the fact that, as the lattice spacing $a$ goes to zero, the bare coupling $g(a)$ also goes to zero. 
Additionally, the gauge field in non-Abelian gauge theories is necessarily compact, proving another motivation for detailed studies of the compact U(1) theory.

The Hamiltonian considered in Ref.~\cite{bauer2021efficient} is formulated in terms of electric rotor and magnetic plaquette operators, given by $\hat{R}(x)$ and $\hat{B}(x)$, respectively.
These operators satisfy
\begin{align}
    \bigl[ \hat B(x), \hat R(y) \bigr] = i \,  \delta^3(x-y)
    \,.
\end{align}
In the charge density $\rho(x)=0$ sector, the rotors are defined such that $\vec{E}(x) = \vec{\nabla} \times R(x)$.
In this way, electric Gauss's law is automatically satisfied.

The lattice version of the theory we consider introduces a periodic lattice of $N_x$ and $N_y$ evenly spaced lattice points in the $\hat x$ and $\hat y$ dimensions with a lattice spacing $a$. 
The lattice version of the continuum Hamiltonian is defined in terms of operators, $\hat{R}_p$ and $\hat{B}_p$, with the index $p$ denoting a specific plaquette in the lattice volume. 
The Hamiltonian can be written in terms of an electric and magnetic component as
\begin{align}
\hat{H} = \hat{H}_E + \hat{H}_B\,.
\label{eq:u1_ham}
\end{align}
After solving the constraint from magnetic Gauss' law, the fully gauge fixed Hamiltonian contains $N_p \equiv N_x N_y - 1$ independent plaquette operators. The fully gauge-fixed electric and compact magnetic Hamiltonians are given by
\begin{align}
\hat{H}_E = \frac{g^2}{2a}\sum_{p=1}^{N_p}
     (\vec{\nabla} \times \hat R_p)^2 
     \,,
\end{align}
and 
\begin{align}
    \hat{H}_B =  \frac{N_p+1}{a \,g^2} -\frac{1}{a\, g^2}\left[\sum_{p=1}^{N_p} \cos \hat B_p + \cos \left(\sum_{p=1}^{N_p} \hat{B}_p \right) \right]\,.
    \label{eq:compactHB_after_constraint}
\end{align}
The non-compact formulation of this theory can be found by making the replacements $\cos \hat{B}_p \to 1 - \frac{1}{2} \hat{B}_p^2$. 
The bases where the operators $\hat{H}_E$ and $\hat{H}_B$ are diagonal are referred to as the electric and magnetic basis. 
Furthermore, operators represented in the electric and magnetic basis will be denoted by superscripts $(e)$ and $(m)$, respectively.

We represent the operators $\hat{R}_p$ and $\hat{B}_p$ using the procedure given in Ref.~\cite{bauer2021efficient}, which involves carefully choosing the maximum value that the $\hat{B}_p$ operators are sampled, denoted by $b_\text{max}$.
The main idea for choosing $b_\text{max}$ is that, in the magnetic basis, the low energy eigenstates of the theory have a typical width proportional to the gauge coupling $g$. 
By choosing $b_\text{max} \sim g$, one only samples the wavefunction where it has support, which results in an efficient representation for all values of $g$.
We first describe how the operators are sampled, and then explain how to choose the precise value of $b_\text{max}$.

Each plaquette is represented using $n_q$ qubits.
Working in magnetic basis, the magnetic operators $\hat{B}^{(m)}_p$ are diagonal and defined by
\begin{align}
    \label{eq:Bdisc}
    \hat{B}^{(m)}_p \bket{b_{p,j}} = \left(-b_\text{max,$p$}+ j \, \delta b_{p}\right)\bket{b_{p,j}}\,,
\end{align}
where $j = 0, 1, \dots, 2^{n_q}-1$ and $\delta b_p = 2b_\text{max,$p$}/2^{n_q}$.
Because $\hat{B}_p$ and $\hat{R}_p$ are conjugate operators, the rotor operator in the magnetic basis can be written as
\begin{align}
    \label{eq:Rdisc}
    \hat{R}^{(m)}_p &= {\rm FT}^\dagger \hat{R}^{(e)}_p {\rm FT}, \\
    \hat{R}^{(e)}_p \bket{r_{p,j}} &= \left(-r_\text{max,$p$}+j\, \delta r_p \right) \bket{r_{p,j}}\nonumber 
    \,,
    \label{eq:sampleR}
\end{align}
where ${\rm FT}$ denotes the usual quantum Fourier transform and
\begin{align}
    r_\text{max,$p$}=\frac{\pi 2^{n_q}}{2b_\text{max,$p$}}\,, \qquad \delta r_p = \frac{\pi}{b_\text{max,$p$}}\,.
\end{align} 
With these definitions, one can implement Trotter time evolution of the Hamiltonian in Eq.~\eqref{eq:u1_ham} in a similar way as in Eq.~\eqref{eq:sho_trotter}.

We now discuss the procedure for choosing $b_\text{max,$p$}$, which can in general be different for different plaquettes $p$. 
In the compact formulation, it was shown that choosing $b_\text{max,$p$}$ according to
\begin{align}
    b_\text{max,$p$} &= \text{min} \left(g \frac{2^{n_q}}{2} \sqrt{\frac{\beta_{R,p}}{\beta_{B,p}}} \sqrt{\frac{2 \pi}{2^{n_q}}}, \pi \right)
    \label{eq:bMax}
\end{align}
reproduces the low-lying spectrum of the theory to per-mille level precision while only sampling the operators a number of times that corresponds to $n_q=3$.
The variables $\beta_{R,p}$ and $\beta_{B,p}$ are found by matching the non-compact magnetic Hamiltonian to a Hamiltonian of the form
\begin{align}
    \widetilde{H} = \frac{g^2}{2} \beta_{R,p}^2 \hat{R}^2_p + \frac{1}{2g^2} \beta_{B,p}^2 \hat{B}^2_p\,,
\end{align}
and ignoring the cross-terms. 
Further details regarding the digitization of the $\hat{R}_p$ and
$\hat{B}_p$ operators in this formulation can be found in Refs~\cite{bauer2021efficient, Kane:2022ejm}.

While this formulation is efficient in terms of the number of qubits required per site $n_q$ to achieve a high precision in the low energy states, it was shown that performing time evolution using Trotter methods has a gate cost that scales exponentially with volume, i.e., the number of plaquettes $N_p$ \cite{Grabowska:2022uos, Kane:2022ejm}. 
Specifically, the exponential scaling was shown to be caused by the $\cos \sum_p \hat{B}_p$ term in the magnetic Hamiltonian, which couples the entire lattice together. 
This exponential volume scaling can be broken, however, by performing a carefully chosen change of operator basis \cite{Grabowska:2022uos}, which we now review.

\begin{figure}[h]
    \centering
    \includegraphics[width=0.35\textwidth]{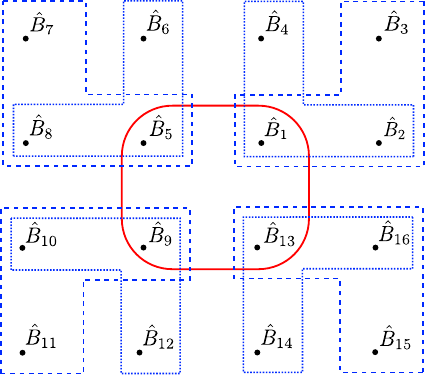}
    \caption{Visual representation of the connectivity of the cosine terms in the weaved magnetic Hamiltonian for $N_p=16$. Operators appearing inside the same box also appear as a sum inside a single cosine term in $\hat{H}_B$. Boxes with different line-styles or colors correspond to different cosine terms. The red solid square shows the reduced connectivity of the $\cos \sum_p \hat{B}_p$ term. The blue dashed and dotted rectangles show the increased connectivity of the previously local $\cos \hat{B}_p$ terms.}
    \label{fig:weaved_basis}
\end{figure}

The rotor and magnetic operators in this so-called weaved basis are given by
\begin{align}
    \hat{B}_p &\rightarrow \mathcal{W}_{p p'} \hat{B}_{p'}, \qquad \hat{R}_p \rightarrow \mathcal{W}_{p p'} \hat{R}_{p'},
\label{eq:opBasisChange}
\end{align}
where $\mathcal{W}$ is an orthogonal matrix of dimension $N_p \times N_p$. 
For any value of $N_p$, there exists an efficient classical algorithm for choosing $\mathcal{W}$ that reduces the gate count scaling from exponential to polynomial in $N_p$ \cite{Grabowska:2022uos}. 
Using this change of basis, the dominant contribution to the gate cost of a single Trotter step was shown to scale as $\mathcal{O}(N_p^{n_q} + N_p (N_p / \log_2 N_p)^{n_q})$, which is polynomial in the volume, with the power of the polynomial determined by $n_q$. 
This scaling arises because the number of terms appearing in a single cosine in the weaved basis scales as $\mathcal{O}(\log_2 N_p)$.
An example demonstrating the connectivity of the magnetic Hamiltonian operators for $N_p=16$ is shown in Fig.~\ref{fig:weaved_basis}.

One important assumption that went into choosing the large coupling limit of $b_\text{max,$p$}$ to be $\pi$ in the original operator basis was that the coefficient of a magnetic field operator $\hat{B}_p$ is equal to 1 anywhere it appears in the compact magnetic Hamiltonian. 
Because this is generally not true when working in the weaved basis, maintaining an efficient representation requires modifying the prescription for choosing $b_\text{max,$p$}$ in the large $g$ limit \cite{Kane:2022ejm}.
To understand this, first notice that, in the weaved basis, some of the $\hat{B}_p$ operators will have coefficients smaller than one.
Consequently, even if $b_\text{max,$p$} = \pi$, an operator $\hat{B}_p$ inside a given cosine will not get sampled between the full range of $[-\pi, \pi]$. 
It was shown that this problem could be fixed by scaling the upper limit for each $b_\text{max,$p$}$ by the smallest coefficient of the operator $\hat{B}_p$ anywhere it appears in the Hamiltonian. 
To demonstrate this procedure, we review the example for the $N_p=3$ case given in Ref.~\cite{Kane:2022ejm}. 
The rotation matrix used is given by
\begin{align}
    \label{eq:weavedNp3}
    \mathcal{W} = \frac{1}{\sqrt{6}} \left(
\begin{array}{ccc}
 \sqrt{2} & -2 & 0 \\
 \sqrt{2} & 1 & -\sqrt{3} \\
 \sqrt{2} & 1 & \sqrt{3} \\
\end{array}
\right)
    \,,
\end{align}
which leads to the following weaved magnetic Hamiltonian
\begin{align}
    \hat{H}_B^{\text{w}} &= \frac{N_p+1}{a\, g^2} -\frac{1}{a\,g^2} \Bigg( \cos\left[\sqrt{3} \hat{B}_1\right] + \cos\left[\frac{\hat{B}_1 - \sqrt{2} \hat{B}_2}{\sqrt{3}}\right] 
    \nonumber\\
    & \quad + \cos\left[\frac{\sqrt{2} \hat{B}_1+\hat{B}_2 - \sqrt{3} \hat{B}_3}{\sqrt{6}}\right] \nonumber \\
    &\quad +\cos\left[\frac{\sqrt{2} \hat{B}_1+\hat{B}_2 + \sqrt{3} \hat{B}_3}{\sqrt{6}}\right] \Bigg)
    \,.
    \label{eq:HExamp}
\end{align}
In order to ensure that each $\hat{B}_p$ operator gets sampled between $[-\pi, \pi]$ in each of the cosine terms, one must scale the upper limit for each $b_\text{max,$p$}$; in this case, the upper limits for $b_\text{max,$1$}, b_\text{max,$2$}$, and $b_\text{max,$3$}$, are chosen as $\sqrt{2} \pi, \sqrt{6} \pi$, and $\sqrt{3}\pi$, respectively. It was shown in Ref.~\cite{Kane:2022ejm} that this procedure for scaling $b_\text{max,$p$}$ results in a precision of the low-lying spectrum similar to that of the original basis.

We conclude this section by discussing the expected behavior of $\Delta$ with $g$ for the compact U(1) gauge theory.
As explained in Ref.~\cite{Bender:2020jgr}, because the untruncated U(1) electric Hamiltonian is unbounded, as one approaches the continuum limit, the physical energy gap $E^\text{phys}_1-E^\text{phys}_0$ diverges; the value $a(E^\text{phys}_1-E^\text{phys}_0)$ approaches a constant as $a \to 0$. This, combined with the fact that the physical energy difference $E^\text{phys}_\text{max}-E^\text{phys}_0$ scales as $\sim 1/a$, implies that $\Delta$ will approach a constant as $a \to 0$.
This scaling is qualitatively different than that of gauge theory with a finite physical energy gap (such as QCD), which is discussed in detail in Sec.~\ref{sec:gsprep}.

\section{Numerical Results}

\subsection{U(1) gauge theory \label{ssec:u1_numerical}}

In this section, we prepare the ground state of the previously described formulation of a compact U(1) lattice gauge theory using QETU.
We study the algorithm cost by determining how its parameters $\Delta$ and $\gamma$ depend on the parameters of the Hamiltonian: the number of plaquettes $N_p$, number of qubits per plaquette $n_q$, and gauge coupling $g$ (or lattice spacing $a$).
We also study how the fidelity of the final state scales with the number of calls to $\me^{-i \tau H}$, where the time evolution operator is implemented both exactly as well as using Trotter methods.
The scaling results from our analysis are summarized in Table.~\ref{tab:cost}. A general discussion regarding the physical reasoning for the observed scaling of these parameters is given in Sec.~\ref{sec:gsprep}.

Preparing the ground state requires knowledge of $aE_0^\text{phys}, aE_1^\text{phys}$, and $a E_\text{max}^\text{phys}$. In a realistic quantum simulation, one would likely be able to estimate $aE_0^\text{phys}$ beforehand.
Calculating $aE_1^\text{phys}$ and $a E_\text{max}^\text{phys}$ is generally more difficult; still, one can often use physical arguments to bound both $a(E_1^\text{phys}-E_0^\text{phys})$ and $E_\text{max}^\text{phys}$.
For the purposes the current study, we calculate $aE_0^\text{phys}$ and $aE_1^\text{phys}$ using exact diagonalization. Regarding $aE_\text{max}^\text{phys}$, we provide arguments for placing upper bounds on its value, and compare our bounds to the exact result. 
One important consideration is that, because $\Delta$ is a ratio of energies, the explicit dependence on the lattice spacing $a$ cancels.
The value of $\Delta$ therefore only depends on the lattice spacing $a$ through discretization effects.

We begin by placing an upper bound on $aE_\text{max}^\text{phys}$.
In the digitization scheme we use, one can write the U(1) Hamiltonian as $H^{\text{phys}, (b)} = \text{FT}^\dagger H^{(e)}_E \text{FT} + H^{(b)}_B$, where the superscript $e\,(b)$ indicates that the matrix is represented in the electric (magnetic) basis, where it is implied the Fourier transform is performed locally at each lattice site.
In this section, it is understood that the symbols $H_E$ and $H_B$ denote the unscaled Hamiltonians.
Using the fact that the Fourier transform is unitary and therefore does not change the eigenvalues of $H_E^{(e)}$, we see
\begin{equation}
\begin{split}
    \max(H^\text{phys}) &\leq \max(\text{FT}^\dagger H^{(e)}_E \text{FT}) + \max(H^{(b)}_B)
    \\
    &= \max(H^{(e)}_E) + \max(H^{(b)}_B)
    \label{eq:upper_bound_maxH}\,.
\end{split}
\end{equation}
Because $H_E^{(e)}$ and $H_B^{(b)}$ are diagonal matrices, the maximum eigenvalue of each matrix is simply the largest entry on the diagonal. To proceed, we must look in more detail at the forms of the two Hamiltonians. 

Because the magnetic Hamiltonian is a constant term $(N_p+1)/(ag^2)$ minus a sum of $N_p+1$ cosine terms, the maximum value any single diagonal entry can take is $2(N_p+1)/(ag^2)$.
This upper bound, however, overestimates the actual value, especially at small values of $g$. To understand this, recall that the magnetic operators are sampled from $-b_\text{max}$ to $b_\text{max}$, where $b_\text{max} \sim g$. As $g \to 0$, the compact magnetic Hamiltonian approaches the non-compact version, with each term of the form $B^2/g^2$. If $B \sim g$, then $B^2/g^2 \sim 1$ is roughly independent of $g$, while our upper bound scaled as $1/g^2$. This issue can be avoided by taking advantage of the structure of the weaved magnetic Hamiltonian. It was shown in Ref.~\cite{Grabowska:2022uos} that, after changing to the weaved basis, each cosine term in the magnetic Hamiltonian contains a sum of no more than $\mathcal{O}(\log_2 N_p)$ magnetic field operators. Thus, the spectrum of each individual cosine term can be found exactly using classical resources that scale only polynomially with $N_p$. Once the maximum entry of each individual term is known, we can then place an upper bound on the maximum energy of $H_B$ through
\begin{equation}
\begin{split}
    \max(H^{(b)}_B) \leq \frac{1}{a\, g^2} \Bigg[&(N_p+1) - \sum_{j=0}^{N_p-1}\max\left(\cos \widetilde{B}_j\right)
    \\
    &- \max\left( \cos \sum_{j=0}^{N_p-1} \widetilde{B}_j \right) \Bigg],
\end{split}
\end{equation}
where $\widetilde{B}_j$ is the $j^\text{th}$ magnetic operator in the weaved basis.

\begin{figure}[h]
    \centering
    \includegraphics[width=0.45\textwidth]{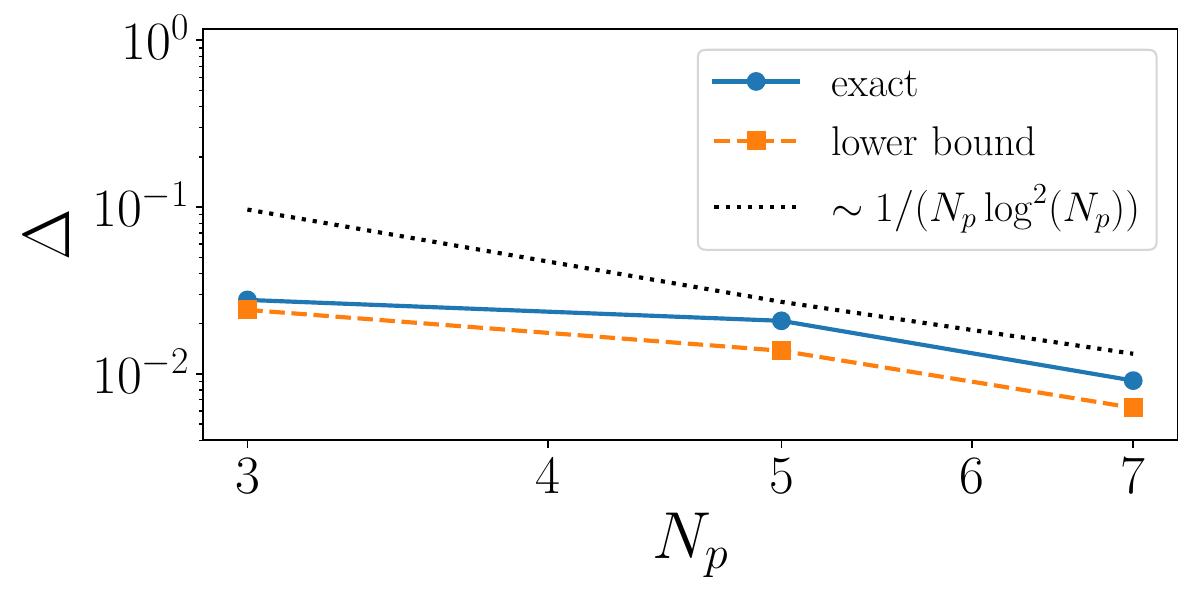}
    \includegraphics[width=0.45\textwidth]{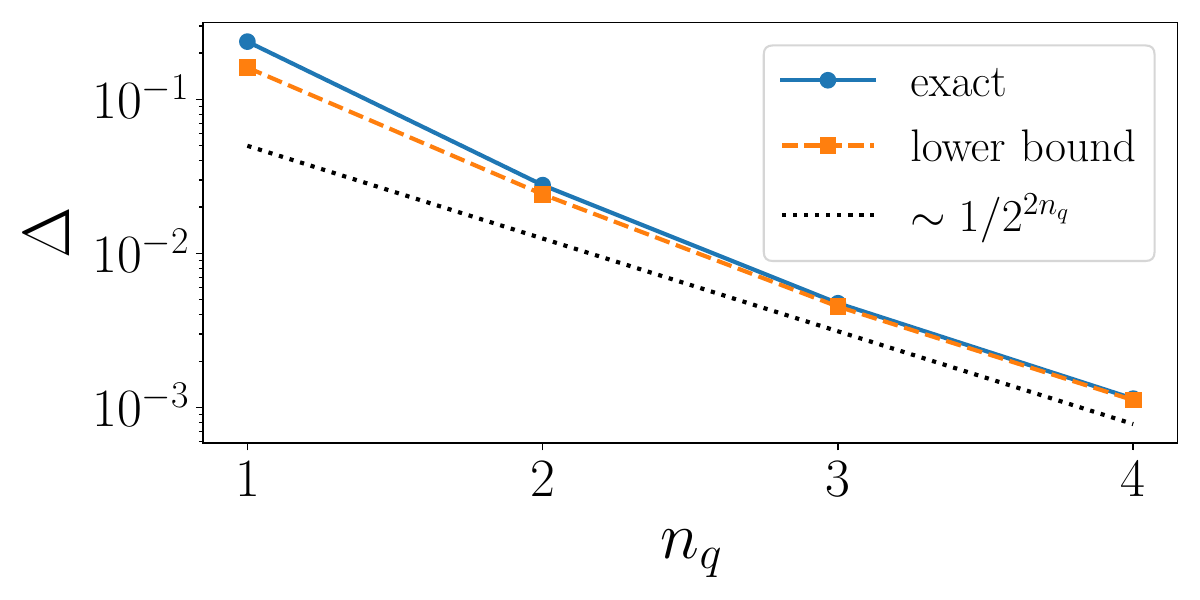}
    \includegraphics[width=0.45\textwidth]{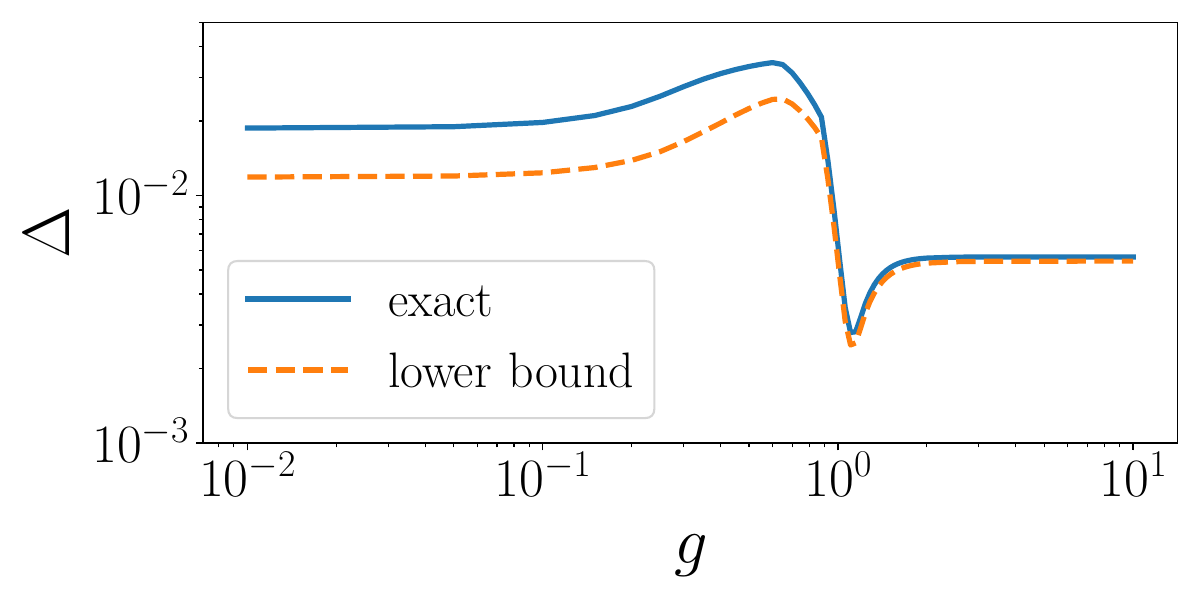}
    \caption{Comparison of the exact value of $\Delta$ to the upper bound calculated using the procedure in Sec.~\ref{ssec:u1_numerical}. Top: $\Delta$ as a function of $N_p$ using $n_q=2$ and $g=1.4$.
    The gap scales asymptotically as $1/(N_p \log^2(N_p))$. Middle: $\Delta$ as a function of $n_q$ using $N_p=3$ and $g=1.4$.
    The gap generally decreases as $1/2^{2n_q}$.
    Bottom: $\Delta$ as a function of $g$ using $N_p=3$ and $n_q=3$.
    The gap is generally independent of $g$, with a large dip near $g\sim 1$, which is due to the gap $a(E_1-E_0)$ becoming small.
    This behavior near $g=1$ is an artifact of using the weaved basis.}
    \label{fig:max_E_upper_bound}
\end{figure}

In a similar way, an upper bound of the maximum value of $H_E$ can also be found.
In both the original and weaved basis, we can write the Hamiltonian generally as $g^2/2 \sum_{i,j=0}^{N_p-1} c_{ij} R_i R_j$ (note that many of the $c_{ij}$'s are zero).
Because each $R_i R_j$ term in this sum is a $2^{2n_q} \times 2^{2n_q}$ diagonal matrix, we can explicitly evaluate the spectrum of each term classically at a cost quadratic in $N_p$. 
The upper bound placed on the maximum energy of $H_E$ is given by 
\begin{equation}
    \max(H^{(e)}_E) \leq  \frac{g^2}{2 a} \sum_{i,j=0}^{N_p-1} \max(c_{ij}R_i R_j).
\end{equation}
Our final upper bound on the maximum energy of the full Hamiltonian is then found using Eq.~\eqref{eq:upper_bound_maxH}. 
Using this upper bound, combined with the exact values for $aE_1$ and $aE_0$, we can place a lower bound on $\Delta$.
This lower bound is compared to the exact value in Fig.~\ref{fig:max_E_upper_bound}, as a function of $N_p$, $n_q$, and $g$.
We discuss each plot individually.

The top plot shows the exact value and upper bound of $\Delta$
as a function of $N_p$ for $n_q=2$ and $g=1.4$. 
Before discussing the results, we point out that $N_p=3,5,7$ correspond to lattices with sites $N_x \times N_y$ of $2\times2$, $2\times 3$, and $2\times 4$, respectively.
Due to the inherent limitations of classical simulation, we only increase the number of sites in a single dimension, and so we expect the finite volume errors to remain roughly the same size for all values of $N_p$.
From the plot we see that the upper bound is larger than the exact value, with the difference generally growing with $N_p$. 
The overall scaling of $\Delta$ is roughly as $1/(N_p \log^2(N_p))$. 
To understand this behavior, we can study how the number of terms in the Hamiltonian grows with $N_p$. 
For $H_B$, the number of terms grows linearly with $N_p$, implying that the maximum entry in $H_B$ scales linearly with $N_p$ as well. 
While the number of terms in $H_E$ depends on the specific weaved matrix used, general statements can be made about the scaling.
As demonstrated in Ref.~\cite{Grabowska:2022uos}, in order to break the exponential volume scaling in the gate count, a single rotor operator in the original basis is generally expressed as $\mathcal{O}(\log N_p)$ operators in the weaved basis.
Because $H_E$ is a sum of $N_p$ terms that are squares of differences of rotors, this leads to the number of terms scaling as $\mathcal{O}(N_p \log_2^2 N_p)$.
For these reasons, we expect gap to scale roughly as $\log_2 N_p$.
However, for smaller values of $g$, we expect that $H_B$ and $H_E$ become of similar magnitude, and cancellations between $H_B$ and $H_E$ could lead to a milder scaling with $N_p$.

The middle plot shows the exact value and upper bound of $\Delta$ as a function of $n_q$ for $N_p=3$ and $g=1.4$. 
These results show that the quality of the upper bound generally increases with $n_q$. 
Furthermore, we see that the lower bound generally scales as $\mathcal{O}(2^{-2n_q})$. 
To understand this scaling, we start with $H_B$, which is a sum of the cosine of sums of magnetic operators.
Even though the number of Pauli-Z operators in each magnetic operator grows exponentially with $n_q$, the maximum value a given cosine can take is 1, regardless of $n_q$. 
For $H_E$ on the other hand, the rotor operators do not appear inside a cosine, and therefore the maximum eigenvalue grows with the number of terms. 
Additionally, the maximum eigenvalue of a single rotor operator scales as $\mathcal{O}(2^{n_q})$, as seen from the relation $r_\text{max,$p$} = \pi 2^{n_q} / (2 b_\text{max,$p$})$. 
Because each term in $H_E$ is bilinear, the maximum eigenvalue of each term grows as $\mathcal{O}(2^{2n_q})$, leading to the observed scaling. 
For smaller values of $g$, however, we expect that $H_B$ and $H_E$ become of similar magnitude, and the scaling with $n_q$ will likely be more mild due to cancellations between $H_B$ and $H_E$.

Lastly, the bottom plot shows $\Delta$ as a function of $g$ (which is a function of the lattice spacing $a$) for $N_p=3$ and $n_q=3$.
First, notice that the quality of the upper bound is higher for large $g$. 
Second, except for the region of $g \sim 1$, the value of $\Delta$ is roughly independent of $g$.  
The roughly constant behavior for small and large $g$ can be understood by the weak and strong coupling limits of the Hamiltonian. 
For large $g$, the electric Hamiltonian dominates. 
This, combined with the fact that $b_\text{max,$p$}$ approaches a constant for large $g$ and $H_E \sim g^2$, leads to the spectrum of $H_E$ increasing as $g^2$ for large $g$. Because $\Delta$ depends only on ratios of energy differences, the $g$ dependence cancels, and we expect $\Delta$ to approach a constant for large $g$. 
Similarly, $\Delta$ approaches a constant for small $g$, which can be understood by recalling that as $g \to 0$, the compact theory approaches the non-compact version.
The non-compact theory is a free theory, i.e., a theory of non-interacting harmonic oscillators, with the gauge coupling $g$ playing the role of the mass $m$ of the canonical quantum harmonic oscillator.
The spectrum of the compact U(1) Hamiltonian in the small $g$ limit is therefore independent of $g$, leading to the observed behavior. 
The large dip near $g \sim 1$ is an artifact of using the weaved basis, and is not present in the original basis. 
It was found in Ref.~\cite{Kane:2022ejm} that controlling digitization errors near $g=1$ required a tuning of the choices of $b_\text{max,$p$}$, which we did not perform here. It is therefore possible that after performing this tuning, the dip near $g=1$ will disappear.

We conclude this discussion by pointing out that, even though one can argue how $\Delta$ scales with various parameters, significant savings can still be achieved by either improving the lower bound, or performing a dedicated calculation to determine the exact value. 
Our studies indicate that the cost reduction of such a study will be more significant as one increases $N_p$ towards realistic values, and at small $g$.

Next, we study how the parameter $\gamma$ scales with $N_p, n_q$ and $g$. 
As will be argued in Sec.~\ref{sec:gsprep}, using direct state preparation methods to implement the initial guess wavefunction $\ket{\psi_\text{init}}$ results in an overlap $\gamma$ that is exponentially suppressed in the number of sites.
In this study, we instead consider implementing $\ket{\psi_\text{init}}$ using adiabatic state preparation, with the objective of demonstrating that, even for a simple adiabatic state preparation procedure, $\gamma$ can be made to decrease only polynomially with the number of sites.
Note that because we use a simple adiabatic procedure, our results likely contain considerable Trotter and adiabatic violation errors. For this reason, the observed scaling of $\gamma$ will not necessarily align with the expectation drawn from physical intuition.
However, we stress once more that the purpose of this study is to demonstrate that $\gamma$ can be made to decrease only polynomially with the number of sites.
A dedicated study comparing more sophisticated adiabatic state preparation procedures is left for future work.

The general strategy is to start in the ground state of the large coupling limit of our theory. 
At large $g$, the electric Hamiltonian dominates and the ground state approaches a state with constant entries; this state can be prepared by applying a Hadamard gate on each qubit. 
Once the large coupling ground state is prepared, the ground state of the target theory at some smaller coupling is prepared by adiabatically evolving between the two Hamiltonians.

We follow closely the procedure and notation given in Ref.~\cite{Kovalsky:2022wcy}.
The initial strong coupling Hamiltonian is denoted by $H_1$ and the target Hamiltonian by $H_2$, with gauge couplings $g_1$ and $g_2$, respectively.
The adiabatic evolution is performed according to a time-dependent Hamiltonian $H[u(t)]=(1-u(t)) H_1+u(t) H_2$, where $u(t) \in [0, 1]$ is the ramping function satisfying $u(0)=0$ and $u(T)=1$. 
The parameter $T$ is the total time the system is adiabatically evolved. 
The exact time-ordered time-evolution operator $U(T) = \mathcal{T} \me^{-i \int_0^T H[u(t)] dt}$ is implemented using Trotter methods in two stages.
First, the integral over $t$ is split into $M$ discrete steps of size $\delta t = T/M$, such that
\begin{equation}
    U(T) \approx \prod_{k=0}^{M-1} U_1(k, \delta t) U_2(k, \delta t)\,,
\end{equation}
where $U_1(k, \delta t) = \me^{-i H_1 \delta t_1}$ and $U_2(k, \delta t) = \me^{-i H_2 \delta t_2}$. 
The values of $\delta t_1$ and $\delta t_2$ are given by
\begin{align}
    \delta t_1 &= \int_{k \delta t}^{(k+1) \delta t} dt (1-u(t))\,,
    \\
    \delta t_2 &= \int_{k \delta t}^{(k+1) \delta t} dt \,  u(t)\,.
\end{align}
Second, for $i=1,2$, each operator $U_i(k, \delta t)$ is approximated using a single step of a first-order Trotter formula, written explicitly as $U_i(k, \delta t) \approx \me^{-i \delta t_i H_E} \me^{-i \delta t_i H_B}$.

For our study, we choose the strong coupling Hamiltonian with $g_1=10$ as $H_1$. 
The ground state is prepared by applying a Hadamard gate on each qubit. 
We choose the simple linear ramp with $u(t) = t/T$.
Furthermore, we set $T=1$ and perform the adiabatic evolution using $M=2$ steps. 
While it is known that, in order for the adiabatic theorem to be satisfied, the parameter $T$ needs to scale as one over the square of the smallest energy gap along the adiabatic trajectory of the unscaled/unshifted Hamiltonian, i.e., $T = \mathcal{O}((aE_1-aE_0)^{-2})$, see, e.g., Ref.~\cite{Cheung_2011}, we find that our simple parameter choices lead to polynomial scaling with all parameters. 
To be conservative, we choose a scaling that is equal to or worse than the observed scaling in our numerical results. 

\begin{figure}[h]
    \centering
    \includegraphics[width=0.45\textwidth]{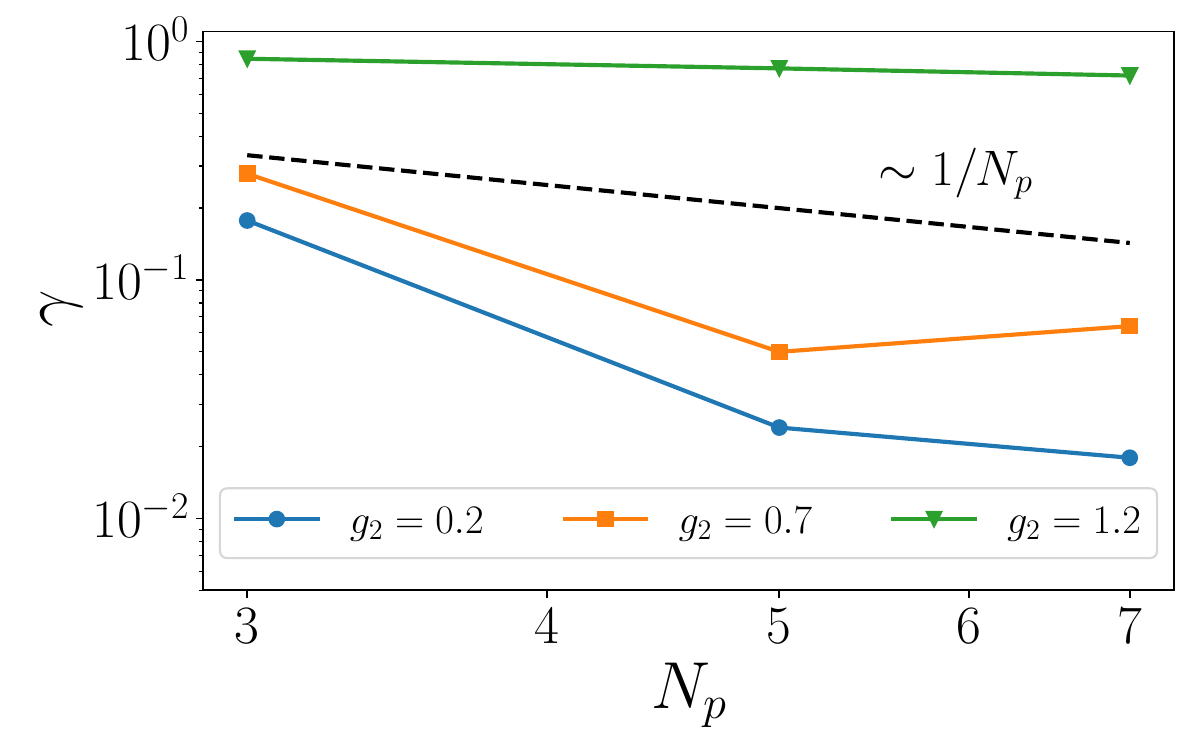}
    \includegraphics[width=0.45\textwidth]{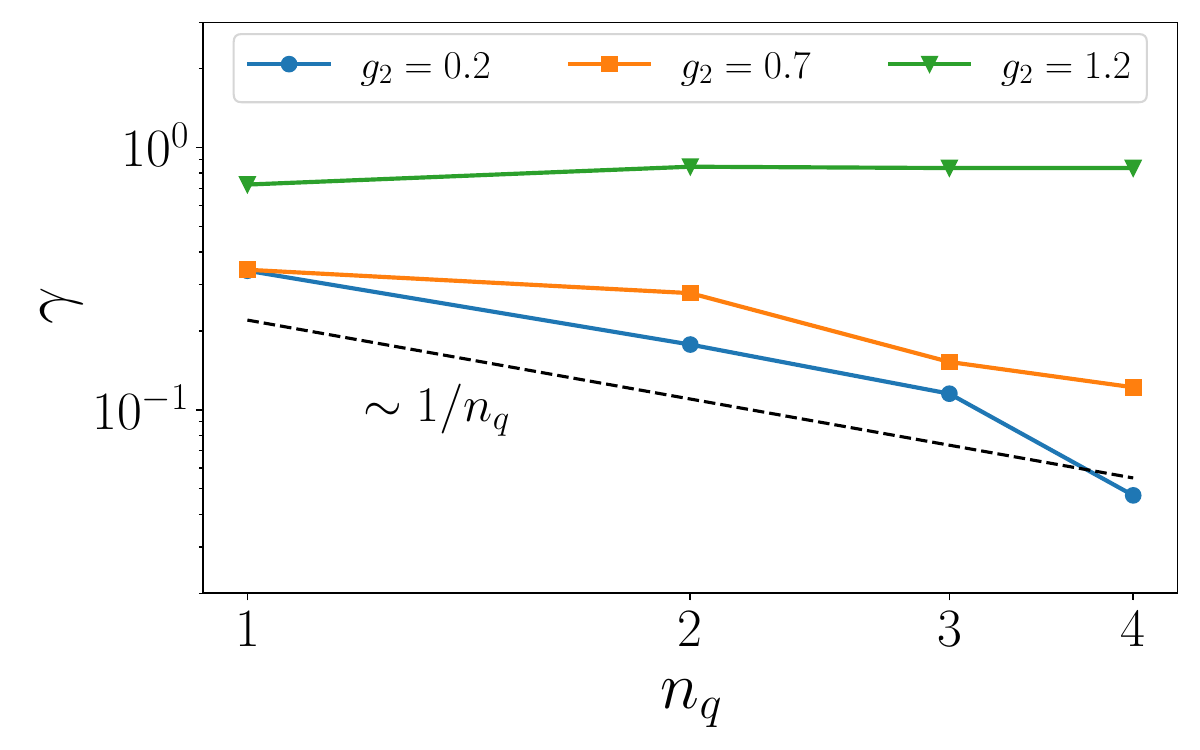}
    \includegraphics[width=0.45\textwidth]{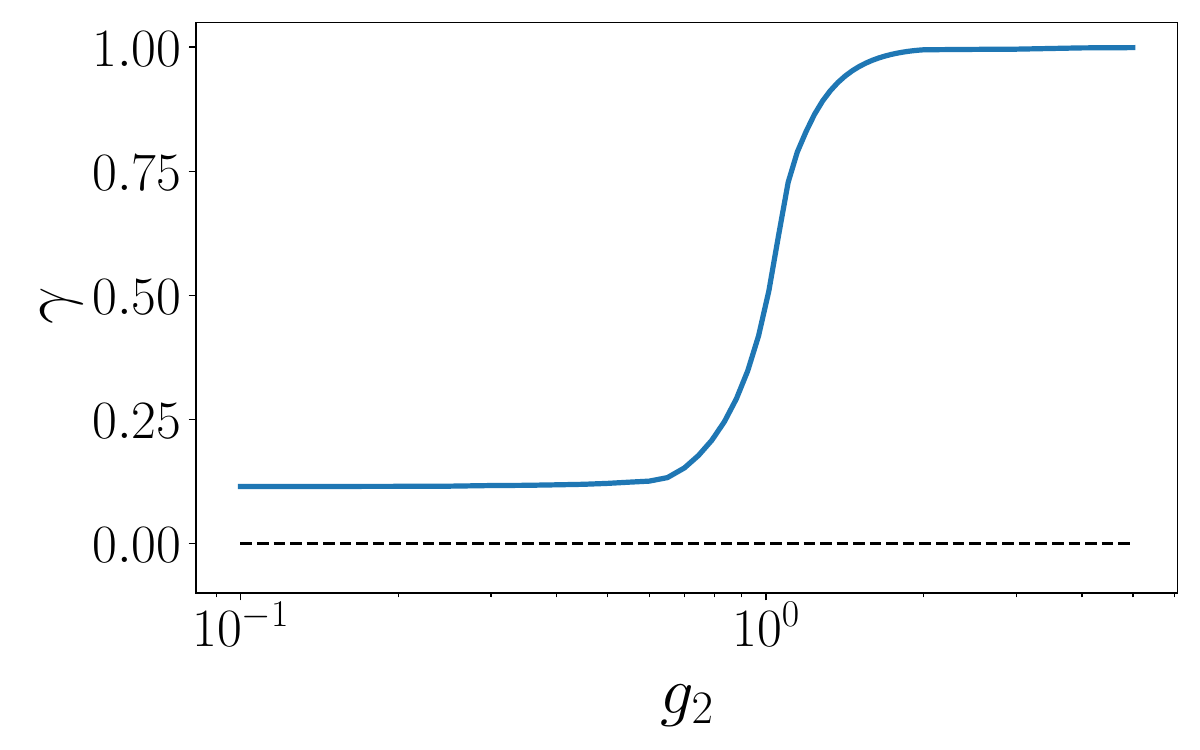}
    \caption{This figure shows how $\gamma$ scales with the parameters of our system when the initial guess is prepared using the adiabatic state preparation procedure described in the main text. Top: $\gamma$ as a function of $N_p$ for $n_q=2$. The different colored lines indicate different values of $g_2$.
    The small-volume results indicate that $\gamma$ is not exponentially decreasing as a function of $N_p$.
    The black dashed line shows a curve scaling as $\sim 1/N_p$. Middle: $\gamma$ as a function of $n_q$ for $N_p=3$. The different colored lines indicate different values of $g_2$. The black dashed line shows a curve scaling as $\sim 1/n_q$. Bottom: $\gamma$ as a function of the gauge coupling $g$ for $N_p=3$ and $n_q=3$. The value is near 1 for large $g$, and approaches a constant value for small~$g$.}
    \label{fig:gamma_vs_params}
\end{figure}

The top plot in Fig.~\ref{fig:gamma_vs_params} shows the value of $\gamma$ as a function of $N_p$ for multiple values of $g_2$. For $g_2=1.2$, $\gamma$ is large and only has a mild dependence on $N_p$. For $g_2=0.2, 0.7$ the value of $\gamma$ decreases from $N_p=3$ to $N_p=5$, but from $N_p=5$ to $N_p=7$ is relatively constant.
We observe a scaling no worse than $\gamma \sim \mathcal{O}(1/N_p)$.
The middle plot in Fig.~\ref{fig:gamma_vs_params} shows the value of $\gamma$ as a function of $n_q$ for multiple values of $g_1$. 
For $g_2=1.2$, we observe a mild dependence on $n_q$. For $g_2=0.2, 0.7$ we observe a general $1/n_q$ scaling. 
In this study we observe a scaling no worse than $\gamma \sim \mathcal{O}(1/n_q)$.
Lastly, the bottom plot in in Fig.~\ref{fig:gamma_vs_params} shows $\gamma$ as a function of $g_2$ for $N_p=3$ and $n_q=3$. 
For $g_2 \gtrsim 2$ we find $\gamma$ to be close to 1. This is expected as the large coupling ground state with $g_1=10$ and the ground state for $g_2 \gtrsim 2$ have reasonable overlap. 
As one decreases $g_2$ further, we observe a steep decrease in $\gamma$ near $g_2=1$.
The value of $\gamma$ approaches a constant in the small $g_2$ limit. 
This behavior is consistent with the decrease in overlap between the ground state of the strong coupling Hamiltonian $H_1$ and the ground state of the target Hamiltonian $H_2$.

For the purpose of quoting how $\gamma$ scales with the parameters of our system, we interpret the numerical results 
in a conservative way; we choose a scaling that is equal to or worse than the observed scaling in our numerical results.
Due to the simplicity of our adiabatic procedure combined with the small volumes that are accessible to classical simulation, predicting the scaling for realistic lattice sizes will require further dedicated studies.

\begin{figure}[h]
    \centering
    \includegraphics[width=0.45\textwidth]{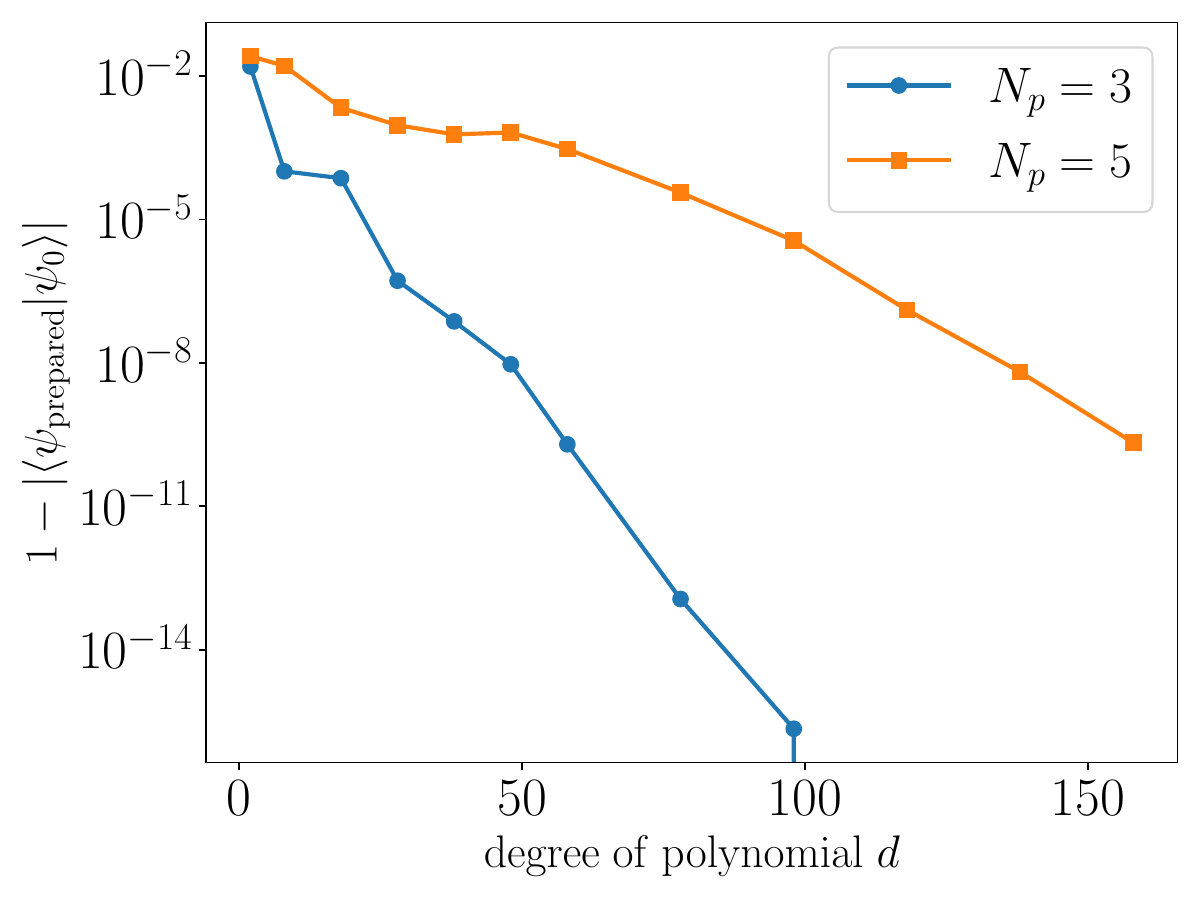}
    \caption{Error of the ground state prepared using QETU as a function of the degree of the Chebyshev approximation $d$, where the time evolution operator was implemented \textit{exactly}. Different colored points correspond to different number of plaquettes $N_p$. All results used $n_q=1$ and $g=1.4$. The error decreases exponentially with $d$, with the rate being slower for $N_p=5$ due to a smaller gap $\Delta$.}
    \label{fig:error_vs_Np_exact}
\end{figure}

We now study the fidelity of the prepared state as a function of the number of calls to the \textit{exact} time evolution operator $\me^{-i \tau H}$.
The spectral norm and energy gap were determined exactly by diagonalizing the Hamiltonian. 
The value of $\Delta = (E_1 - E_0)/1.5$ was used, with $\tau = 1$.
We denote the state prepared using QETU by $\ket{\psi_\text{prepared}}$, and the exact ground state by $\ket{\psi_0}$. The error is defined to be $1 - |\bra{\psi_\text{prepared}}\ket{\psi_0}|$.
Figure~\ref{fig:error_vs_Np_exact} shows the error as a function of the degree of the polynomial $d$ used to approximate the shifted sign function for $N_p = 3, 5$, $n_q=1$, and $g=1.4$.  
The error decreases exponentially as we increase $d$, with the rate of convergence being slower for $N_p=5$ due to the smaller value of $\Delta$.

\begin{figure}[h]
    \centering
    \includegraphics[width=0.48\textwidth]{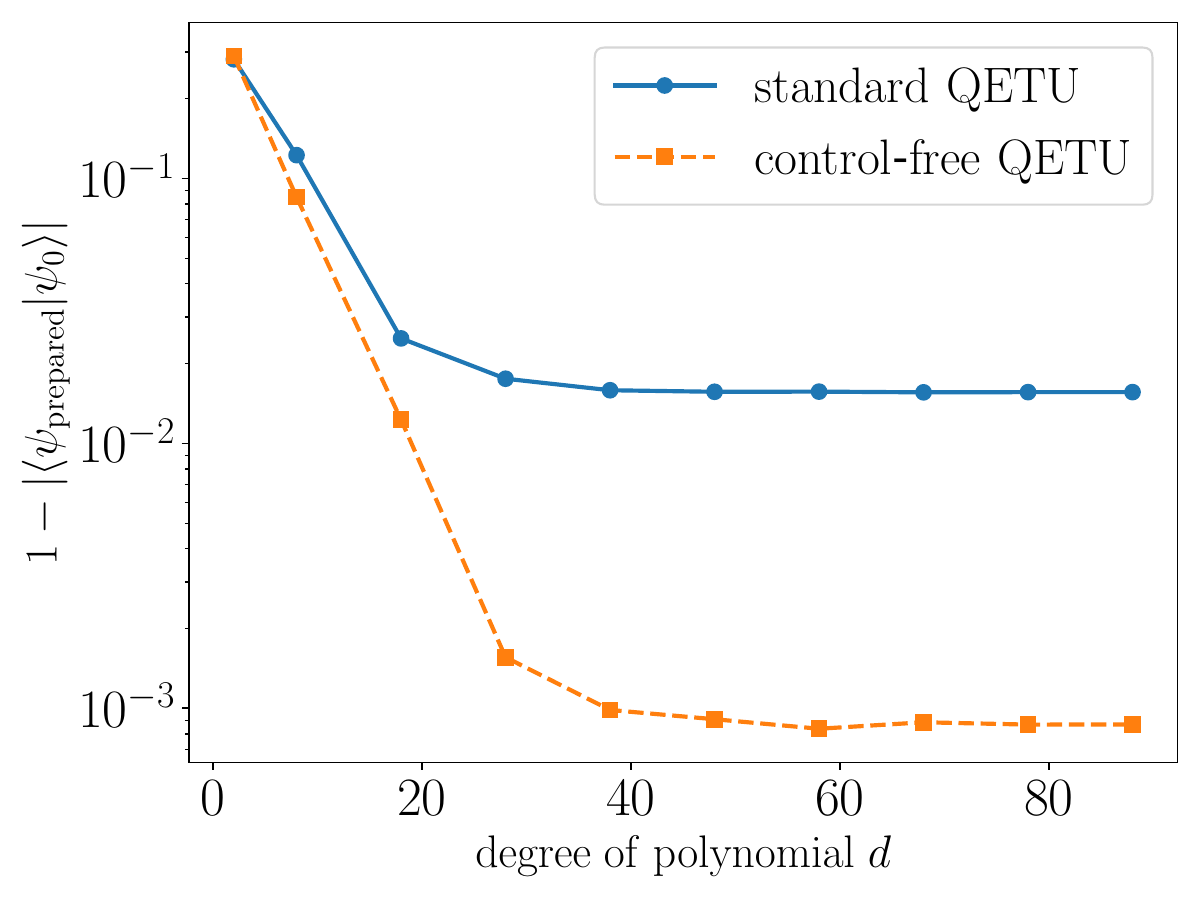}
    \caption{Error as a function of $d$ using both the standard and control-free versions of QETU. For both cases, a single Trotter step was used to approximate $\me^{-i \tau H}$ with $\tau = 1.78$. The step sizes used were $\delta \tau = \tau$ and $\delta \tau = \tau/2$ for the standard and control-free versions, respectively. The Trotter error in the control-free version is smaller than the standard version by an order of magnitude.}
    \label{fig:compare_ctrl_free_to_std}
\end{figure}

We now move to the studies of \textit{approximate} implementations of the time-evolution operator with the aid of Trotter methods. 
The first study compares the overall Trotter error between the standard controlled version of QETU and the control-free version.
As explained in Sec.~\ref{ssec:qetu_review}, the control-free version of QETU algorithm is designed to avoid using controlled calls to the time-evolution circuit.
This method requires a Hamiltonian dependent procedure; the details for the U(1) case are given in App.~\ref{app:ctrl_free}. 
By repeatedly calling $\me^{-i \tau H}$, the standard version returns $F(\cos(\tau H/2))$, while the control-free version of QETU returns $F(\cos(\tau H))$. 
This implies that one can use instead $\me^{-i \tau H/2}$ as a building block for control-free QETU. 
Because one only has to time evolve by half the total time $\tau/2$, for the same number of Trotter steps, one can use a step size half as large, leading to a smaller overall error relative to the standard controlled version of QETU. 
Figure~\ref{fig:compare_ctrl_free_to_std} shows the error as a function of $d$ when using both the standard and control-free versions of QETU. 
The system parameters used are $N_p=3, n_q=2$ and $g=0.6$. 
We approximated $\me^{-i \tau H}$ with $\tau = 1.78$ using a single Trotter step of $\delta \tau = \tau$ and $\delta \tau = \tau/2$ for the standard and control-free versions of QETU, respectively.
As $d$ is increased, the error of both methods leveled out, with the error of the control-free version being an order of magnitude smaller due to the smaller time-step used.
We see that, in addition to requiring less gates, the control-free version of QETU also results in a smaller Trotter error for the same number of calls to the time-evolution circuit.
The remainder of the results in this section were obtained using the control-free version of QETU.

\begin{figure}[h]
    \centering
    \includegraphics[width=0.48\textwidth]{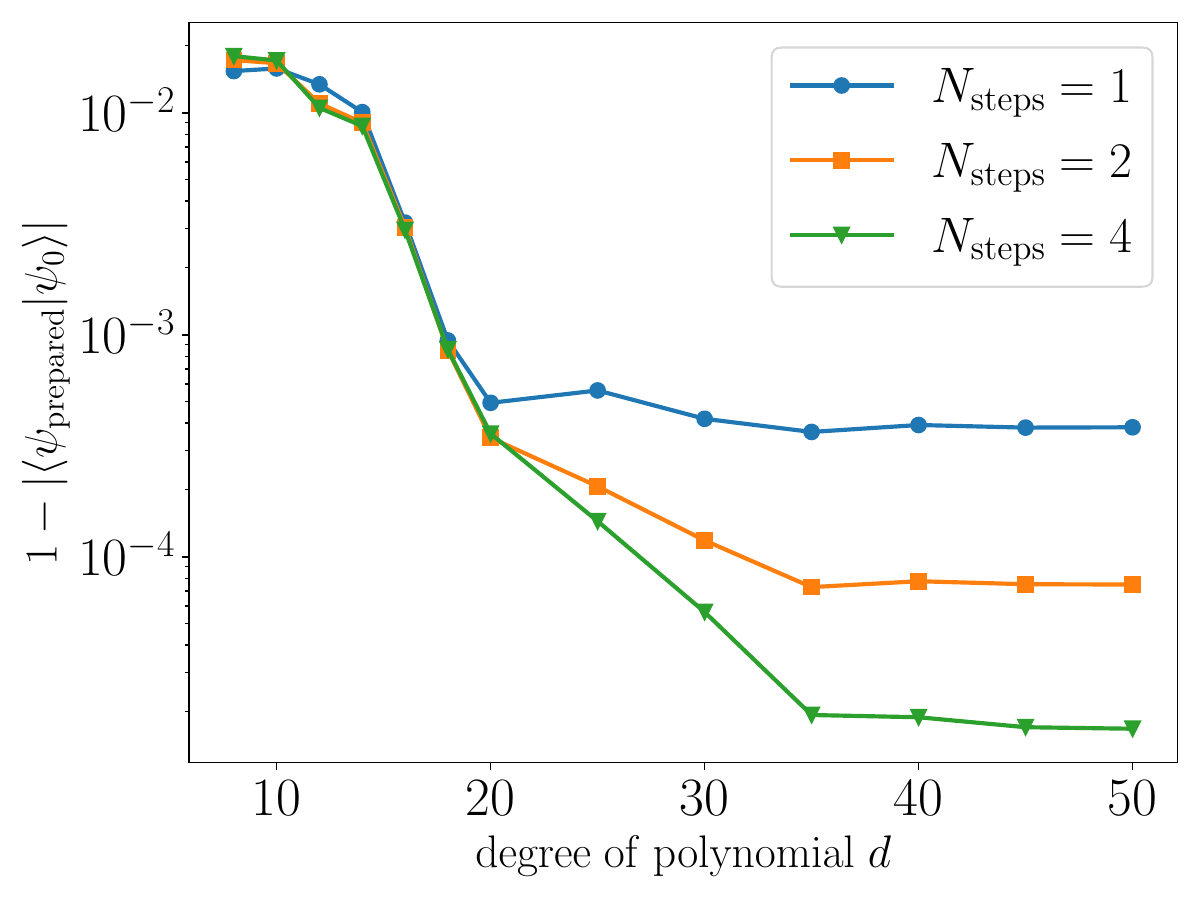}
    \caption{Error of the state prepared using the control-free version of QETU as a function of $d$, where different colored and shaped data points correspond to different numbers of Trotter steps used to approximate the time-evolution circuit. The results are shown for $N_p=3$, $n_q=2$, $g=0.6$, and $\tau = 1.5$. As the number of steps is increased, the error saturates at large $d$ to smaller values due to the reduced Trotter errors.}
    \label{fig:error_vs_deg_many_dtau}
\end{figure}

We now study how the maximum achievable precision depends on the Trotter step size $\delta \tau$. 
Figure~\ref{fig:error_vs_deg_many_dtau} shows the error as a function of calls to the time evolution operator $\me^{-i \tau H}$ for $\tau = 1.5$, approximated using a first order Trotter formula with $N_\text{steps} = 1,2,4$. 
The parameters of the system studied are $N_p=3, n_q=2$ and $g=0.6$.
We see that the error decreases exponentially at first and then levels out for large number of calls.
This occurs because the error is now dominated by the Trotter error, and improving the quality of the approximation of the projector is no longer beneficial. 
Even though we use the first order Trotter formula, the maximum precision achievable scales as $\mathcal{O}(\delta \tau^2)$. This is due to the fact that the leading order error term of the form $\delta \tau [H_E, H_B]$ is zero for this system.

\begin{figure}[h]
    \centering
    \includegraphics[width=0.48\textwidth]{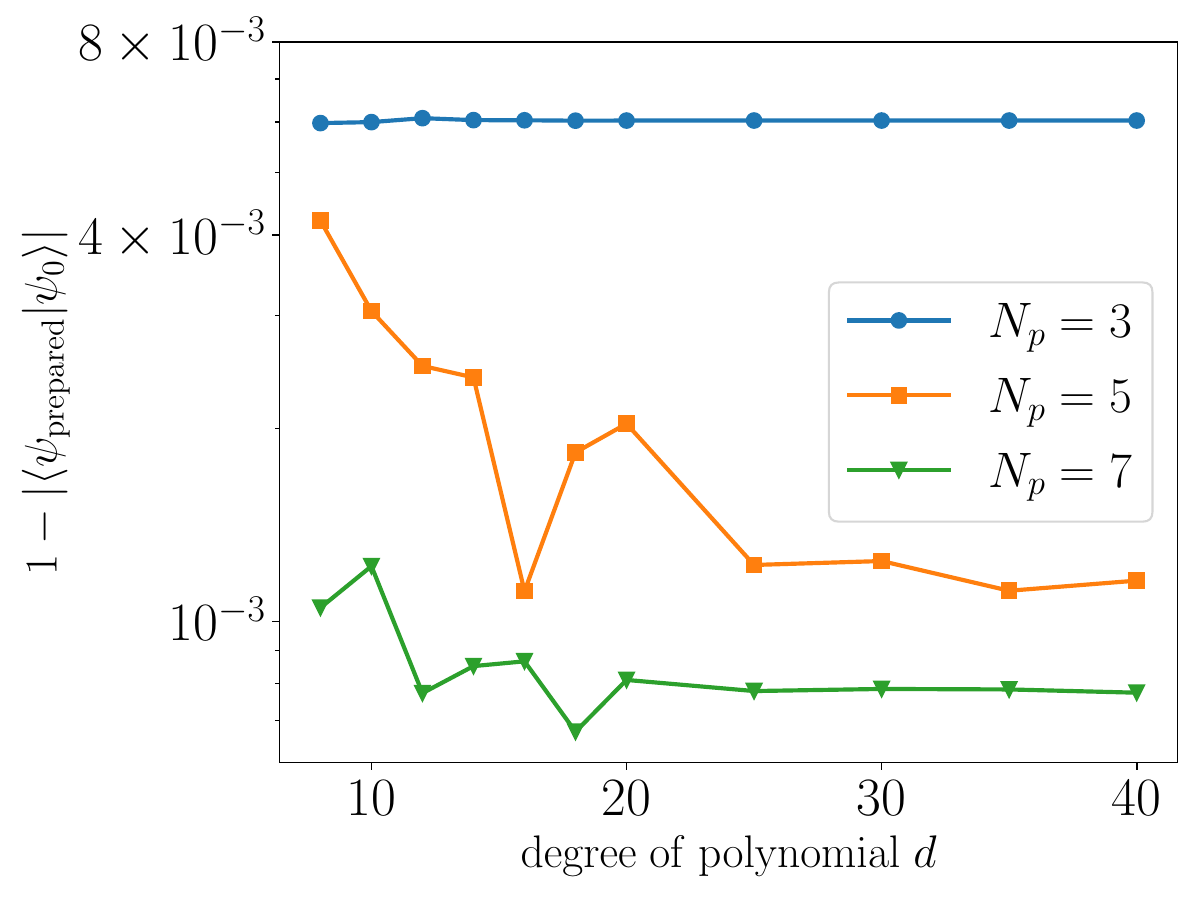}
    \caption{Error of the state prepared with the control-free version of QETU as a function of $d$, where different colored/shaped data points correspond to different values of $N_p$. The results are shown for $n_q=1$, $g=1.4$, and $\tau = 1.5$. A single Trotter step was used to approximate the time-evolution circuit. As the number of plaquettes $N_p$ increases, the Trotter error decreases. This behavior is due to the effective Trotter step size decreasing with~$N_p$.}
    \label{fig:trotter_error_vs_Np}
\end{figure}

The final study we perform regarding the Trotter error is how, for some fixed total precision, the Trotter step size $\delta \tau$ must scale with the volume.
Typically, as one increases the number of terms in a Hamiltonian, for fixed step size, the Trotter error increases due to the increased number of terms that do not commute.
From this argument, as we increase $N_p$ or $n_q$, we expect that $\delta \tau$ must be decreased accordingly if one wants to maintain a constant level of precision. 
However, scaling $H_E$ and $H_B$ by the parameter $c_1$ is equivalent to scaling the Trotter step size by $c_1$. 
Because $c_1$ generally increases with $N_p$ and $n_q$, the effective Trotter step size decreases with $N_p$ and $n_q$. 
If the decrease in error from the smaller effective step size is more significant than the increase in error from the extra non-commuting terms, then the Trotter error will \textit{decrease} as we increase $N_p$ or $n_q$.
We observe this to be the case, and show an example of this counter-intuitive behavior in Fig.~\ref{fig:trotter_error_vs_Np}.
In this plot, we show the error as a function of $d$ for three values of $N_p=3,5,7$ using $n_q=1$ and $g=1.4$.
For each value of $N_p$, we use a single Trotter step with $\delta \tau = 1.5$. 
Notice that we can use a value of $\delta \tau>1$ and still see convergence due to the fact that the effective Trotter step size is scaled by $c_1$.
Looking at Fig.~\ref{fig:trotter_error_vs_Np}, we see that as we increase $N_p$, the maximum achievable precision increases.
This behavior was also observed as $n_q$ was increased. 
Because this behavior is expected to continue as one increases $N_p$ and $n_q$ towards realistic values, the Trotter error will eventually become negligible for any realistic precision requirements. 
Practically speaking, this implies that for a realistic calculation, one can approximate the time evolution operator using a single Trotter step of a first order Trotter formula with $\delta \tau = \tau_\text{max}$, independent of $N_p$ or $n_q$. 
What at first seemed like a technical feature of QETU, turns out to offer a powerful protection against further $N_p$ or $n_q$ scaling. 
We conclude by stressing that, even though the effective step size is scaled by $c_1$, the un-scaled $\delta \tau$ must still satisfy $\delta \tau \leq \tau_\text{max}$ in order to guarantee isolation of the ground state when applying the filter operator.

We conclude with the classical computational cost of calculating the angles $\{\varphi_j\}$ needed in the QETU circuit. We found numerically that the cost scales quadratically with the number of angles. Because the number of angles is proportional to $\Delta^{-1}$, the associated classical cost scales as $(\Delta^{-1})^2 = \mathcal{O}(N_p^2 2^{4 n_q})$.

\begin{table}[]
    \centering
    \begin{tabular}{ccc}
        Parameter & & Scaling 
        \\
        \hline \hline
        $\gamma$ & & $\mathcal{O}(N_p^{-1} n_q^{-1})$  
        \\
        $\Delta$ & & $\mathcal{O}(N_p^{-1} 2^{-2n_q})$
        \\
        $N_\text{steps} = \tau / \delta \tau$ & & $\mathcal{O}(1)$
        \\
        Gates$(\me^{-i \delta \tau H})$ & &$\mathcal{O}(N_p^{n_q})$
    \end{tabular}
    \caption{Scaling of parameters in the cost of state preparation using QETU in terms of $N_p$, $n_q$, and $g$. 
    The $\gamma$ parameter defines the number of measurements needed to measure the ancillary qubit in the zero state, which scales as $O(1/\gamma^2)$. This scaling can be can be improved to $O(1/\gamma)$ using amplitude amplification at the cost of increasing the circuit depth by a factor of $\gamma^{-1}$.
    The $\Delta$ parameter defines the query depth of the time evolution circuit $\me^{-i \tau H}$, which scales as $\mathcal{O}(\Delta^{-1})$.
    Note that while the gauge coupling $g$ does not appear in the asymptotic scaling of the $\gamma$ and $\Delta$ parameters, their values still depend on on $g$.
    The $\tau / \delta \tau$ parameter defines the number of Trotter steps used when approximating $\me^{-i \tau H}$, which, surprisingly, does not scale with $N_p$ or $n_q$.
    The Gates$(\me^{-i \delta \tau H})$ parameter defines the number of gates required to implement a single Trotter step for the particular formulation of U(1) gauge theory we consider~\cite{Grabowska:2022uos, Kane:2022ejm}.
    This unusual scaling is due to the fact that the theory we consider is highly non-local.
    }
    \label{tab:cost}
\end{table}

\subsection{Wavepacket construction \label{ssec:numerics_wp}}

In this section, we use QETU to prepare a Gaussian state, defined in Eq.~\eqref{eq:wavepacket}, with $x_0=0$ and $\sigma_x=0$.
We first show an example of how na\"ively applying QETU according to Eq.~\eqref{eq:minmax_wp_naive} leads to the error decreasing only polynomially with the number of Chebyshev polynomials. 
Next, we show how the modifications described in Sec.~\ref{ssec:wp_intro} achieve an exponential scaling in the error for any desired value of the width, including a method that avoids the costly implementation of LCU to add the even and odd components of $\Fex(x)$.
From there, we compare the gate count cost of our method to that of exact state preparation methods, and find that our method requires less gates than exact state preparation methods for states represented by $>2-5$ qubits.

\begin{figure}[h]
    \centering
    \includegraphics[width=0.48\textwidth]{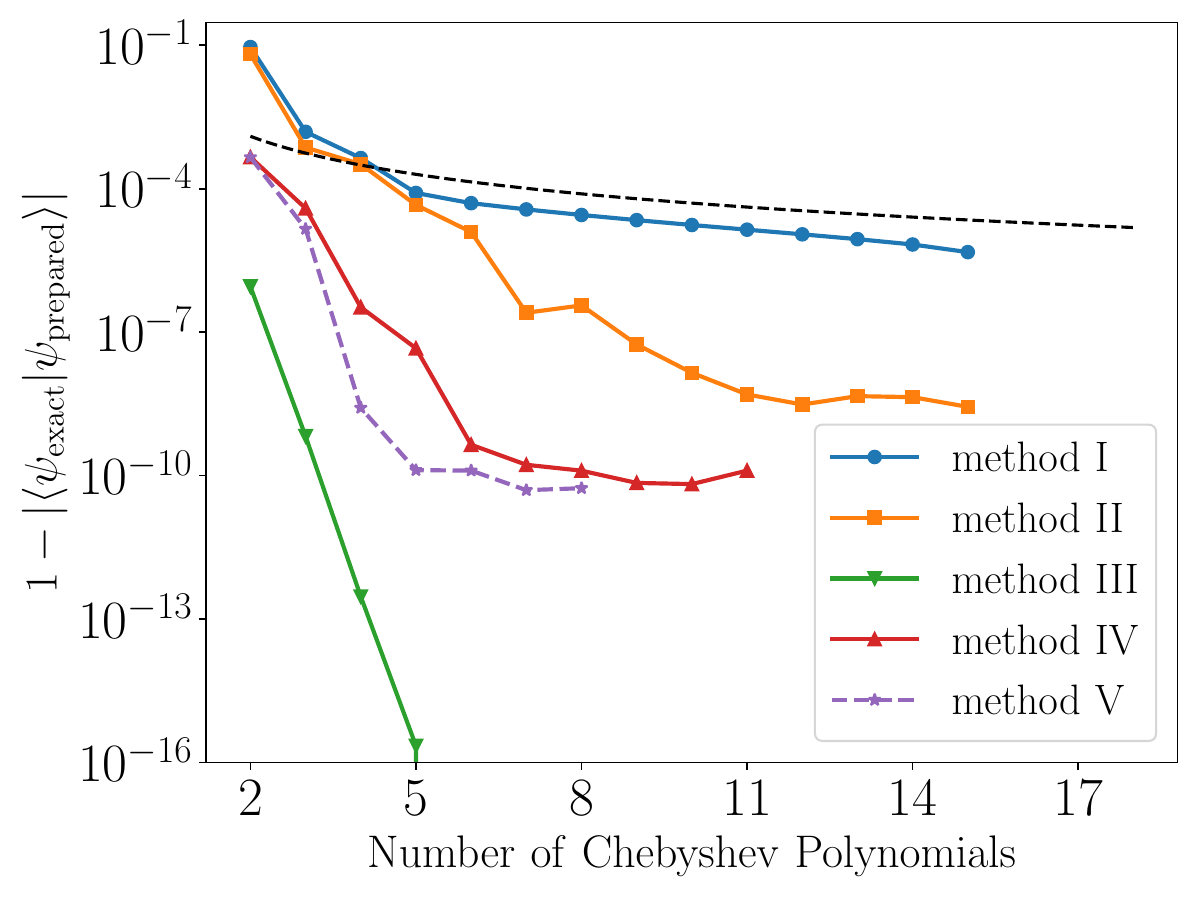}
    \caption{Error of the prepared Gaussian state using QETU as a function of the number of Chebyshev polynomials included in both the even and odd components. 
    Different colored points correspond to different methods of construction. 
    Method I determined the Chebyshev expansion by solving the optimization problem in Eq.~\eqref{eq:minmax_wp_naive}, using $\eta = 0$ and $\tau = 1$.
    Method II shows results solving the optimization problem in Eq.~\eqref{eq:minmax_allX_vary_eta_tau} for $\tau$ fixed to $\tau = 1$. 
    Method III used values for $\eta$, $\tau$, and the Chebyshev coefficients by solving the optimization problem in Eq.~\eqref{eq:minmax_allX_vary_eta_tau}.
    Method IV first set $\tau=2$ to make $\Fex(x)$ purely even, and used values for $\eta$ and the Chebyshev coefficients by solving the optimization problem in Eq.~\eqref{eq:minmax_allX_vary_eta_tau}.
    Method V also set $\tau=2$, and used values for $\eta$ and the Chebyshev coefficients by solving the optimization problem in Eq.~\eqref{eq:minmax_only_sampled} where one only samples the function at points $\tilde{x}_j$.
    The black dashed line shows the curve $\sim 1/n^2$. 
    The results are shown for $n_q=4$ and $\sigma_x / x_\text{max} = 0.4$. 
    The error using Method I decreases quadratically with the number of Chebyshev polynomials. 
    Using Method II, the error appears to at first decrease exponentially, then decreases only polynomially.
    The error using Method III decreases exponentially, and reaches floating point precision with only five Chebyshev polynomials.
    Using methods IV and V, the error first decreases exponentially, then levels out at $\sim 10^{-10}$, with method V generally outperforming method IV.}
    \label{fig:wp_compare_methods}
\end{figure}

As explained in Sec.~\ref{ssec:wp_intro}, due to the presence of the $\arccos(x)$ term, we expect the error of the approximation to decrease only polynomially with the number of Chebyshev polynomials. 
An example of the polynomial convergence is shown in Fig.~\ref{fig:wp_compare_methods}, using parameters $n_q = 4$, $\sigma_x/x_\text{max}=0.4$, $\tau = 1$ and $\eta = 0$.
Looking at the data labeled Method I, we observe that the error converges quadratically as the number of Chebyshev polynomials is increased. 
We now discuss modifications that can improve the scaling.

The first modification we study is to determine the parameter $\eta$ and the Chebyshev coefficients according to the optimization problem in Eq.~\eqref{eq:minmax_allX_vary_eta_tau}, except with $\tau$ fixed to $\tau = 1$. Looking at the data labeled Method II in Fig.~\ref{fig:wp_compare_methods}, we see that while varying $\eta$ improves the scaling to a point, the error eventually starts to decrease polynomially.

To improve upon the previous method, we now allow the parameter $\tau$ to be chosen, along with $\eta$ and the Chebyshev coefficients, according to the optimization problem in Eq.~\eqref{eq:minmax_allX_vary_eta_tau}. 
During our numerical tests, we found that standard minimization techniques were highly sensitive to the initial values of $\eta$ and $\tau$.

The next modification we study is now $\tau$ to determine the parameter $\eta$ and the Chebyshev coefficients according to the optimization problem in Eq.~\eqref{eq:minmax_allX_vary_eta_tau}, except with $\tau$ fixed to $\tau = 1$. Looking at the data labeled Method II in Fig.~\ref{fig:wp_compare_methods}, we see that while varying $\eta$ improves the scaling to a point, the error eventually starts to decrease polynomially. 
To avoid this problem, we first determine good starting values for $\eta$ and $\tau$ using a brute force approach, and then use these values as inputs into a standard minimization procedure.
While brute force approaches are technically inefficient, for our two dimensional parameter space and cheap cost function, we find this method to be an appropriate choice.
Note that negative values of $\eta$ are allowed, and often result in the smallest error.
Our numerical studies routinely found that using a value of $\tau = 4$ and varying $\eta$ leads to the smallest errors. 
One possible explanation for this is that, for $\tau = 4$, the parameter $\eta$ can be varied to change the shape of the function $\Fex(\tilde{x}_i)$ to be approximately linear and quadratic, for the odd and even components, respectively; functions resembling linear and quadratic dependence are approximated well using only a few Chebyshev polynomials.
Figure~\ref{fig:wp_compare_methods} shows the error using this procedure as a function of the number of Chebyshev polynomials, labeled as Method III. 
The error is five orders of magnitude smaller than using the previous methods, and decreases exponentially.

While we have demonstrated that QETU can be used to implement Gaussian states with a cost $\mathcal{O}(n_q \log(1/\epsilon))$, the cost of performing LCU to add the even and odd pieces introduces a large overall coefficient. 
If one could modify the procedure to avoid using LCU, the gate count would be reduced by a factor of 10 or more. 
As discussed in Sec.~\ref{ssec:wp_intro}, because of our specific digitization of the $\hat{x}$ operator, choosing a value of $\tau=2$ results in the function $\Fex(x)$ being purely even.
After setting $\tau = 2$, the parameter $\eta$ and Chebyshev coefficients are determined by solving the optimization problem in Eq.~\eqref{eq:minmax_allX_vary_eta_tau} with $\tau$ set to $\tau = 2$.
Figure~\ref{fig:wp_compare_methods} shows the error as a function of the degree of the Chebyshev polynomial used.
The error decreases exponentially at first, levels off, then reaches zero as the number of parameters equals the number of points. 
Importantly, the error levels off at a value of $\sim 10^{-10}$, which will be completely negligible for a realistic simulation.
While the error for the same number of Chebyshev polynomials is generally larger than if one implemented the even and odd pieces separately and then added them using LCU, the cost of performing LCU leads to over an order of magnitude larger gate count. 
For this reason, converting $\Fex(x)$ into an even function to avoid the cost of LCU results in the best precision for a given gate cost. 

The final improvement studied exploits the fact that, because the eigenvalues of the $\hat{x}_\text{sh}$ operator are known exactly, we must only reproduce the function $\Fex(x)$ at those points.
After fixing $\tau = 2$, the parameter $\eta$ and the Chebyshev coefficients are determined according to the optimization problem in Eq.~\eqref{eq:minmax_only_sampled}.
Looking at Fig.~\ref{fig:wp_compare_methods}, we find the error using this method results in smaller errors than the previous method of sampling all $x$ values.
In a similar way, the error decreases exponentially, approaches a constant of $\sim 10^{-10}$, then becomes zero when the number of Chebyshev polynomials equals the number of points the function $\Fex(x)$ is sampled. 
This method results in the smallest error per gate count, and is used to produce the rest of the results in this section.

\begin{figure}[h]
    \centering
    \includegraphics[width=0.48\textwidth]{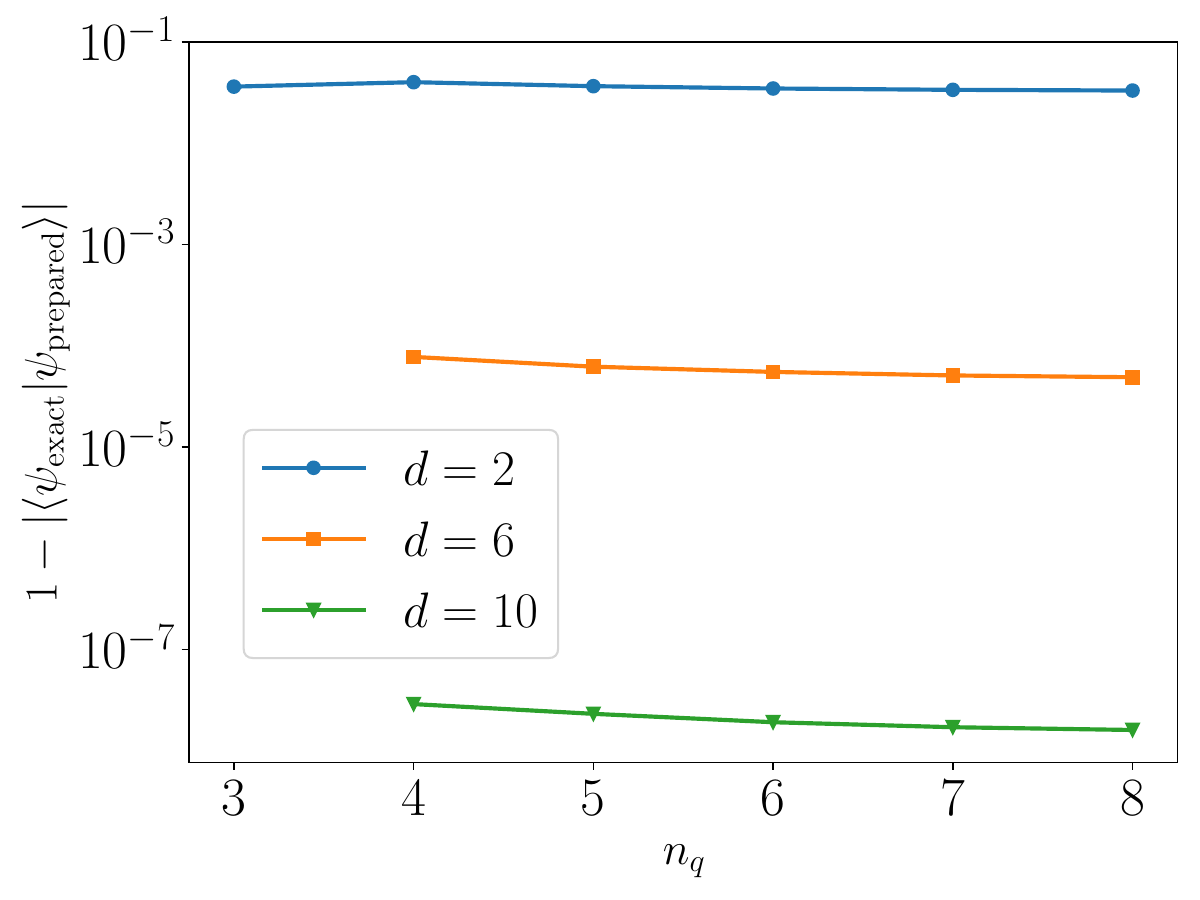}
    \caption{Error of the prepared Gaussian state using QETU as a function of the number of qubits $n_q$ used to represent the state. 
    The state was prepared by setting $\tau = 2$ and solving the optimization problem in Eq.~\eqref{eq:minmax_only_sampled}.
    Different colored points show different degree Chebyshev expansions.
    The results for $n_q=3$ using $d=6, 10$ are not shown because the error was zero. 
    Except for small values of $n_q$ where the state is prepared exactly, the error is independent of $n_q$. 
    Results are shown for $\sigma_x/x_\text{max} = 0.2$.
    }
    \label{fig:wp_even_error_vs_nq}
\end{figure}

\begin{figure}[h]
    \centering
    \includegraphics[width=0.48\textwidth]{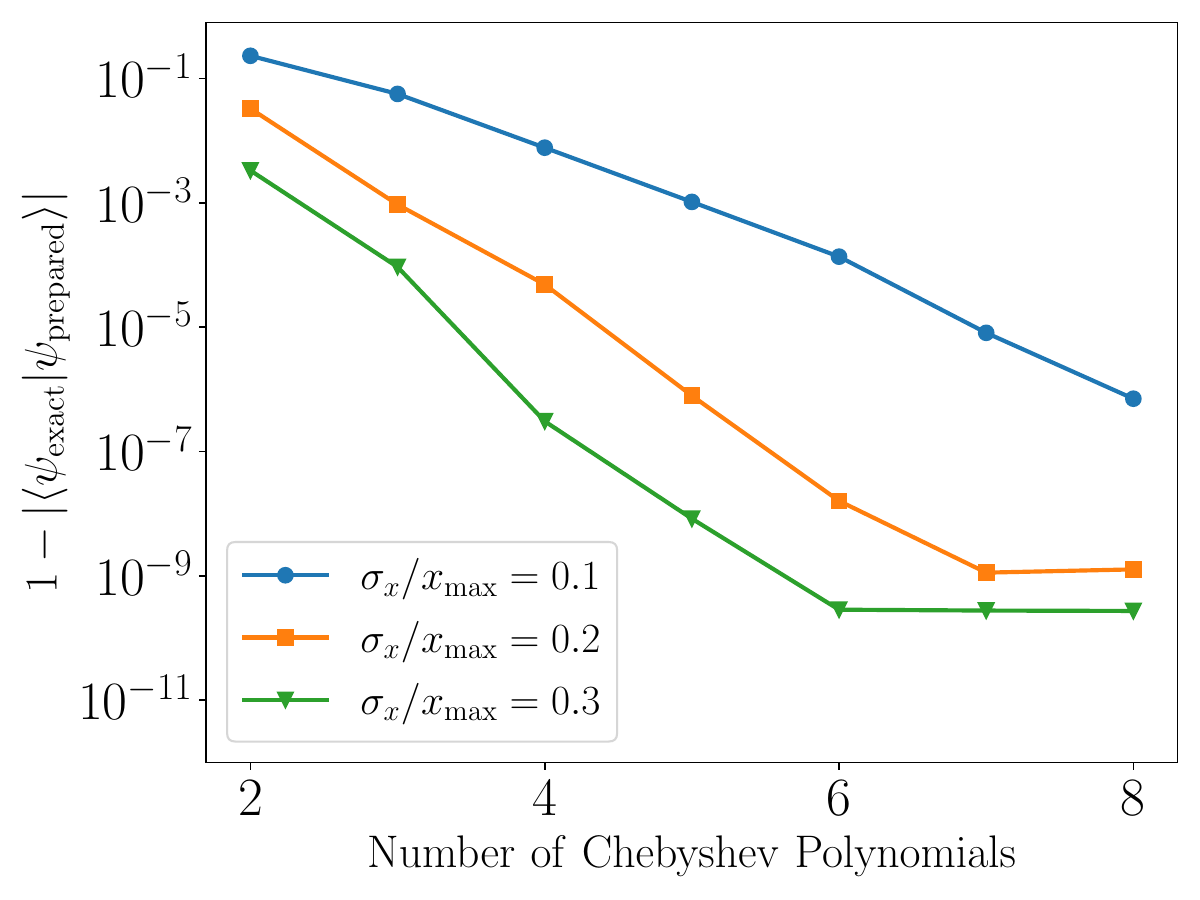}
    \caption{Error of the prepared Gaussian state using QETU as a function of the number of Chebyshev polynomials used to construct the Guassian filter operator. 
    The state was prepared by setting $\tau = 2$ and solving the optimization problem in Eq.~\eqref{eq:minmax_only_sampled}.
    Different colored points show different values of the wavepacket width $\sigma_x / x_\text{max}$.
    In all cases, the error decreases exponentially.
    More sharply peaked wavepackets require more Chebyshev polynomials to achieve the same level of precision.
    Results are shown for $n_q = 5$.
    }
    \label{fig:wp_even_error_many_sigma}
\end{figure}

We now study how the precision varies with the number of qubits $n_q$. Figure~\ref{fig:wp_even_error_vs_nq} shows the error of the prepared state as a function of $n_q$ for different degree polynomial approximations, using $\sigma_x / x_\text{max} = 0.2$. Our results indicate that the precision is independent of $n_q$, except in the cases where the small number of sample points results in an exactly prepared Gaussian state. From this we learn that the number of Chebyshev polynomials required to achieve some desired precision is independent of $n_q$. This result will be important when comparing the gate cost to exact state preparation methods. 

The final component of our precision study is to understand how the precision scales as the width $\sigma_x/x_\text{max}$ of the wavepacket is varied.
Figure~\ref{fig:wp_even_error_many_sigma} shows the error of the prepared state as a function of the number of Chebyshev polynomials used for different values of $\sigma_x/x_\text{max}$, using $n_q=5$ qubits.
We find that the error decreases exponentially for all values of the width, with more sharply peaked Gaussian states requiring more Chebysev polynomials to achieve the same precision as more broadly peaked states.
This intuitive result is similar to the fact that the cost of approximating the shifted error function increases as the energy gap $\Delta$ is decreased.

We conclude this section with a comparison of the gate cost when preparing a Gaussian state using QETU to using an exact state preparation procedure. 
For this comparison, we compile the resulting quantum circuits using a universal gate-set consisting of CNOT, $R_z$, and $R_x$ gates using QISKIT~\cite{Qiskit}. 
Before discussing precise gate counts, we argue the expected scaling for both methods. 
Starting with QETU, one controlled call to the $\me^{-i \tau \hat{x}_\text{sh}}$ operator requires $\mathcal{O}(n_q)$ CNOT and $R_z$ gates. 
The number of $R_x$ gates is simply equal to the number of Chebyshev polynomials included in the approximation. 
Exact state preparation methods on the other hand require $\mathcal{O}(2^{n_q})$ CNOT, $R_z$ and $R_x$ gates. 
Even though QETU has a better asymptotic scaling, because of the relatively large coefficient in the overall QETU cost, we expect that for small values of $n_q$, exact state preparation will be less costly than using QETU.
The question we now answer is at what value of $n_q$ does the cost of QETU become cheaper than exact state preparation methods.

For the gate count comparison, we compare the number of gates and neglect the scaling from the $\gamma$ factor for the QETU cost. 
We do this because algorithms with shorter depths for a single run, rather than total gates required, when using NISQ and early-fault tolerant devices are preferred. 
For the gate count comparison, because the cost of $R_z$ and $R_x$ gates are similar, we chose to compare the number of total rotation gates, as well as the number of CNOT gates. It is important to note that the gate count using QETU for a fixed number of calls to the $\me^{-i \tau \hat{x}_\text{sh}}$ is independent of value of $\sigma_x / x_\text{max}$. What is important, however, is the precision one can achieve for that particular value of $\sigma_x / x_\text{max}$. The cost for exact state preparation methods is independent of $\sigma_x / x_\text{max}$.

\begin{figure}[h]
    \centering
    \includegraphics[width=0.48\textwidth]{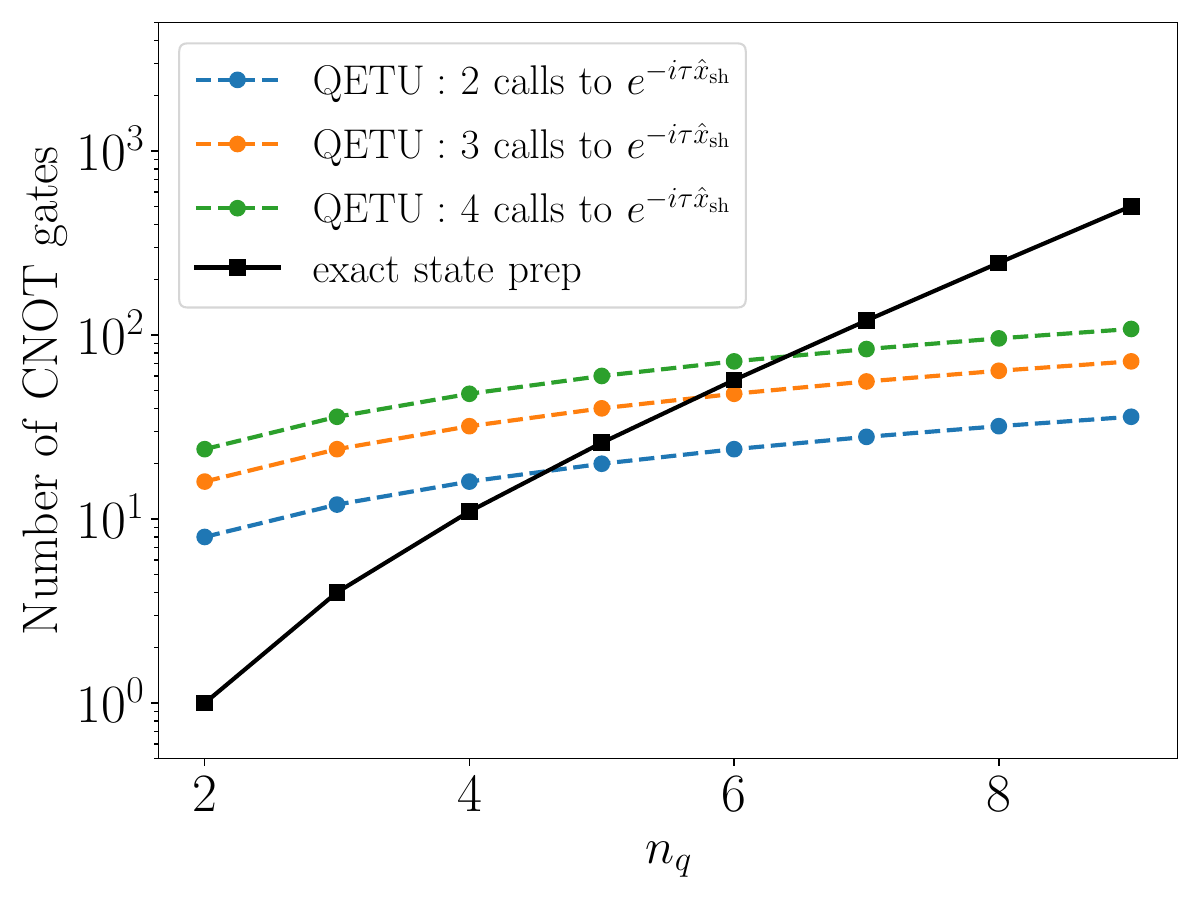}
    \includegraphics[width=0.48\textwidth]{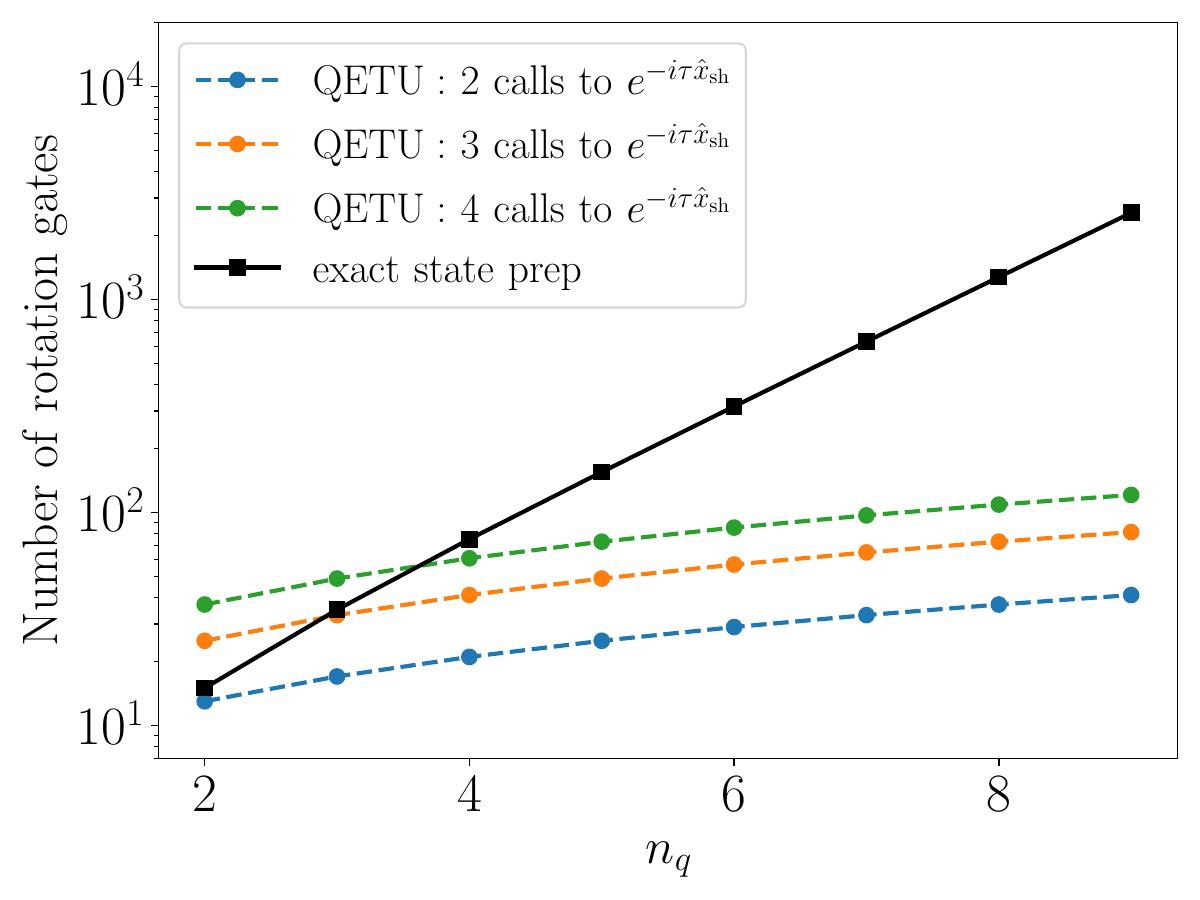}
    \caption{Top: Number of CNOT gates required to prepare a Gaussian state using both QETU and exact state preparation methods as a function of the number of qubits used to represent the state $n_q$. The factor of $\gamma$ is not included in the QETU count. The black squares with a solid line show the count for exact state preparation method. The different colored circles with dashed lines show the cost of QETU for different number of calls to the $\me^{-i \tau \hat{x}_\text{sh}}$ operator. Exact state preparation methods become more expensive around $n_q=5$. Bottom: Same as top but for the number of rotation gates required. Exact state preparation becomes more expensive around $n_q=2$.}
    \label{fig:wp_cost_comparison}
\end{figure}

Figure~\ref{fig:wp_cost_comparison} shows the gate counts required to prepare the Gaussian state, using both QETU and exact state preparation methods, as a function of $n_q$. The top and bottom plots show the number of CNOT and rotation gates, respectively. For QETU, the gate count is shown for 2, 3, and 4 calls to the $\me^{-i \tau \hat{x}_\text{sh}}$ operator. 
As expected, the scaling for the number of CNOT and rotation gates is linear in $n_q$ when using QETU, and exponential in $n_q$ for exact preparation methods. 
Comparing first the CNOT count, looking at the top plot in Fig.~\ref{fig:wp_cost_comparison}, we notice that for $n_q=5$, the CNOT count when using QETU starts to be cheaper than exact state preparation methods.

We now compare the number of rotation gates for both methods. Looking at the bottom plot in Fig.~\ref{fig:wp_cost_comparison}, we see that already for $n_q=2,3$, QETU requires less rotation gates than exact state preparation methods. 
As already shown in Fig.~\ref{fig:wp_even_error_many_sigma}, for certain values of $\sigma_x/x_\text{max}$, using only 2-3 Cheybshev polynomials can achieve a sub-percent precision on the prepared state. Additionally, we showed in Fig.~\ref{fig:wp_even_error_vs_nq} that for fixed degree Chebyshev approximations, the precision is independent of the value of $n_q$.

Taken together, all of these results imply that, depending on the desired error tolerance, value of $\sigma_x/x_\text{max}$, and whether noisy or error-corrected qubits are used, it makes sense to consider using QETU for values of $n_q$ as small as $n_q\gtrsim2-5$.
Note that it is in principle possible to reduce the cost of exact state preparation methods by dropping gates with small rotation angles below some threshold.
The gate count savings and the error introduced will require a dedicated study, which will be interesting to perform in future work.

We conclude by discussing how the cost of our method compares to the cost of the Kitaev-Webb (KW) algorithm for preparing Gaussian states \cite{kitaev2009wavefunction}. 
While the KW algorithm can implement Gaussian states with a cost polynomial in the number of qubits, the algorithm comes with a large overall prefactor due to the need to perform arithmetic on the quantum computer.
A detailed study comparing the gate cost of using the KW algorithm to the gate cost of exact state preparation methods was performed in Ref.~\cite{Deliyannis:2021che}.
It was found that, due to a large prefactor, state preparation using the KW algorithm for a single Guassian wavepacket was more expensive than exact state preparation methods up to $n_q \sim 14$ qubits. 
This, combined with the fact that the cost of KW scales more quickly than linear in $n_q$, implies that using QETU to prepare a one dimensional Gaussian wavefunction will also be cheaper than using the KW algorithm for any value of $n_q$. 

In principle, it is also possible to use QETU to also prepare multi-dimensional Gaussian states.
Because of the probabilistic nature of using QETU, one will have to ensure that the $\gamma$ factor does not decrease exponentially with the dimension of the Gaussian state being prepared.
This would be nontrivial to achieve, given the non-unitary nature of the Gaussian filter transformation~\eqref{eq:wavepacket}.
It will be interesting to explore this application of QETU in a future work.

The final study we perform is on how $\gamma$ depends on $n_q$, the number of Chebyshev polynomials, and the width $\sigma_x / x_\text{max}$. Recall that $\gamma$ is defined to be the magnitude of the final state prepared using QETU. Because the initial state guess and implemented operator $F(\cos(\tau \hat{x}_\text{sh}/2))$ are known exactly, we can directly evaluate $\gamma$ in the limit where we reproduce $\Fex(x)$ exactly. Doing so gives
\begin{equation}
\begin{split}
    \gamma &= | \bra{\psi_\text{init}} \Fex(\cos(\tau \hat{x}_\text{sh}/2))^2 \ket{\psi_\text{init}}|
    \\
    &= \frac{c^2}{2^{n_q}} \sum_{j=0}^{2^{n_q}-1}  \me^{-x_j^2/\sigma_x^2}\,,
\end{split}
\end{equation}
where going to the second line we used $\ket{\psi_\text{init}} = 2^{-n_q/2} \sum_{j=0}^{2^{n_q}-1} \ket{x_j}$ and $F(\cos(\tau \hat{x}_\text{sh}/2)) = c\, \me^{-\hat{x}^2/(2\sigma_x^2)}$.
If we replace $\Fex$ with the approximation $F$ with error $\epsilon$, the above expression for $\gamma$ is expected to be correct up to $\mathcal{O}(\epsilon)$ corrections. 
In the limit of $n_q \to \infty$, the value of $\gamma$ is simply the area under the Gaussian curve. 
We therefore expect $\gamma$ to approach a constant as $n_q$ is increased. 
The value of this constant will depend on $\sigma_x/x_\text{max}$, and is given by $c^2 \int_{-x_\text{max}}^{x_\text{max}} dx\, \me^{-x^2 / \sigma_x^2}$. 
From this we expect that $\gamma$ decreases as the ratio $\sigma_x / x_\text{max}$ decreases.

Using this expression for $\gamma$, we can identify a scenario where $\gamma$ can be prohibitively small. 
Suppose that one uses a small value of $n_q$ to implement a sharply peaked Gaussian state. 
It is possible that one only samples at values of $x$ where the function $\me^{-x^2/(2\sigma_x^2)}$ is exponentially suppressed, resulting in an exponentially small value of $\gamma$. 
However, because sampling a sharply peaked function a small number of times will introduce large digitization errors, this situation is not likely to occur in practice; in a realistic scenario, one will sample the wavepacket using a large enough value of $n_q$ to avoid a small value of $\gamma$.

\begin{figure}[h]
    \centering
    \includegraphics[width=0.45\textwidth]{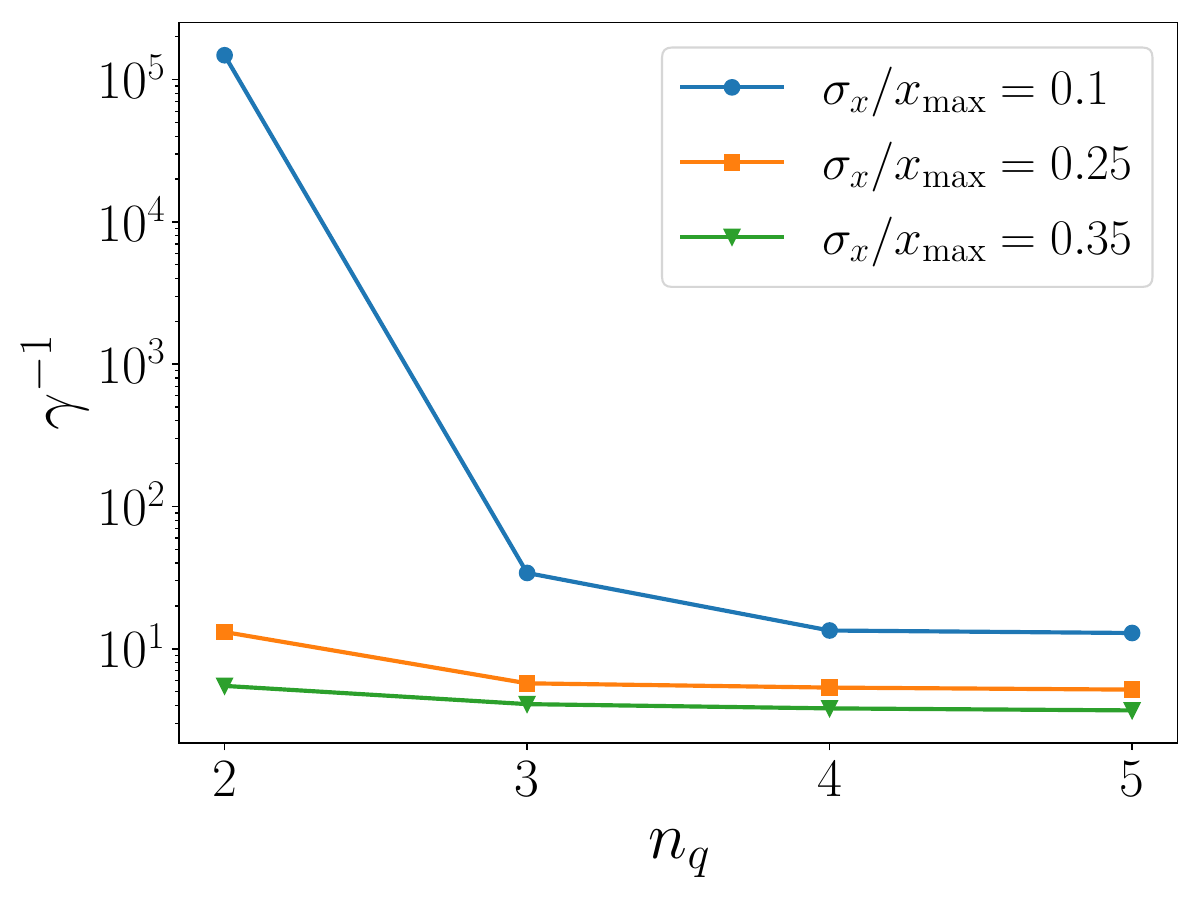}
    \includegraphics[width=0.45\textwidth]{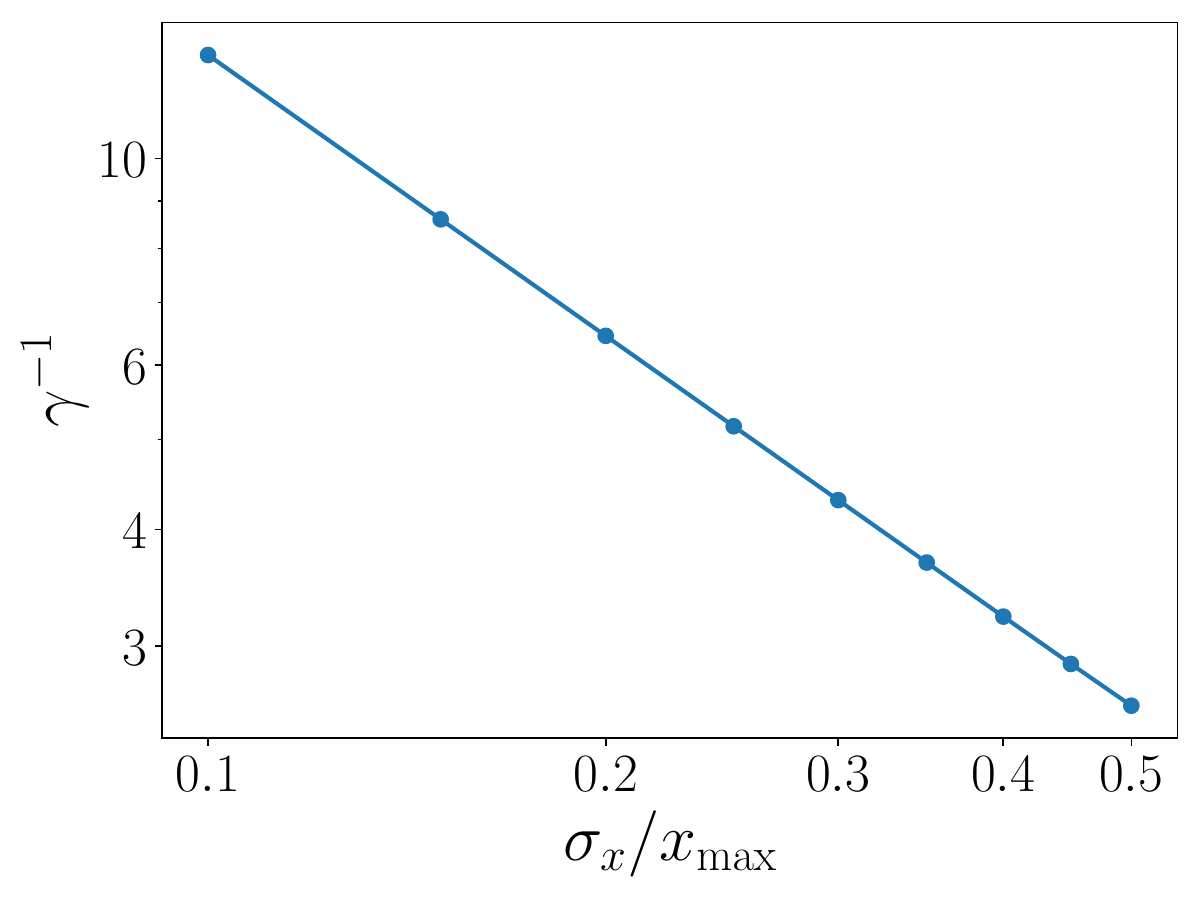}
    \caption{Dependence of $\gamma$ on various parameters when constructing the Gaussian state by setting $\tau = 2$ and solving the optimization problem in Eq.~\eqref{eq:minmax_only_sampled}. 
    Top: $\gamma^{-1}$ as a function of $n_q$. Different colored points indicate different values of $\sigma_x / x_\text{max}$. The value of $\gamma^{-1}$ is large for $n_q=2$ and $\sigma_x / x_\text{max} = 0.1$ due to sampling the sharply peaked Gaussian only at points where the value is exponentially suppressed. Bottom: $\gamma^{-1}$ as a function of $\sigma_x / x_\text{max}$ for $n_q=5$. The value of $\gamma$ is proportional to $\sigma_x / x_\text{max}$.}
    \label{fig:wp_gamma}
\end{figure}

Figure~\ref{fig:wp_gamma} shows the dependence of $\gamma^{-1}$ on $n_q$ and $\sigma_x/x_\text{max}$. All results were calculated using a degree 18 even Chebyshev approximation. For the data shown, the largest error from the finite Chebyshev approximation of the state produced using QETU was $\sim 10^{-9}$.
The top plot in Fig.~\ref{fig:wp_gamma} shows $\gamma^{-1}$ as a function of $n_q$ for different values of $\sigma_x / x_\text{max}$. Using $n_q=2$ for $\sigma_x / x_\text{max} = 0.1$ results in $\gamma^{-1} \sim 10^5$ because the wavepacket is only sampled at values where the value is exponentially suppressed. 
This problem can be avoided by increasing $n_q$, and is less severe for larger values of $\sigma_x / x_\text{max}$.
We notice that by $n_q = 5$ the value of $\gamma^{-1}$ levels out.
The bottom plot in Fig.~\ref{fig:wp_gamma} shows $\gamma^{-1}$ as a function of $\sigma_x / x_\text{max}$ for $n_q=5$. We see that $\gamma$ is directly proportional to $\sigma_x / x_\text{max}$, with more sharply peaked states having smaller values of $\gamma$. 
For a sharply peaked value of $\sigma_x/x_\text{max} = 0.1$, we find $\gamma^{-1} \sim 13$. 

The main value of our scrupulous numerical investigations in this section is the demonstrated ability to prepare wavepackets with exponential precision, for arbitrary values of the distribution parameters.
This required several optimization procedures and could not have been trivially predicted based on the construction proposed in~\cref{ssec:wp_intro}.
Our result proves an important point that, despite the presence of $\arccos(x)$ in the QETU approximation, it can be used for representing functions of Hermitian operators, with exponential precision.
(In Ref~\cite{dong2022ground} this issue was avoided by constructing a polynomial approximation to the error function which, in turn, was approximating the cosine-transformed original step function.)

\section{Ground state preparation cost for a general gauge theory\label{sec:gsprep}}

In this section, we discuss how the cost of ground state preparation using QETU is expected to scale with the parameters of a general lattice pure gauge theory in $d$ spatial dimensions.
Understanding the cost in terms of a total gate count requires detailed information regarding the specific formulation of the lattice gauge theory in question, as well as the algorithm used to implement $\me^{-i \tau H}$. 
To keep the discussion in this section as general as possible, we consider the cost in terms of the number of calls to the time evolution circuit.
For ease of notation, throughout this section, we denote energies of the unscaled Hamiltonian of the general lattice gauge theory by $E_i$.

We consider a pure gauge theory on a hypercubic spatial lattice with $N_s$ sites in each dimension, and a lattice spacing $a$ between neighboring sites. 
The $d$ dimensional volume is given by $V = (a N_s)^d$. 
To truncate the infinite dimensinoal Hilbert space of this bosonic theory, we represent each lattice site using $n_q$ qubits. 
We consider a general gauge theory sharing the same qualitative features as QCD.
First, the theory is assumed to have a mass gap between the vacuum and the first excited state.
Second, we assume that the theory is asymptotically free, i.e., that as the lattice spacing $a$ goes to zero, the bare coupling $g(a)$ also goes to zero. 
The true, continuous, infinite volume theory is recovered while simultaneously taking $(a N_s)^d \to \infty$, $n_q \to \infty$, and $a \to 0$ (while appropriately adjusting $g$ according to the chosen renormalization scheme).
We work in the units of $\hbar = c = 1$.

One important consideration is that, in a lattice gauge theory, one does not directly choose the lattice spacing $a$.
One can only choose the value of the bare gauge coupling $g$. 
The two parameters are related via the renormalization group; one should view the gauge coupling as a function of the lattice spacing, i.e, $g = g(a)$. 
What this means in practice is that one can only calculate dimensionless quantities in a lattice gauge theory. 
For example, if one considers an energy $E$, one can only calculate the dimensionless value $a E$.
Once the lattice spacing $a$ has been determined using some renormalization scheme, the dimensionful energy $E$ can be extracted. 
However, this consideration can be avoided for our discussion using the fact that the cost of QETU depends only on ratios of energies. 
The explicit lattice spacing dependence cancels, and we can consider directly energy differences $E_1-E_0$ and $E_\text{max}-E_0$ in physical units.

While the explicit dependence on the lattice spacing cancels, the energies still have implicit dependence on $a$. 
The energies also have an implicit dependence on the physical volume $V$ and the number of qubits per lattice site $n_q$. 
To better understand this dependence, it is useful to recall that the lattice acts as both a high and low energy regulator of our quantum field theory.
Using a finite lattice spacing introduces an energy cutoff $\sim 1/a$, while using a finite volume provides a low energy cutoff of $\sim 1/V$.
In addition, using a finite value for $n_q$ provides another high energy cutoff, denoted as $\Lambda_{n_q}\sim2^{n_q}$.
Assuming one has properly renormalized the theory, as these regulators are removed, i.e., sending $a \to 0$, $V \to \infty$ and $\Lambda_{n_q} \to \infty$, one should obtain the correct physical energies. 
We proceed with the discussion assuming access to (early) fault-tolerant quantum computers, where finite volume errors $\varepsilon_V$, finite lattice spacing errors $\varepsilon_a$, and finite $n_q$ errors $\varepsilon_{n_q}$ are controlled.

We now discuss the scaling of $E_1-E_0$ and $E_\text{max}-E_0$ with $N_s, a$ and $n_q$. 
The energy difference $E_1 - E_0$ in a continuous theory with a mass gap has a finite value.
Therefore, as one removes the regulators, the gap $E_1-E_0$ is expected to approach a constant.
However, $E_\text{max} - E_0$ diverges in this limit.
We now consider how quickly this term grows as we remove each of the regulators.
Starting with the volume, it is helpful to think from the perspective of fixed $a$ and $n_q$.
Increasing the volume is achieved by increasing the number of sites $N_s$.
In the simplest possible case of a free theory, each additional site increases the maximum possible energy by the maximum energy of a single site.
For a local gauge theory, the maximum energy is expected to increase in a similar way when using more lattice sites.\footnote{Note that, because gauge fixing generally leads to some non-locality in the resulting Hamiltonian, this argument does not necessarily apply to such formulations. If one does use a non-local Hamiltonian, understanding how $E_\text{max}-E_0$ scales with the number of sites will likely require a dedicated study.} 
The maximum energy of our interacting theory is therefore expected to grow linearly with the total number of sites $N_s^d$. 
For fixed $a$, this is equivalent to growing linearly with the volume $V$.
Turning now to the lattice spacing, recall that a finite value of $a$ imposes a high energy cutoff given by $\sim 1/a$. 
We therefore expect the maximum energy to scale in the same way as $1/a$. Lastly, it is known that the maximum energy of a bosonic theory generally increases exponentially with $n_q$~\cite{jordan2011quantum, klco2019digitization}. 

In terms of these energy gaps, we know $\Delta \sim (E_1-E_0)/(E_\text{max}-E_0)$. Combining our scaling arguments, we find
\begin{equation}
    \Delta^{-1} = \mathcal{O}\left(\frac{E_\text{max}-E_0}{E_1-E_0}\right)= \mathcal{O}\left(a^{-1} N_s^d 2^{n_q \alpha}\right),
\end{equation}
where the parameter $\alpha \in \mathbb{R}_{> 0}$ is both theory- and formulation dependent.
Numerical studies of this scaling were performed for a compact U(1) lattice gauge theory in 2 spatial dimensions in Sec.~\ref{ssec:u1_numerical}.

We now turn our attention to how the overlap $\gamma$ between the initial guess and exact ground state scales. For this discussion, we assume that the vacuum is translationally invariant, which is true for theories in the Standard Model. There are many ways to prepare this initial guess, including adiabatic state preparation~\cite{farhi2000quantum,farhi2001quantum,aharonov2003adiabatic}, variational methods~\cite{peruzzo2014variational,mcclean2016theory,mcardle2019variational,kokail2019self,Gomes:2021ckn,Liu:2021otn}, and direct preparation~\cite{vartiainen2004efficient,mottonen2004quantum,khaneja2001cartan,bullock2004canonical,earp2005constructive,drury2008constructive,iten2016quantum,cortese2018loading,sun2021asymptotically,zhang2021low,zhang2022quantum}.
We now argue that preparing such an initial guess wavefunction using direct state preparation will lead to exponential scaling in the cost of preparing the ground state with QETU.

Consider a situation where one prepares regions of the lattice using exact state preparation, where for simplicity we assume that each region contains $N_r$ sites in each dimension.
The number of these regions is given by $(N_s/N_r)^d$.
The gate cost of preparing the state of a single region, each with overlap $\gamma_i$, scales as $2^{(N_r)^d}$, with the total overlap given by $\gamma \sim (\gamma_i)^{(N_s/N_r)^d}$.
Any choice of $N_r$ that breaks the exponential scaling in one of these costs necessarily introduces an exponential scaling in the other. 
Even though each overlap $\gamma_i$ can in principle be improved using amplitude amplification from $\gamma_i$ to $\sqrt{\gamma_i}$~\cite{Brassard_2002,lin2022lecture}, the overall $\gamma$ parameter still decreases exponentially as one increases the number of lattice sites.
Creating a trial state using direct state preparation methods is therefore inefficient. 
As a proof of principle that this exponential volume scaling can be overcome, we studied how $\gamma$ scales with the volume if one uses a simple adiabatic state preparation procedure to prepare the initial guess wavefunction $\ket{\psi_\text{init}}$ for the U(1) formulation we considered. We found that doing so results in $\gamma \sim \mathcal{O}((N_s^d)^{-1})$. 
We conclude by noting that it is possible that more sophisticated adiabatic preparation procedures, or the use of variational state preparation, could result in a better scaling of $\gamma^{-1}$ with the volume. 
It will be interesting to explore this in future work.

Another component of the cost is how the number of Trotter steps used to approximate the time evolution circuit scales with the system size. 
As one increases $N_s$ or $n_q$, the number of non-commuting terms in the Hamiltonian increases, and we therefore expect the number of steps required to maintain a constant precision to increase accordingly.
However, as discussed in Sec.~\ref{ssec:u1_numerical}, scaling the Hamiltonian such that the spectrum is in the range $[0, \pi]$ can be equivalently viewed as scaling the Trotter step size. 
Because the maximum energy generally grows with $N_s$ and $n_q$, the effective Trotter step size will decrease with $N_s$ and $n_q$.
If this decrease in the Trotter step size results in a smaller error that the increase from the additional non-commuting terms, the overall Trotter error would actually \textit{decrease} with $N_s$ and $n_q$. 
As shown in Sec.~\ref{ssec:u1_numerical}, this was indeed found to be the case for the U(1) gauge theory we considered.
Due to the similarity of Hamiltonian structure of general lattice gauge theories, it is possible that this trend will also be present for other lattice gauge theories.

Lastly, we point out an observation that may help dampen the scaling of $\Delta^{-1}$ as one increases $N_s$ and $n_q$.
The main idea is that in general, an initial guess with good overlap with the ground state will generally have smaller overlap with excited states, with the overlap continuing to decrease for higher excited states. As a simple example, consider the quantum harmonic oscillator, with operators sampled using $n_q$ qubits. One possible initial guess is a constant wavefunction for all $x$. Because the $n^\text{th}$ excited state has $n$ nodes, the overlap of this initial guess and higher excited states will decrease with $n$, due to the fact that one sums more highly oscillatory functions. Let us denote the overlap of the highest energy state and the initial guess as $\gamma_{n_q, \text{min}}$, with associated energy $E_{2^{n_q}-1}$.
Now suppose that one increases the number of qubits to $m_q > n_q$. This increases the size of the Hilbert space, but the key point is that only states with higher energies are added. Because these higher energy states also have more oscillations, the overlap the initial guess has with these additional states are all smaller than $\gamma_{n_q, \text{min}}$. Depending on the error threshold, it is conceivable that, for a small enough value of $\gamma_{n_q, \text{min}}$, one can simply ignore the states with energy larger than $E_{2^{n_q}-1}$. With regards to using QETU, this implies we only have to divide our energy gap by $E_{2^{n_q}-1}-E_0$ instead of $E_{2^{m_q}-1}-E_0$. Because this argument is true for any $m_q > n_q$, the amount one must scale down the physical energy gap eventually becomes independent of $n_q$. 

It is possible that these arguments can be applied in a similar way to lattice gauge theories in order to argue that the scaling with $N_s$ and $n_q$ could be milder than previously argued.
In the best case scenario where the dependence on $N_s$ and $n_q$ vanishes for some values of the parameters, $\Delta$ would only depend on physical energies $E_0$, $E_1$, and some $E_{n^*}$, where $E_{n^*}$ is the highest energy state that must be filtered out using QETU. In this way, $\Delta$ is no longer explicitly dependent on the lattice spacing, only through finite lattice spacing errors. If such a scenario is true, then $\Delta$ would become independent of $N_s$, $n_q$, and $a$. The cost would then be some overall large pre-factor, dependent on the value of $\Delta$ where it becomes independent of $N_s, n_q$ and $a$, multiplied by the cost of implementing $\me^{-i \tau H}$ using Trotter methods.
Our preliminary numerical investigations showed that the validity of such a hypothesis is heavily dependent on both the initial guess state and the value of the coupling constant.
We leave further investigations for future work.

Importantly, the above-described method of dampening the scaling with $N_s$ or $n_q$ is specific to the QETU approach, and would not be possible if one instead used the Hamiltonian input model, as in Ref.~\cite{lin2020}.
In the Hamiltonian input model, one performs repeated calls to a block encoding of $H$ and uses the Quantum Eigenvalue Transformation to implement the projector.
Already at the stage of constructing the block encoding circuit for $H$, this method requires the Hamiltonian to be scaled so that $||H|| \leq 1$.
Scaling down the Hamiltonian, and therefore the gap, by a factor of $E_\text{max}-E_0$ is thus unavoidable in this scenario.
On the contrary, in the QETU case, where one performs repeated calls to the unitary operator $\me^{-i \tau H}$, there are no fundamental obstacles for constructing the time-evolution circuit for some $H$ with a large spectral norm. 
The spectrum of $H$ is shifted solely to avoid the problems associated with the periodic nature of the matrix function which QETU implements.
Consequently, this technical difference between the Hamiltonian and time-evolution input models could result in the cost of QETU having better asymptotic scaling with $N_s$ or $n_q$.

\section{Discussion and Conclusion \label{sec:conclusion}}

In this work, we performed an extensive study of the QETU algorithm and its applications to state preparation in simulations of quantum field theory.
By modifying the original algorithm in Ref.~\cite{dong2022ground}, we were able to achieve significant cost savings when the time evolution circuit is implemented both exactly as well as approximately using Trotter methods.

We applied our improved procedure to prepare a ground state in a particular lattice formulation of U(1) gauge theory in 2 spatial dimensions.
To avoiding the costly controlled calls to the time evolution circuit, we based our circuits on the control-free version of the QETU algorithm.
The considered control-free implementation of QETU generalizes to any Hamiltonian of the form $H = H_x + H_p$, where the bases of the kinetic piece $H_p$ and the potential piece $H_x$ are related by a Fourier transformation; a form that is common to many types of lattice field theories and their different formulations~\cite{jordan2011quantum, klco2019digitization, kaplan2020gauss, haase2021resource, bauer2021efficient, bauer2023new, Bender:2020jgr}.

We studied how the cost of the QETU-based state preparation algorithm scales with parameters of our physical system.
In particular, we discussed how scaling down the spectrum of the physical Hamiltonian ensues the scaling of the energy gap $\Delta$, and placed upper bounds on this parameter for arbitrary system sizes.
Next, we discovered that scaling down the Hamiltonian spectrum leads to unexpected positive consequences; namely, it makes the Trotter error decrease as the number of sites $N_p$ or the number of qubits per site $n_q$ are increased. 
This behavior is due to the fact that scaling down the Hamiltonian spectrum is equivalent to scaling down the Trotter step size.
Additionally, using a simple adiabatic state preparation procedure, we demonstrated that one can achieve a value of $\gamma^{-1}$ that scales only polynomially with the system size.
These studies lay the basis for the further studies on applications of QETU-based state preparation techniques for alternative formulations of gauge theories.

We followed our numerical results with a discussion of asymptotic costs of using QETU for state preparation in general lattice field theories with properties similar to QCD.
Assuming that the scaling of $\Delta$ and $\gamma$ are similar to the scaling we found for the test theory, shown in Tab.~\ref{tab:cost}, QETU can be used to prepare the ground state of a general lattice gauge theory with a volume scaling 2 powers higher than the cost of implementing a single Trotter step. 
If one considers a local theory, this leads to the cost scaling as the cube of the number of sites.
We then argued that the scaling of $\Delta$ with the number of sites or qubits per site can be reduced by exploiting the fact that, for a good initial guess, highly excited states will have a negligible overlap, and do not need to be filtered out.

In this work, we also developed a novel application of the QETU algorithm for the preparation of Gaussian states. 
The main idea is to use QETU to implement the Gaussian filter operator $\me^{-\hat{x}^2/(2\sigma_x^2)}$.
We argued that a na\"ive application of QETU to this problem results in the error decreasing only polynomially with the degree of the Chebyshev expansion. 
We showed, however, that one could instead achieve an exponential scaling, for any value of the width of the Gaussian state, by performing simple modifications to the QETU procedure. 
With these improvements, we presented a procedure that allows one to prepare Gaussian states while avoiding the costly step of adding the even and odd pieces using LCU.
By performing a gate cost analysis, we showed that preparing Gaussian states with QETU using our improved procedure outperforms exact state preparation methods for as few as $n_q \gtrsim 2-5$ qubits, depending on whether one is using noisy or error-corrected qubits.

This work leads naturally to a number of additional interesting applications. 
While we used QETU to prepare the vacuum state of a pure gauge theory, it is in principle possible to extend QETU to prepare hadronic states in QCD, another important step towards simulating QCD. 
For concreteness, consider the task of preparing the quantum state of the pion.
While the pion is not the ground state of the QCD Hamiltonian, it is, however, the lowest energy state with the quantum numbers of the pion.
Additionally, the gap between the pion state and the next excited state with the desired quantum numbers is at least twice the pion mass. 
If one prepares an initial state $\ket{\psi_\text{init}}$ with the quantum numbers of the pion, e.g., using the high quality interpolating fields that have been developed for use in lattice QCD (see, \textit{e.g.}, Ref.~\cite{Gattringer:2010zz} for a pedagogical introduction) then one can construct a filter operator using the QCD Hamiltonian and isolate the pion state.
Because these interpolating fields are not unitary but Hermitian operators, a dedicated study to how best implement these operators will be required.
It is possible, in principle, to use QETU to construct interpolating fields.
An interesting question to ask is if one needs to apply the interpolating field to the interacting vacuum prepared using, e.g., QETU, or if one can simply apply the interpolating field to any state with the quantum numbers of the vacuum. 
If one could avoid preparing the interacting ground state while avoiding $\gamma$ scaling exponentially poorly with the number of sites, this would result in a significant cost reduction.

Another interesting followup is exploring the use of QETU for preparing multi-dimensional Gaussian states, which are relevant for lattice gauge theories as well. 
The general form of such a state is $\me^{-\sum_{ij} c_{ij} x_i x_j}$.
A simple procedure for constructing this state using QETU is to set $\fex(x) = \me^{-x}$ and use $\me^{i \sum_{ij} c_{ij} \, \hat{x}_i \hat{x}_j}$ as a building block in the QETU circuit.
One important consideration using this method, or any variation, is to ensure that the $\gamma$ parameter does not decrease exponentially as one increases the dimension of the Gaussian.
If this can be avoided, it will be interesting to compare the cost of QETU to the Kitaev-Webb algorithm for preparation of such states.

Our studies provide the foundation for further investigations on the applicability of QETU in the context of highly efficient preparation of various functions of Hermitian operators.
In our following work~\cite{hariprakash2023strategies}, we investigate whether the QETU algorithm can be utilized for efficiently implementing block encodings, which paves the way toward synergizing QETU with state-of-the-art simulation strategies~\cite{rossi2022multivariable,Motlagh:2023oqc}.

\FloatBarrier

\section*{Acknowledgements}
CFK would like to thank Stefan Meinel for access to their computing cluster for some of the calculations. CFK would also like to thank Dorota Grabowska for useful discussions regarding adiabatic state preparation for the U(1) formulation considered in the paper. 
We thank Neel Modi and Siddharth Hariprakash for access to their code and Christian Bauer for useful discussions.
This material is based upon work supported by the U.S. Department of
Energy, Office of Science, Office of Advanced Scientific Computing Research, Department of
Energy Computational Science Graduate Fellowship under Award Number DE-SC0020347.
MK acknowledges support from the DOE grant PH-HEP24-QuantISED, B\&R KA2401032/34. 
This work was supported by DOE Office of Advanced Scientific Computing Research (ASCR) through the ARQC program (NG).
Support is also acknowledged from the U.S. Department of Energy, Office of Science, National Quantum Information Science Research Centers, Quantum Systems Accelerator.

\bibliographystyle{apsrev4-1}
\bibliography{bibi}

\clearpage
\newpage
\onecolumngrid

\appendix

\section{Optimal choice of time step $\delta \tau$}
\label{app:optimal_dtau}
In this appendix, we discuss the optimal choice of the time step $\delta \tau$ that minimizes the total number of calls to the Trotter circuit $N_\text{tot}$, given by
\begin{equation}
    N_\text{tot} = \frac{1}{b \Delta \delta \tau} \log \left( \frac{a}{\epsilon-c(\delta \tau)^p} \right).
\end{equation}
We start by showing that $\frac{dN_\text{tot}}{d\delta \tau}$ always has a zero. From there, we present a perturbative solution to the value of $\delta \tau$ that minimizes $N_\text{tot}$. We conclude by showing that $\frac{d^2N_\text{tot}}{d\delta \tau^2}$ is positive definite, which shows that the extreme value is always a minimum. 

The first derivative is given by
\begin{equation}
    \frac{dN_\text{tot}}{d\delta \tau} = \frac{1}{b \Delta \delta \tau^2} \left[ c\, p \frac{ \delta \tau^{p}}{\epsilon-c \delta \tau^p} - \log \left( \frac{a}{\epsilon-c (\delta \tau^*)^p} \right) \right].
\end{equation}
For $\delta \tau = 0$, $\frac{dN_\text{tot}}{d\delta \tau} = -\log(a/\epsilon)$. Because $a > \epsilon$ in general, the first derivative is negative at $\delta \tau = 0$. This, combined with the fact that the function diverges to $+ \infty$ when $\delta \tau \to (\epsilon/c)^{1/p}$ from the left, implies the first derivative always has a zero for some value of $\delta \tau$.

Setting the first derivative to zero gives
\begin{equation}
    0 = c\, p \frac{ (\delta \tau^*)^{p}}{\epsilon-c (\delta \tau^*)^p} - \log \left( \frac{a}{\epsilon-c (\delta \tau^*)^p} \right).
\end{equation}
To obtain approximate analytic values for $\delta \tau^*$, one can expand the logarithm.
If we define $x = \frac{c}{\epsilon} \delta \tau^p$, then the equation becomes
\begin{equation}
    0 = p \frac{x}{1-x} + \log\left(\frac{\epsilon}{a}\right) + \log(1-x).
    \label{eq:dtau_star_exact}
\end{equation}
By expanding the log to lowest order, we obtain the following approximate equation for $x$ 
\begin{equation}
    0 = p \frac{x}{1-x} + \log\left(\frac{\epsilon}{a}\right)\,,
\end{equation}
which can be solved analytically to give the following expression 
\begin{equation}
    \delta \tau^* \approx \left(\frac{\epsilon}{c}\left(1 - \frac{p}{p+\log(\frac{a}{\epsilon})}\right)\right)^{1/p}\,.
    \label{eq:dtau_star_approx}
\end{equation}
Because $a>\epsilon$ in general, we see that this approximate solution for $\delta \tau^*$ is slightly smaller than the maximum value of $(\epsilon/c)^{1/p}$. Using the values $\epsilon=10^{-3}, p=1, a=1, c=0.1$, the approximate expression in Eq.~\eqref{eq:dtau_star_approx} gives a value of $\delta \tau^*$ which differs from the exact value obtained by solving Eq.~\eqref{eq:dtau_star_exact} numerically by only $3.2\%$.

The second derivative is given by
\begin{equation}
    \dv[2]{N_\text{tot}}{\delta \tau} = \frac{1}{b \Delta (\delta \tau)^3}\left[c p (\delta \tau)^p \frac{(p-3)\epsilon + 3 c (\delta \tau)^p}{(\epsilon-c (\delta \tau)^p)^2} + 2 \log \left( \frac{a}{\epsilon-c (\delta \tau)^p} \right) \right].
\end{equation}
The goal is to show that when $\delta \tau = \delta \tau^*$, the second derivative is positive.
This will be done by implicitly solving Eq.~\eqref{eq:dtau_star_exact} to replace the log term, which gives
\begin{equation}
    \dv[2]{N_\text{tot}}{\delta \tau}\Big|_{\delta \tau = \delta \tau^*} = \frac{1}{b \Delta (\delta \tau^*)^3}\left[c p (\delta \tau^*)^p \frac{(p-3)\epsilon + 3 c (\delta \tau^*)^p}{(\epsilon-c (\delta \tau^*)^p)^2} + 2 c\, p \frac{ (\delta \tau^*)^{p}}{\epsilon-c (\delta \tau^*)^p} \right].
\end{equation}
By combining the fractions, we obtain the following result
\begin{equation}
    \dv[2]{N_\text{tot}}{\delta \tau}\Big|_{\delta \tau = \delta \tau^*} = \frac{c\,p (\delta \tau^*)^p}{b \Delta (\delta \tau^*)^3} \frac{1}{(\epsilon-c (\delta \tau^*)^p)^2}\left[\epsilon(p-1) + c (\delta \tau^*)^p \right].
\end{equation}
Now, for any $p\geq 1$, and assuming all parameters are positive, we see $\dv[2]{N_\text{tot}}{\delta \tau}\Big|_{\delta \tau = \delta \tau^*}>0$.

\section{Control-free implementation of ground state preparation in lattice field theories \label{app:ctrl_free}}

In this appendix, we show how to prepare the ground state of the U(1) gauge theory described in Sec.~\ref{sec:u1} using the control-free version of QETU. 

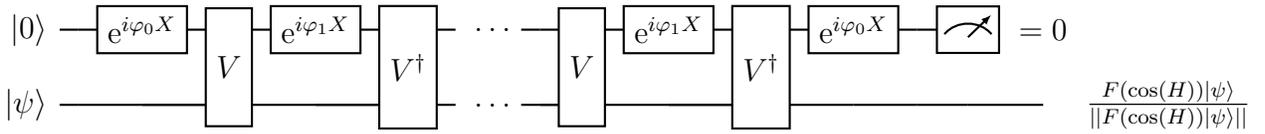
\begin{figure*}[h]
    \large
    \centering
    \begin{quantikz}[row sep={1cm,between origins}, column sep=0.25cm]
        \lstick{$\ket{0}$} & \qw & \gate{\me^{i \varphi_{0} X}} & \gate[2]{V} & \gate{\me^{i \varphi_{1} X}} & \gate[2]{V^\dagger} & \qw & \ldots\ & \qw & \gate[2]{V} & \gate{\me^{i \varphi_{1} X}} & \gate[2]{V^\dagger} & \gate{\me^{i \varphi_{0} X}} &\qw & \meter{} & = 0
        \\
        \lstick{$\ket{\psi}$} & \qw & \qw & \qw & \qw & \qw & \qw & \ldots\ & \qw & \qw & \qw & \qw & \qw &\qw & \qw & \qw & \frac{F(\cos(H)) \ket{\psi}}{||F(\cos(H)) \ket{\psi}||}
    \end{quantikz}
    \caption{Control-free QETU circuit diagram. The top qubit is the control qubit, and the bottom register is the state we apply a matrix function to.
    Here $V$ is the operator given in Eq.~\eqref{eq:ctrl_free_V} that implements simultaneous controlled forward and backward time evolution.
    By applying alternating calls to $V$ and $V^\dagger$, upon measuring the ancillary qubit to be in the zero state, one prepares the normalized quantum state $F(\cos(H))\ket{\psi}/||F(\cos(H)) \ket{\psi}||$, where $F(\cos(H))$ can be a general even polynomial.
    For symmetric phase factors $\{\varphi_j\} = (\varphi_0, \varphi_1, \dots, \varphi_1, \varphi_0) \in \mathds{R}^{d+1}$, then $F(\cos(H))$ is an even polynomial of degree $d$. 
    The probability of measuring the control qubit in the zero state is $p = ||F(\cos(H)) \ket{\psi}||^2$.}
    \label{fig:qetu_circ_ctrl_free}
\end{figure*}

We first describe the general procedure for implementing the control-free version of QETU originally described in Ref.~\cite{dong2022ground}. Suppose that one has access to an oracle that implements a controlled call to both forward and backward time evolution simultaneously,
\begin{equation}
    V = \begin{pmatrix}
            \me^{i \tau H} & 0 \\
            0 & \me^{-i \tau H}
        \end{pmatrix}
    \label{eq:ctrl_free_V}.
\end{equation}
Instead of a controlled call to $U = e^{-i \tau H}$, as in the original QETU circuit, the oracle $V$ can then be used as a building block.
The control-free QETU Theorem~\cite{dong2022ground} assumes the access to a circuit implementing $V$ for an $n$-qubit Hermitian operator $H$. 
It states that for any even real polynomial $F(x)$ of degree $d$ satisfying $|F(x)| \leq 1, \, \forall x \in [-1,1]$, one can find a sequence of symmetric phase factors $\{\varphi_j\} = (\varphi_0, \varphi_1, \dots, \varphi_1, \varphi_0) \in \mathds{R}^{d+1}$, such that the circuit in Fig.~\ref{fig:qetu_circ_ctrl_free} denoted by $\mathcal{U}$ satisfies $\left( \bra{0} \otimes \mathds{1}_n \right) \mathcal{U} \left( \ket{0} \otimes \mathds{1}_n \right) = F(\cos(H))$ \cite{dong2022ground}.

\begin{figure*}[h]
    \large
    \centering
    \begin{quantikz}[row sep={1cm,between origins}, column sep=0.5cm]
        \qw & \gate[2]{V^{(j)}} & \qw
        \\
        \qw & \qw & \qw
    \end{quantikz}
    \quad
    =
    \centering
    \begin{quantikz}[row sep={1cm,between origins}, column sep=0.5cm]
        \qw & \ctrl{1} & \octrl{1} & \qw
        \\
        \qw & \gate{\me^{-i \tau H^{(j)}}}  & \gate{\me^{i \tau H^{(j)}}} & \qw
    \end{quantikz}
    \quad
    =
    \begin{quantikz}[row sep={1cm,between origins}, column sep=0.25cm]
        \qw & \octrl{1} & \qw & \octrl{1} & \qw
        \\
        \qw & \gate{K_j} & \gate{\me^{-i \tau H^{(j)}}} & \gate{K_j} & \qw
    \end{quantikz}
    \caption{Circuit relations demonstrating how the relation $K_j \me^{-i \tau H^{(j)}} K_j = \me^{i \tau H^{(j)}}$ can be leveraged to implement $V^{(j)}$ while only having to control the $K_j$ Pauli operators.}
    \label{fig:UUdag_from_K}
\end{figure*}
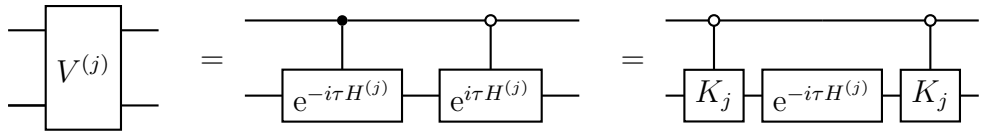

We now describe a general procedure for preparing $V$ assuming one has access to $U = \me^{-i \tau H}$. 
The procedure for implementing $V$ involves grouping the terms of $H$ into $l$ groups $H=\sum_{j=1}^l H^{(j)}$ such that each term in $H^{(j)}$ anticommutes with the Pauli operator $K_j$, i.e., $K_j H^{(j)} K_j = -H^{(j)}$. 
We will now show that, once this grouping is found, only the $K_j$ operators must be controlled.
We now define
\begin{equation}
    V^{(j)} = \begin{pmatrix}
            \me^{i \tau H^{(j)}} & 0 \\
            0 & \me^{-i \tau H^{(j)}}
        \end{pmatrix},
    \label{eq:ctrl_free_Vj}
\end{equation}
such that $V = \prod_{j=1}^l V^{(j)}$.
Using the key relation $K_j \me^{-i \tau H^{(j)}} K_j = \me^{i \tau H^{(j)}}$, each $V^{(j)}$ will be implemented using the circuit identity in Fig.~\ref{fig:UUdag_from_K}, in which only the $K_j$ operators need to be controlled.
If one finds $l$ such groups, instead of having to control each term in $\me^{-i \tau H}$, the control-free version of QETU requires only an additional $\mathcal{O}(l)$ controlled operations.

We now explain how to prepare $V$ for a general theory where the Hamiltonian is a sum of a kinetic and potential piece.
Furthermore, we assume that the basis in which the kinetic piece is diagonal is related to the basis in which the potential piece is diagonal by a Fourier transformation.
Under these assumptions, the Hamiltonian is written as $H = H_x + \text{FT}^\dagger H_p \text{FT}$, where $H_x$ and $H_p$ are the potential and kinetic components, respectively, and $\text{FT}$ is the Fourier transform.
Note that $H_x$ and $H_p$ as written are diagonal matrices.
A single Trotter step is implemented as $U(\delta \tau) = \text{FT}^\dagger \me^{-i \delta \tau H_p} \text{FT} \me^{-i \delta \tau H_x}$.
The procedure we show below can be used for any system of this type, including the U(1) gauge theory described in Sec.~\ref{sec:u1}.

Because the controlled implementation of a product of unitary operators is equal to the product of the controlled versions of the individual operators, we only need to consider a single Trotter step; the total control-free time evolution operator can then be built from combining multiple Trotter steps.
First, we show that the Fourier transform does not need to be controlled.
We denote by $U_\text{ctrl}(\delta \tau)$ the controlled version of a single Trotter step. 
Explicit evaluation gives
\begin{equation}
    \begin{split}
    U_\text{ctrl}(\delta t) &= \left( \unit \otimes \ket{0} \bra{0} + \text{FT}^\dagger \me^{-i \delta t H_p} \text{FT} \otimes \ket{1} \bra{1}\right) \left(\unit \otimes \ket{0} \bra{0} + \me^{-i \delta t H_x } \otimes \ket{1} \bra{1} \right)
    \\
    &= \left( \text{FT}^\dagger \text{FT} \otimes \ket{0} \bra{0} + \text{FT}^\dagger \me^{-i \delta t H_p} \text{FT} \otimes \ket{1} \bra{1}\right) \left(\unit \otimes \ket{0} \bra{0} + \me^{-i \delta t H_x } \otimes \ket{1} \bra{1} \right)
    \\
    &= \left( \text{FT}^\dagger \otimes \unit \right) \left(  \unit \otimes \ket{0} \bra{0} +  \me^{-i \delta t H_p} \otimes \ket{1} \bra{1}\right) \left( \text{FT} \otimes \unit \right) \left(\unit \otimes \ket{0} \bra{0} + \me^{-i \delta t H_x } \otimes \ket{1} \bra{1} \right)\,, 
\end{split}
\end{equation}
where we used $\text{FT}^\dagger \text{FT} = \unit$. From this we see that the Fourier transform does not need to be controlled.

Next, using the following circuit identity
\begin{figure}[h]
    \centering
    \begin{quantikz}[row sep={0.75cm,between origins}, column sep=0.4cm]
        \qw & \ctrl{1} & \octrl{1} & \qw
        \\
        \qw & \gate{U_1(\delta \tau) U_2(\delta \tau)}  & \gate{U_1(-\delta \tau) U_2(-\delta \tau)} & \qw
    \end{quantikz}
    \quad 
    =
    \begin{quantikz}[row sep={0.75cm,between origins}, column sep=0.4cm]
        \qw & \ctrl{1} & \octrl{1} & \ctrl{1} & \octrl{1} & \qw
        \\
        \qw & \gate{U_1(\delta \tau)} & \gate{U_1(-\delta \tau)} & \gate{U_2(\delta \tau)} & \gate{U_2(-\delta \tau)} & \qw
    \end{quantikz},
\end{figure}
\FloatBarrier
and replacing $U_1(\delta \tau) = U_x(\delta \tau) \equiv \me^{-i \delta t H_x}$ and $U_2(\delta \tau) =  \text{FT}^\dagger U_p(\delta \tau) \text{FT} \equiv \text{FT}^\dagger \me^{-i \delta t H_p} \text{FT}$, we find 
\begin{figure}[h]
    \centering
    \begin{quantikz}[row sep={0.75cm,between origins}, column sep=0.4cm]
        \qw & \ctrl{1} & \octrl{1} & \qw
        \\
        \qw & \gate{U(\delta \tau)}  & \gate{U(-\delta \tau)} & \qw
    \end{quantikz}
    =
    \begin{quantikz}[row sep={0.75cm,between origins}, column sep=0.4cm]
        \qw & \ctrl{1} & \octrl{1} & \qw & \ctrl{1} & \octrl{1} & \qw & \qw
        \\
        \qw & \gate{U_x(\delta \tau)} & \gate{U_x(-\delta \tau)} & \gate{\text{FT}^\dagger} & \gate{U_p(\delta \tau)} & \gate{U_p(-\delta \tau)} & \gate{\text{FT}} & \qw
    \end{quantikz}.
\end{figure}
\FloatBarrier
Note that $H_x$ and $H_p$ are both diagonal matrices. If we can find a general procedure for decomposing a general diagonal unitary matrix into $l$ groups $H^{(j)}$ that all anticommute with the same $K_j$ Pauli operator, we will have a control-free implementation of QETU for Hamiltonians of this form. 
We now present a general method applicable to arbitrary Hamiltonians; applying this method to the specialized case of diagonal Hamiltonians is straightforward.

A general Hamiltonian $\tilde{H}$ acting $n$ qubits can be decomposed into a sum of $4^n$ Pauli strings containing $\unit, X, Y, Z$ matrices 
\footnote{Note that in a practical setting, to implement the exponential of an arbitrary Hamiltonian, one will first split the Hamiltonian into groups of commuting Pauli strings. Constructing the $V$ operator for control-free QETU will then require using our method for each set of commuting Pauli strings}.
The method for choosing the $H^{(j)}$ groups and their associated $K_j$ Pauli strings contains two main steps.
The first step considers only Pauli strings containing $\unit$ and $Z$ matrices.
The process starts by defining $H^{(1)}_Z$ as the sum of all Pauli strings with a $Z$ gate acting on the first qubit.
The associated $K^{(Z)}_1$ operator is simply an $X$ gate acting on the first qubit.
From the remaining pool of Pauli strings we now define $H^{(2)}_Z$ as a sum of all terms with a $Z$ gate acting on the second qubit, and $K^{(Z)}_2$ is an $X$ gate acting on the second qubit.
By repeating this method for all $n$ qubits, we will construct $n$ such groups $H_Z^{(j)}$ and their associated $K^{(Z)}_j$ operators (if the Hamiltonian $\tilde{H}$ is diagonal, one can stop here and implement the control-free version of QETU).

The second step focuses on the remaining Pauli strings that contain only $\unit, X, Y$ matrices.
In a similar way, we start by defining $H^{(1)}_{XY}$ as the sum of all Pauli strings with either an $X$ or $Y$ gate acting on the first qubit.
Because $Z$ anti-commutes with both $X$ and $Y$, the associated $K_1^{(XY)}$ operator is simply a $Z$ gate acting on the first qubit.
From the remaining pool of operators, we can now choose $H^{(2)}_{XY}$ as a sum of all terms with either an $X$ or $Y$ gate acting on the second qubit, and $K_2^{(XY)}$ will be a Z gate acting on the second qubit.
After repeating this for all $n$ qubits, we will have $n$ groups $n$ groups $H_{XY}^{(j)}$ and their associated $K_j^{(XY)}$ operators.
Using this method, the operator $V$ needed to implement the control-free version of QETU for arbitrary Hamiltonians $\tilde{H}$ can therefore be implemented with the same cost as $e^{i \tau \tilde{H}}$ plus an extra $\mathcal{O}(n)$ CNOT gates.
For many physical systems of interest, this additional cost is negligible compared to implementing the uncontrolled time evolution circuit.

As a demonstration of this method, we work through a two qubit example considering a diagonal Hamiltonian. 
A general two qubit diagonal Hamiltonian can be written as 
\begin{equation}
    \tilde{H} = a_{0} \unit \otimes \unit + a_{1}\unit \otimes Z + a_{2} Z \otimes \unit + a_{3} Z \otimes Z\,,
\end{equation}
for general coefficients $a_i$. The term associated with $a_0$ is a global phase, and can be ignored for this argument. 
Using the previously described method, the groups $H^{(j)}$ are given by
\begin{align}
    H^{(1)} &= a_{2} Z \otimes \unit + a_{3} Z \otimes Z\,.
    \\
    H^{(2)} &= a_{1} \unit \otimes Z\,,    
\end{align}
For this choice of groupings, the associated $K_j$ operators are 
\begin{alignat}{9}
    K_1 &= X \otimes \unit\,&.
    \\
    K_2 &= \unit \otimes X\,&,
\end{alignat}
With these choices, the circuit for the $V$ operator is 
\begin{figure}[h]
    \centering
    \begin{quantikz}[row sep={0.55cm,between origins}, column sep=0.4cm]
        \qw & \ctrl{1} & \octrl{1} & \qw
        \\
        \qw & \gate[2]{\me^{-i \delta \tau \tilde{H}}}  & \gate[2]{\me^{i \delta \tau \tilde{H}}} & \qw
        \\
        \qw & & & \qw
    \end{quantikz}
    \quad
    =
    \begin{quantikz}[row sep={0.55cm,between origins}, column sep=0.4cm]
        \qw & \octrl{1} & \qw & \octrl{1} & \octrl{1} & \qw & \octrl{1} & \qw
        \\
        \qw & \gate[2]{K_1} & \gate[2]{\me^{-i H^{(1)} \delta \tau}} & \gate[2]{K_1} & \gate[2]{K_2} & \gate[2]{\me^{-i H^{(2)} \delta \tau}} & \gate[2]{K_2} & \qw
        \\
        \qw & & & & \qw & & & \qw
    \end{quantikz}
    \\[6ex]
    =\begin{quantikz}[row sep={0.55cm,between origins}, column sep=0.4cm]
        \qw & \octrl{1} & \qw & \octrl{1} & \octrl{2} & \qw & \octrl{2} & \qw
        \\
        \qw & \gate{X} & \gate[2]{\me^{-i H^{(1)} \delta \tau}} & \gate{X} & 
        \qw & \gate[2]{\me^{-i H^{(2)} \delta \tau}} & \qw & \qw
        \\
        \qw & \qw & \qw & \qw & \gate{X} & \qw & \gate{X} & \qw
    \end{quantikz},
\end{figure}
\FloatBarrier
\noindent which requires an additional 4 CNOT gates, instead of controlling on the entire time evolution operator $e^{-i \delta \tau \tilde{H}}$.

\end{document}